\documentclass[a4paper,fleqn,usenatbib,useAMS]{mnras}
\usepackage{color}
\usepackage{amsmath}	
\usepackage{amssymb}	
\usepackage{multicol}        
\usepackage{bm}		
\usepackage{textcomp,amssymb}
\usepackage{leftidx}
\usepackage{array}
\usepackage{supertabular}
\usepackage{hhline}
\usepackage{multirow}
\usepackage{balance}
\usepackage{times}
\usepackage{graphicx}
\usepackage{soul}
\usepackage{bm}
\usepackage{balance}
\usepackage[draft=true]{hyperref} 

\newcommand{\hagn}{\mbox{{\sc \small Horizon-AGN\,\,}}}

\def \apj {{\it ApJ}}
\def \apjl {{\it ApJ Let.}}
\def \apjs {{\it ApJ Sup.}}
\def \pasj {{\it PASJ}}

\def \aap {{\it A\&A}}

\def \mnras {{\it MNRAS}}
\def \nat {{\it Nature}}

\definecolor{lblue}{rgb}{0.1,0.7,1.}
\definecolor{grey}{rgb}{0.75,0.75,0.75}
\definecolor{Orange}{rgb}{1.0,0.5,0.15}
\definecolor{brown}{rgb}{0.7,0.25,0.0}
\definecolor{pink}{rgb}{1.0,0.5,0.5}
\definecolor{darkerred}{rgb}{0.8,0,0}
\definecolor{darkerblue}{rgb}{0,0,0.8}
\definecolor{Blue}{rgb}{0,0.08,0.65}
\definecolor{Red}{rgb}{0.65,0.08,0.05}
\definecolor{Green}{rgb}{0.35,0.45,0.25}

\begin{document}

\author[C. Laigle, C.~Pichon, S.~Arnouts, et al.]{
\parbox[t]{\textwidth}{C. Laigle$^{1}$\thanks{E-mail: clotilde.laigle@physics.ox.ac.uk}, C.~Pichon$^{2,3}$,  
  S.~Arnouts$^{4}$, H.~J.~McCracken$^{2}$,   Y.~Dubois$^{2}$,
  J.~Devriendt$^{1}$, 
  A.~Slyz$^{1}$, 
   D.~Le Borgne$^{2}$,
    A.~Benoit-L{\'e}vy$^{2}$,
    Ho~Seong~Hwang$^{5}$,
    O.~Ilbert$^{4}$,
      K.~Kraljic$^{4}$, \\
  N.~Malavasi$^{6}$,
Changbom~Park$^{3}$ and D.~Vibert$^{4}$
}
\vspace*{6pt} \\ 
$^{1}$ Sub-department of Astrophysics, University of Oxford, Keble Road, Oxford OX1 3RH\\
$^{2}$ Sorbonne Universit{\'e}s, UPMC Univ Paris 6 et CNRS, UMR 7095, Institut d'Astrophysique de Paris, 98 bis bd Arago, 75014 Paris \\
$^{3}$ School of Physics, Korea Institute for Advanced Study (KIAS), 85 Hoegiro, Dongdaemun-gu, Seoul, 02455, Republic of Korea\\
$^{4}$ Aix Marseille Universit\'e, CNRS, Laboratoire d'Astrophysique de Marseille, UMR 7326, 13388, Marseille, France\\
$^{5}$ Quantum Universe Center, Korea Institute for Advanced Study, 85 Hoegiro, Dongdaemun-gu, Seoul 02455, Republic of Korea\\
$^{6}$ Department of Physics and Astronomy, Purdue University, 525 Northwestern Avenue, West Lafayette, IN 47907, USA
}
\date{Accepted . Received ; in original form }

\title[The impact of filaments on galaxy properties with COSMOS2015]{
COSMOS2015 photometric redshifts probe  the impact of filaments on galaxy properties
  }

\maketitle

\begin{abstract}
The variations of  galaxy stellar masses and  colour-types with the distance to  projected cosmic  filaments are quantified  using the precise photometric redshifts  of the {COSMOS2015} catalogue extracted from COSMOS field (2 deg$^{2}$). Realistic   mock catalogues
are  also extracted  from the lightcone of the cosmological hydrodynamical simulation {\sc Horizon-AGN}. They show that the  photometric redshift accuracy of the observed catalogue ($\sigma_z<0.015$ at $M_*>10^{10}{\rm M}_{\odot}$ and $z<0.9$) is sufficient to provide 2D filaments that closely match  their projected 3D counterparts.  Transverse stellar mass gradients are measured in projected slices of thickness 75 Mpc between $0.5< z <0.9$,   showing that the most massive galaxies are statistically closer to their neighbouring filament.  
At  fixed stellar mass, passive galaxies are also found  closer to  their filament while active star-forming galaxies statistically lie further away. The contributions of nodes and  local density are removed from these gradients to highlight the specific  role  played by the geometry of the filaments. We find that the measured signal  does persist after this removal, clearly demonstrating that proximity to a filament is not equivalent to proximity to an over-density.
 These findings are in agreement with  gradients measured both in 2D or 3D in the {\sc Horizon-AGN} simulation and those observed in the  spectroscopic VIPERS survey (which rely  on  the identification of 3D filaments). They  are consistent with a picture in which the influence of the geometry of the large-scale environment drives anisotropic tides which  impact the  assembly history of galaxies, and hence their observed properties. 
\end{abstract}


\section{Introduction}
%
A challenge for current galaxy formation models is to understand the role of the environment, its anisotropy and its dynamics in shaping the properties of galaxies across cosmic time.  Galaxies form in the gravitational potential wells of dark matter haloes.  Their mass assembly can thus be approached at first order by the gravitational history of halo assembly \citep{Blumenthal1984,NFW1995}, while at a second order, the evolution of  baryonic matter (gas accretion and cooling, star formation and feedback) induces  additional levels of complexity \citep[e.g.][and reference therein]{Hopkins2014}. 
Starting from the dark matter only, the theory of cosmological structure formation predicts a strong dependency of  halo mass assembly on its environment, driven by  gravitational physics. 

On the  one hand,  halo mass assembly history is  encoded in the initial conditions of the matter that will end up in the halo (its so-called initial Lagrangian patch). At the simplest level, it is a function of its mass which determines its collapse time encoded in the spherical collapse model \citep{pressetschechter1974}.  
An environmental dependency will occur when also considering the impact of
 the long wavelength density mode corresponding to the large-scale cosmic web.
As encompassed  by the peak-background split theory \citep{ShethTormen1999}, a large-scale overdensity induces a density boost, allowing the proto-halo to pass earlier the critical threshold of collapse and biasing the mass function in the vicinity of the  large-scale structure  \citep[shifted Press-Schechter,][]{Kaiser1984,efstathiouetal1988,ColeetKaiser1989,bondetal1991,Moetwhite1996,shethettormen1999}: the abundance of massive halos will thus be  enhanced in overdense regions \citep{Alonso2015}, and these objects will be more strongly clustered. An observational consequence of this prediction is the mass-density relation \citep{oemler74,davisetgeller1976,baloghetal1997}, where the local density at a given redshift remains a tracer of the initial density boost. 

On the other hand, in complement to the local density, the initial geometric distribution of neighbouring structures will  also explicitly impacts the accretion and merger history of the proto-halo via the tides it applies onto the patch. It has been found that at a given mass, denser environments are on average populated by older haloes, where the age of a halo is defined as the epoch by which the halo has assembled one half of its current mass. This effect is called ``assembly bias" \citep[e.g.][]{shethettormen2004,gaoetal2005,wechsleretal2006,crotonetal2007} 
and can be interpreted as follows. At a given mass, if the halo is isolated, it can continue to grow by accreting matter from its neighbourhood, via  the accretion of larger shells in the initial density field. If this halo lies in the vicinity of a massive structure, it will stop to grow earlier because the matter will recede towards that massive structure. Hence the large-scale structure around the proto-halo  will affect the halo history via external tides \citep[e.g.][]{dalaletal2008,Hahnetal2009,wangetal2011} captured by the ellipsoidal collapse model \citep{bondetmyers1996}. Those tides span a large range of scales. At small scale they may induce mass losses for haloes in the vicinity of a more massive halo, while on larger scale they imply statistically distinct mass assembly histories in different large-scale environment \citep[see also][]{Castorina:2016vg}, e.g. by delaying or quenching the mass inflow.
 An important  ingredient at large scale is the anisotropic structure of the matter distribution, which is organised as the cosmic web \citep{bondetal1996},
made of walls, filaments and nodes. The shape of the tidal tensor\footnote{Specifically, the traceless part of the tidal tensor, which orientation via its eigenvectors
 points along the main axes of the  embedding cosmic web. Recall that gradients of the potential displace halos, while the tides distort and rotate them
 along the directions set by those vectors. } will be precisely determined by this anisotropy. It will affect not only the accretion history of the proto-halo as described above, but also its dynamics via tidal torquing.

  This dynamical impact of the anisotropic cosmic web has  been investigated in dark matter simulations in relation to spin acquisition: \cite{Aubertetal200,aragoncalvoetal2007,hahnetal2007,codisetal2012,aragoncalvoetal2013}  report  angular momentum alignment of low-mass haloes with  filaments and a perpendicular orientation for more massive haloes. This  alignment can be understood when measuring the vorticity content in walls and filaments inherited from  flow-crossing \citep{libeskind13a,laigle2015}. 
  Structures which are forming in these large scale flows will end up with a spin preferentially aligned with that vorticity and consequently with the filament. They will accrete matter through secondary infall with a coherent rotational motion, up to a specific transition mass corresponding to the Eulerian size of the quadrant, beyond which inflow will start to advect misaligned angular momentum.  
  This dynamical connection can also be predicted from the initial conditions and its variation with  the geometry of the environment is now  well established in the context of the constrained tidal torque theory \citep{codisetal2015}.  Interestingly, this dependency still holds for galaxies,  despite  the scale difference and the more complex baryonic physics  in the circumgalactic-medium. \cite{Tempel13} have observed the  alignment of galaxy's spin with their closest filament in the SDSS survey \citep[see also e.g. ][]{hirvetal2016} and \cite{duboisetal14} have first confirmed this dependency for galaxies in hydrodynamical simulations \citep[see e.g. also][]{codisetal2014, chisarietal2015,chisarietal2016,gonzalezetal2017}.
To summarise, it is  expected that the large scale environment impacts both the mass  and spin assembly history of the host haloes. Observations and simulations have shown that the dynamical dependency  on the large-scale environment remains true also for galaxy spin. The key question now is: does it also affect scalar quantities such as the mass, colour and sSFR of galaxies?

 If the effect of the tidal field on accretion is not completely erased by baryonic physics and feedback, the physical properties which depend on the accretion history should therefore be impacted by the corresponding anisotropic infall, and a specific dependency of galaxy masses, colours and star-formation rates on the geometry of the environment should be observable.  However very few observational evidences sustain the prediction that the geometry of the environment impacts significantly  galaxy accretion history.  
\\
The  ``galactic conformity" \citep{weinmannetal2006} is a first measured evidence of this effect. It quantifies a correlation between the quenching of the central and the quenching of its satellite galaxies: the fraction of quiescent satellites is higher around quiescent centrals than around star forming centrals. 
This effect has been detected up to redshift 2.5 for both low and high mass satellite galaxies \citep{kawinetal2016}. 
It suggests that galaxy properties depend not only on  halo mass but also on the larger scales, and is likely connected to the halo assembly history. \cite{hearingetal2015} observed this galactic conformity up to very large separation (4~Mpc): this effect is difficult to explain in semi-analytical models where galaxy properties depend essentially on halo mass. Attempting to make sense of it, they use the term of ``halo accretion conformity" and connect this effect to the mutual evolution of haloes in the same large-scale tidal field. 
The effect of these large scale tides  are not captured by measurements of the local density. 
\\
More recently, results from spectroscopic surveys have reported the dependencies of galaxy masses and types on the geometry of the environment  using SDSS \citep[][]{Yan2013,martinezetal2016,poudeletal2016,chenetal2017}, GAMA \citep[][]{alpaslanetal2016} and, at higher redshift, VIPERS \citep[][]{Malavasi2016b},  based on the three-dimensional  reconstruction of the cosmic web. In particular, \cite{Malavasi2016b} find a significant trend for galaxies with different stellar masses and type to segregate near the filaments: the most massive and passive galaxies are found closer to the filament center than the star-forming, low mass ones.  Can photometric redshift surveys  unravel such signal at higher redshift?
 
The recently achieved accuracy of current photometric redshifts opens the prospect of working now with projected two-dimensional slices which are sufficiently thin to unambiguously study the  projected filaments. One of the assets of photometric surveys is their ability to probe different epochs of cosmic evolution to leverage their relative importance in building up galaxies, with complete sample  at much lower cost than spectroscopic surveys.  The potential of photometric surveys  may have been  initially underestimated because of their insufficient precision along the line-of-sight, making it seemingly too difficult to reconstruct the three-dimensional field at the required scale to trace the cosmic web.  
With the large and deep multi-wavelength surveys such as the Cosmological Evolution Survey\footnote{\url{http://cosmos.astro.caltech.edu/}} \citep[COSMOS;][]{scovilleetal2007}, the upcoming Hyper Suprime-Cam survey \citep[HSC;][]{miyazakietal2012} and in the long term the Large Synoptic Survey Telescope \citep[LSST;][]{ivezicetal2008}, it is timely and  interesting to estimate the requested accuracy to achieve this goal via simulations; it is also timely to exploit the latest {COSMOS2015} catalogue \citep{Laigle2016} 
as a prototype of these upcoming surveys to try and detect the dependencies of galactic properties such as stellar mass and galaxy colours with respect to their position within the  cosmic web. Previous studies on COSMOS have investigated the effect of local environment on galaxy properties, either relying on isotropic density estimators \citep[e.g.][]{scovilleetal2013,darvishetal2016b}, or dividing the field in morphological components \citep[field, filament or cluster;][]{darvishetal2016}. 
However, we stress that our aim in this work is  to investigate the specific impact of the  anisotropic tides  on galaxy properties, therefore disentangling explicitly the effect of the geometry of the large-scale environment on driving these tides (i.e the traceless part of the tidal tensor) from  effects purely driven by the local density (i.e. the trace of the tidal tensor). 
In order to achieve this goal,  we will focus on mass and colour gradients specifically towards filaments.

The paper is organised as follows. Section~\ref{Sec:dataset} describes the observed and simulated datasets, and
 the tool to extract the density and the skeleton. Section~\ref{Sec:Qualification} assesses the reliability of  the skeleton reconstruction using photometric redshifts, while relying on the realistic virtual photometric mocks built from the \hagn hydrodynamical cosmological simulation~\citep{duboisetal14}. Section~\ref{Sec:Results} presents results from  the data and our interpretation. Section~\ref{Sec:Conclusion}
wraps up and outlines future works. Appendix~\ref{Ap:persistence}, \ref{Ap:Dmgal}, \ref{Ap:thickness} investigate how the reconstruction and the signal vary as a function of the thickness of the slice, the photometric redshift accuracy and the  completeness of the sample, hence providing guidelines for computing a similar reconstruction in surveys with different redshift uncertainties. Appendix~\ref{Ap:robdens} assesses the robustness of the gradients towards filaments when removing the contribution of the nodes. 
We use a standard $\Lambda$CDM cosmology with Hubble constant $H_{0}=70.4$ km$\cdot$s$^{-1}\cdot {\rm Mpc}^{-1}$, total matter density $\Omega_{\rm m}=0.272$ and dark energy density $\Omega_{\Lambda}=0.728$. All magnitudes are expressed in the AB system. Throughout this work, uncertainties on the distributions are computed from bootstrap resampling. The shaded area around the distributions correspond  to 1$\sigma$ deviation to the mean for a bootstrap resampling with 100 realisations.   
%
%
\section{Datasets and methods}
\label{Sec:dataset}
%
Before analyzing the relative distribution of galaxies within the projected two-dimensional cosmic web, let us first introduce the {COSMOS2015} photometric catalogue, the mock catalogue extracted from the \hagn simulation, and the ridge tracing algorithm used to trace  the cosmic web.
%
%
\subsection{Photometric catalogue from COSMOS}
\label{Sec:COSMOS}
%
Let us first summarise the main properties of the {COSMOS2015} photometric catalogue that we built in \citet{Laigle2016}.  This catalogue includes apparent 
magnitudes in 30 bands from ultra-violet (UV) to infra-red (IR): 0.25--8\,$\mu$m. 
In optical, it contains the same broadband images as previous releases \citep{capak2007,ilbert2009}.
The COSMOS-20 survey provides most of the optical coverage: six broad bands: $B,V,r,i,z^{++}$, twelve medium bands and two narrow bands taken with Subaru Suprime-Cam \citep{taniguchietal2007,taniguchietal2015}. The $u^{*}$-band data is obtained from the Canada-Hawaii-France Telescope (CFHT/MegaCam). Intermediate bands are very useful to increase the spectral resolution of the measured spectral energy distribution (SED) in the apparent optical wavelength range, contributing in this way to the accuracy of photometric redshifts especially at low redshift \citep[e.g.][]{cardamoneetal2010}.
In the near-infrared (NIR), we rely on the new $Y,J,H,K_s$ images from the UltraVISTA survey \citep[DR2,][]{mccracken12} and the $Y$ band of the Hyper Suprime-Cam at Subaru telescope \citep{miyazaki2012}. The photometry is extracted using {\sc SExtractor} \mbox{\citep{bertinetal96}} in dual image mode. Following a similar reduction procedure as described in \citet{mccracken12}, the detection image is a $\chi^{2}$ sum of the four NIR images of UltraVISTA DR2 and the $z^{++}$ band taken with Subaru Suprime-Cam. This combination ensures to increase the completeness of the catalogue.
Mid-IR data in the four IRAC channels (i.e.~in a wavelength range between $\sim3$ and $8\,\mu$m) come from the SPLASH program (PI: Capak). 
\\
Photometric redshifts ($z_\mathrm{phot}$) are computed using {\sc LePhare} 
\citep{arnoutsetal2002,ilbert06} with a configuration similar to \citet{ilbert13}.  In particular, the principal nebular emission lines are  implemented  using an empirical relation between the UV
light and the emission line  fluxes as described in \cite{ilbert2009}.
From a comparison with the spectroscopic samples zCOSMOS bright \citep{lilly07} 
and faint (Lilly et al., in prep.) the uncertainty of our photometric redshifts turns out to be  
$\sigma_{\Delta z}/(1+z) =0.007$ at $i^{+} <22.5$ and $\sigma_{\Delta z}/(1+z) =0.03$ at $1.5<z<4$ .
Stellar masses, absolute magnitudes, and star formation rates (SFR) are also derived with the {\sc LePhare} code.   
\\
We remove also from the catalogue all the objects which are flagged as belonging to a bad area  or for which the photometry is possibly contaminated by the light of saturated stars, meaning that we keep only objects in $\cal{A}^{\rm UVISTA}$\&$\cal{A}^{\rm !OPT}$\&$\cal{A}^{\rm COSMOS}$, according to the notations in Table 7 of \cite{Laigle2016}.   
\\
Passive galaxies are identified using their locations in the colour-colour plane $NUV$-$r$/$r$-$J$ \citep{williamsetal2009,ilbert13}. Passive objects are identified as those with
$M_{NUV}-M_{r}>3(M_{r}-M_{J})+1$ and $M_{NUV}-M_{r}>3.1$ \citep[Figure 16 in][]{Laigle2016}.   In
particular, this technique avoids mixing the red dusty galaxies and 
passive ones. This criterion is expected to be more robust than a segregation based on the SFR derived from the SED-fitting. In the following, we call galaxy colour-type the galaxy type (passive or star-forming) determined based on the colour diagram. 
\\
The stellar mass completeness of the sample is estimated from the $K_{\rm s}$ magnitude following \cite{pozzettietal2010}. The stellar mass limits up to redshift 1.3 are shown in Table~\ref{Tab:thick} for the entire population of galaxies. In the following, our results are based on a galaxy sample with $M_{\rm lim}>10^{10} {\rm M}_{\odot}$ and $0.5<z<0.9$. This choice is justified in Section~\ref{subsec:slice}. The final galaxy catalogue contains 11 284 objects.
%
 \begin{table*}
\begin{center}
\begin{tabular}{|c|cc|cc|cc|}
\hline
bin   & $M_{\rm lim}$&  $D_{\rm trans}$ 
       & $\Delta z \left[ M_*\sim 10^{10} \right]$
      & $D_{\Delta z} \left[ M_*\sim 10^{10}\right]$ & $\Delta z \left[ D=75{\rm Mpc}\right]$ & $D_{\Delta z}$\\
      
 & &   (Mpc) 
      & ($2\times 1\sigma$)
      &  (Mpc) & & (Mpc)
       \\
  \hline
  0.5$< z <$0.6  & 8.72 & 54   & 0.019 & 58.3 & 0.024 & 75\\
 0.6$< z <$0.7 & 8.92 & 62 & 0.021 & 61.6  & 0.025 & 75\\
 0.7$< z <$0.8   & 8.95 & 70  & 0.024 & 68.0  & 0.027 & 75\\
 0.8$< z <$0.9 & 9.17 & 75  & 0.029& 75.2 & 0.029 & 75 \\
0.9$< z <$1.0  & 9.19 & 82  & 0.032 & 80.4 & -- &  --\\
1.0$< z <$1.1   & 9.23 & 87  &  0.045 & 105.1 & -- & -- \\
1.1$< z <$1.2  & 9.37  & 93  &0.059 & 131.4 & -- &  --\\
1.2$< z <$1.3 & 9.44 & 98 & 0.059 & 128.1 & -- & -- \\
     \hline
\end{tabular}
  \caption{
Limiting mass ($M_{\rm lim}$), estimation of the comoving transverse width of the slice ($D_{\rm trans}$) , redshift errors  (${\Delta z} \left[ M_*\sim 10^{10}\right]$) and corresponding thickness in Mpc  ($D_{\Delta z} \left[ M_*\sim 10^{10}\right]$), and chosen thickness of the slices in redshift ($\Delta z \left[ D=75{\rm Mpc}\right]$) and in Mpc ($D_{\Delta z}$).  
 The limiting mass is estimated as in \protect\cite{Laigle2016}. The comoving transverse distance is estimated at the upper limit of each redshift bin and corresponds to 1.4~deg (side of the COSMOS field). $M_*>10^{10}{\rm M}_{\odot}$ is the conservative limiting mass that we choose in this work to build a mass-selected sample at $0.5<z<0.9$. $\Delta z \left[M_*\sim 10^{10} M_{\odot}\right]$ corresponds to twice the median $1\sigma$ redshift error determined by {\sc LePhare}  for the galaxies with masses $10^{10}  <M_*/M_{\odot}< 10^{10.5}$ (the faintest galaxies in the bin, for which the redshift uncertainties are the highest). $D_{\Delta z}\left[M_*\sim 10^{10} M_{\odot}\right]$ is the corresponding thickness in comoving Mpc at the upper limit of the redshift bin.  Finally, ${\Delta z}\left[ D=75{\rm Mpc}\right]$ is the maximum in each redshift bin of the redshift thicknesses chosen in this work corresponding to $D_{\Delta z}=75$ Mpc. It has been chosen such as to encompass two times the median redshift error in the lowest mass bin at $z\sim 0.9$.
}
\label{Tab:thick}
\end{center}
\end{table*}
%
\subsection{Virtual  catalogue from \hagn}
%
The {\sc Horizon-AGN}\! simulation\footnote{\url{http://www.horizon-simulation.org/}} \citep[see][for more details]{duboisetal14} is run with a $\Lambda$CDM cosmology with total matter density $\Omega_{\rm  m}=0.272$, dark energy density $\Omega_\Lambda=0.728$, amplitude of the matter power spectrum $\sigma_8=0.81$, baryon density $\Omega_{\rm  b}=0.045$, Hubble constant $H_0=70.4 \, \rm km\,s^{-1}\,Mpc^{-1}$, and $n_s=0.967$ compatible with the WMAP-7 data~\citep{komatsuetal11}.
The simulation is run with the {\sc ramses} code~\citep{teyssier02}, and the initially coarse $1024^3$ grid is adaptively refined down to $\Delta x=1$ proper kpc, with refinement triggered in a quasi-Lagrangian manner: if the number of dark matter particles becomes greater than 8, or the total baryonic mass reaches 8 times the initial dark matter mass resolution in a cell. The size of the simulation box is $L_{\rm  box}=100 \, h^{-1}\rm\,Mpc$ on a side, and the volume contains $1024^3$ dark matter particles, corresponding to a dark matter mass resolution of $M_{\rm  DM, res}=8\times 10^7 \, \rm M_\odot$.
\\
Heating of the gas from a uniform UV background takes place after redshift $z_{\rm  reion} = 10$ following~\cite{haardt&madau96}. 
Gas can cool down to $10^4\, \rm K$ through H and He collisions with a contribution from metals using rates tabulated by~\cite{sutherland&dopita93}. 
Star formation occurs in regions of gas number density above $n_0=0.1\, \rm H\, cm^{-3}$ following a Schmidt law: $\dot \rho_*= \epsilon_* {\rho_{\rm g} / t_{\rm  ff}}$,  where $\dot \rho_*$ is the star formation rate mass density, $\rho_{\rm g}$ the gas mass density, $\epsilon_*=0.02$ the constant star formation efficiency, and $t_{\rm  ff}$ the local free-fall time of the gas. The stellar mass resolution is $M_{{\rm*, res}}\simeq 2 \times 10^{6} {\rm M}_{\odot}$. The simulation follows
feedback from stellar winds, supernovae type Ia and type II, 
as well as the formation of black holes, which can grow by gas accretion at a Bondi-capped-at-Eddington rate and coalesce when they form a tight enough binary.
Black holes release energy either in quasar or radio mode when the accretion rate is respectively above and below one per cent of Eddington, with efficiencies tuned to match the black hole-galaxy scaling relations at $z=0$~\citep[see][for details]{duboisetal12agnmodel}.
\\
To better mimic observed surveys, a lightcone is produced on-the-fly at every coarse time step of the simulation.
  The lightcone generation is described in \cite{2010Pichon}. For the lightcone extraction, we have replaced gas cells by gas particles, and treated them as for the stars and dark matter particles. The area of the lightcone is 5 deg$^{2}$ below $z=1$, and 1 deg$^{2}$ above. 
\\
We identify galaxies using the {\sc AdaptaHOP} structure finder \citep{aubert04,tweed09}, applied to the distribution of star particles in the simulated lightcone. Because {\sc AdaptaHOP} works at fixed redshift, the extraction is done in thin slices of the lightcone, each slice beeing kept at fixed redshift. To avoid cutting galaxies, slices are overlapping, then duplicated galaxies are removed. 
Structures are selected using a local threshold of 178 times the average matter density, the local density of individual particles being computed from the 20 nearest neighbours. Only structures that have more than 50 particles are considered. We compute galaxy fluxes and magnitudes using the single stellar population (SSP) models from \citet[][]{bruzual&charlot03}, using a \cite{Salpeter1955} initial mass function. We assume that each star particle behaves as an SSP, and compute its contribution to the total SED by logarithmically interpolating the models in metallicity and age and multiplying by the initial mass of the particle. In this work attenuation by dust is not taken into account.
The computation of the photometry includes redshift distortions due to the galaxy peculiar velocities.
\\
 A critical assessment of the \hagn simulation in term of the evolution of  its galaxy populations is presented in \cite{Kaviraj16} through comparison of colours, luminosity functions and mass functions. When taking into account uncertainties and systematics, they found that the simulation captures reasonably well the evolutionary trend of galaxy mass assembly as seen through the statistical quantities.  The remaining point of tensions concern the high redshift Universe (underestimation of the mass function at the high mass end) and the low mass galaxies (overproduction of low mass galaxies in the simulation). These small discrepancies in these specific ranges of mass and redshift will not affect the conclusions drawn in our work.
The separation between star-forming and passive galaxies is also less straightforward in {\sc Horizon-AGN} than in data: a large fraction of passive galaxies are indeed moved towards the star-forming region in the colour diagram due to residual star formation, which is not completely extinguished in passive systems. For this reason relying on the colour diagram itself is not sufficient to properly separate passive and star-forming galaxies,  and an additional dynamical or morphological criterium could be required \citep[e.g. the $V/\sigma$ parameter, see][]{duboisetal16}. In this work we will rely on the specific star-formation-rate (sSFR) to characterise galaxy types. The sSFR in the simulation is computed from  the stellar mass formed over the last 100 Myr. The sSFR-mass relation is in broad agreement with the observations \citep{duboisetal16}, although the low-mass galaxies at low redshift are generally too passive, and the high-mass galaxies contain residual star formation.
 \\
We extract from the \hagn simulated lightcone (see Figure~\ref{fig:3d} below) a mock catalogue which has the same characteristics in terms of photometry and photometric redshift accuracy as {COSMOS2015}\!. 
In particular, we divide the simulated galaxies in bins of redshift and magnitude, and we perturb their redshift using a random gaussian error corresponding to the observed one in the same redshift and magnitude bins. This mock catalogue is used to test the two-dimensional skeleton reconstruction and our ability to recover mass gradients in two dimensions.
\begin{figure*}
\begin{center}
 \includegraphics[scale=0.51]{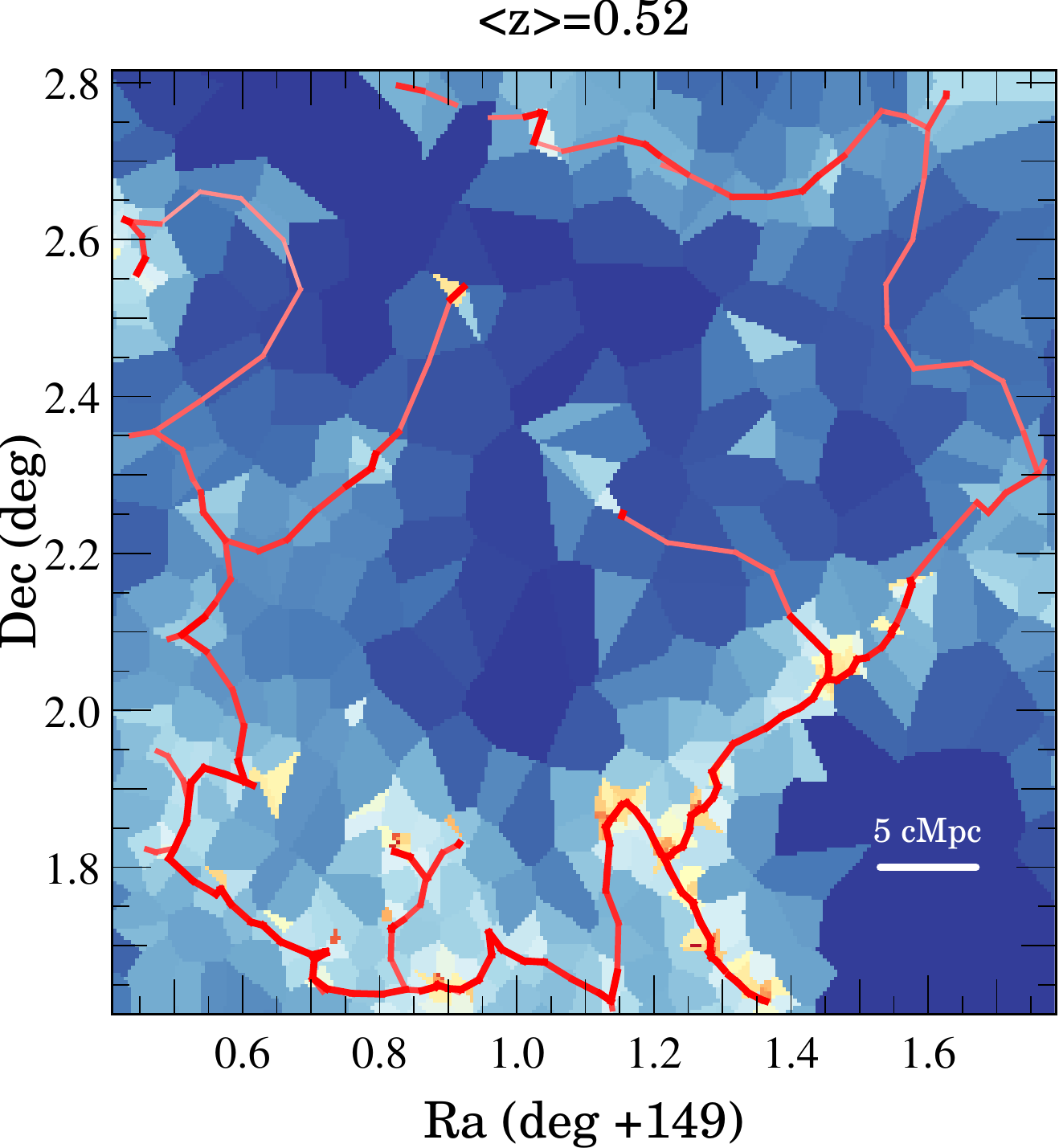}
 \hspace{0.2cm}
  \includegraphics[scale=0.51]{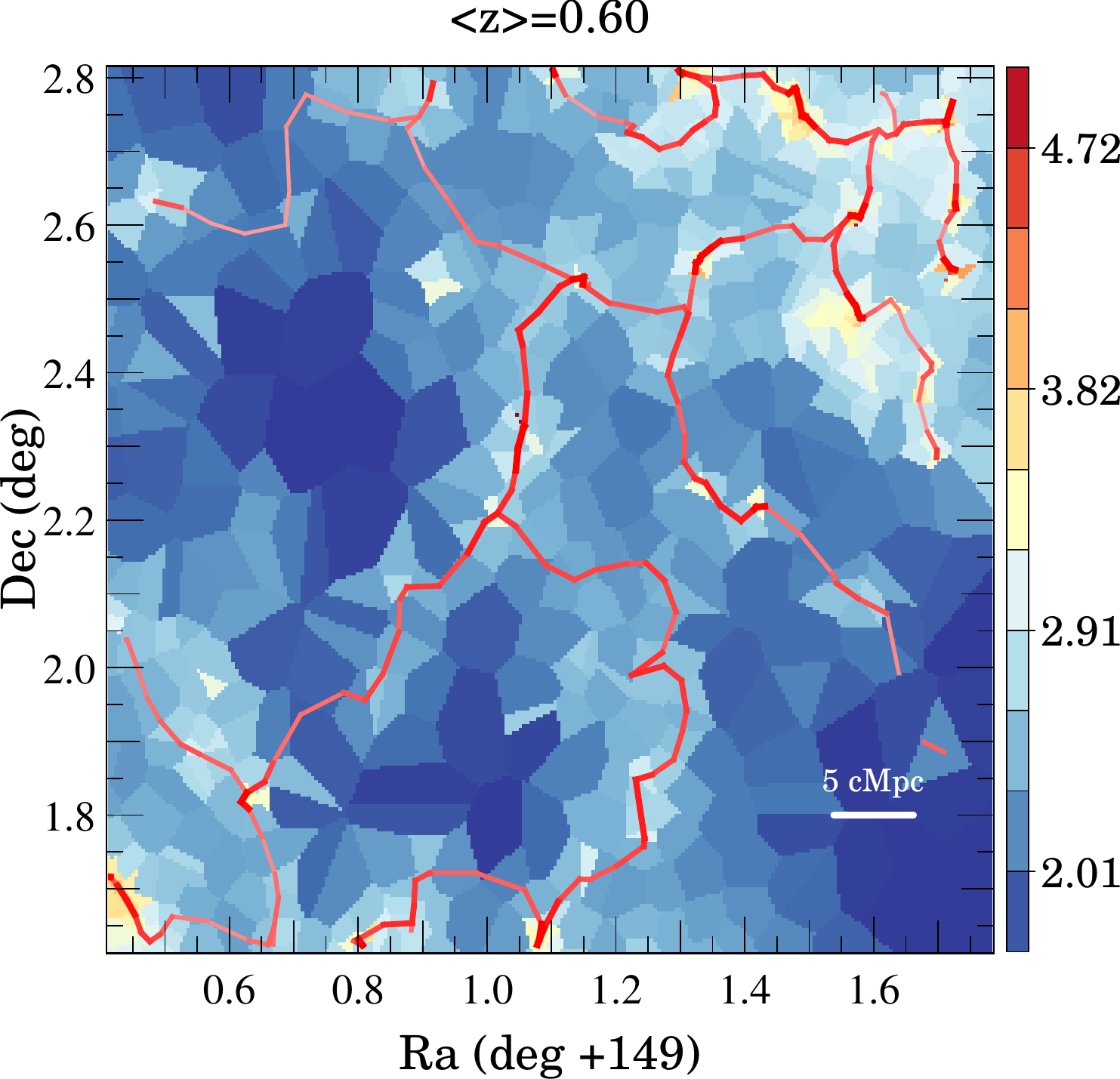}
    \par
   \vspace{0.4cm}
   \includegraphics[scale=0.51]{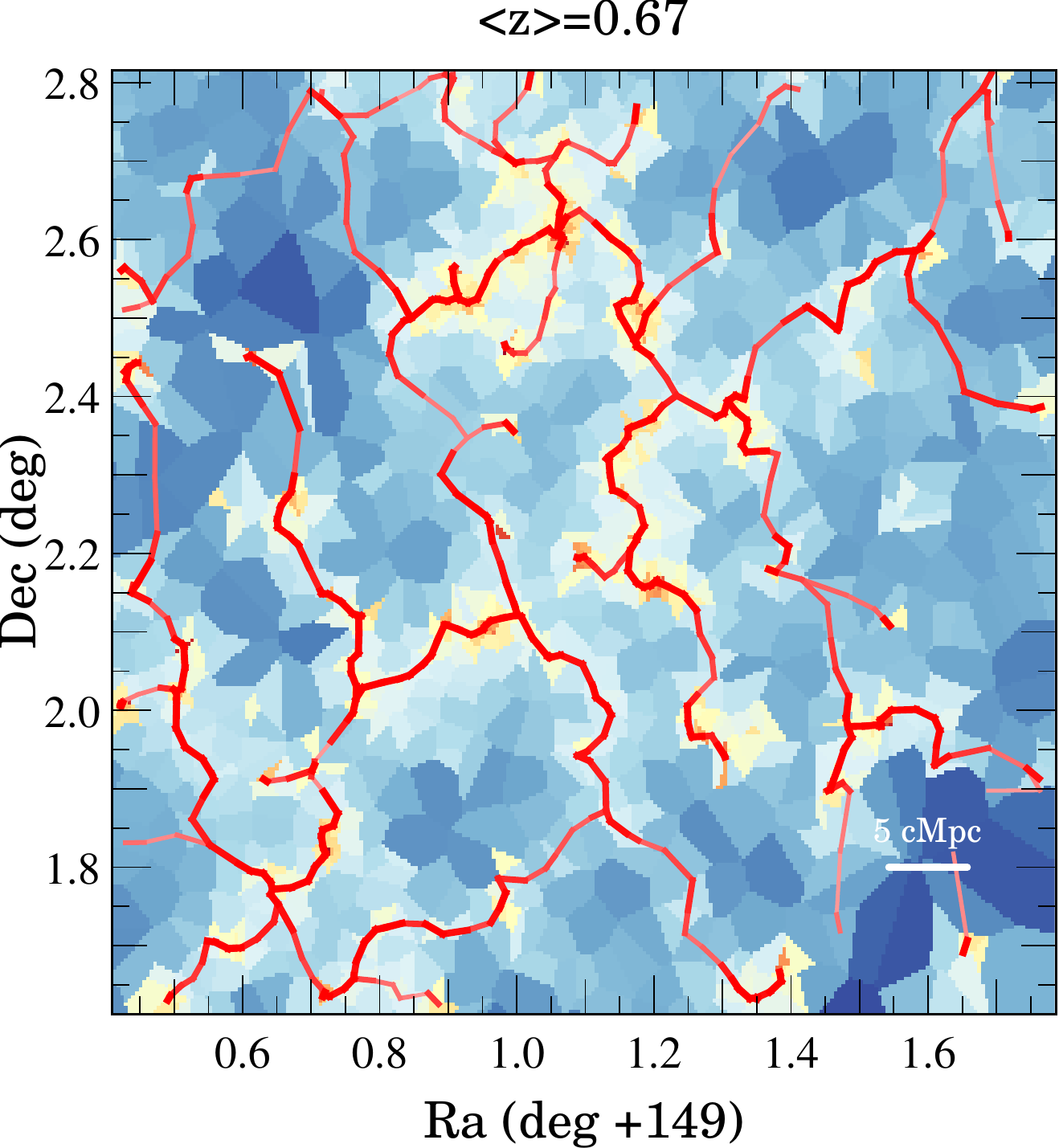}
    \hspace{0.2cm}
    \includegraphics[scale=0.51]{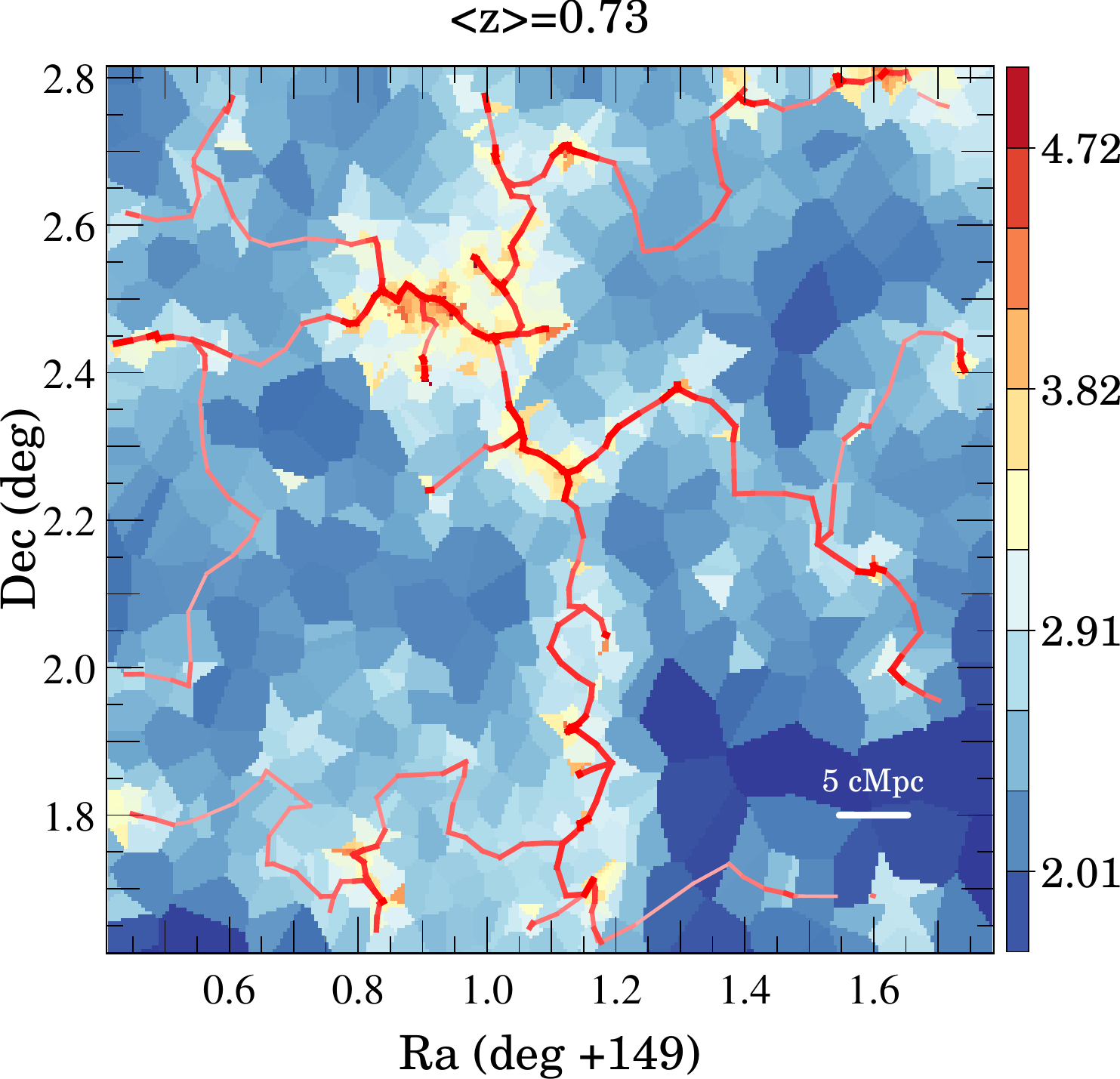}
 \caption{The density  field estimated from the Delaunay tessellation at different redshifts in slices of thickness 75 Mpc in the COSMOS field. The density refers to the galaxy number density and is down-weighted by the photometric errors. The red skeleton has been computed with a persistence threshold of $2\sigma$. Both the thickness  and the intensity of the red colour reflect the robustness of the segments, the more robust segments being thicker and darker. The
horizontal white bar in the lower right indicates the comoving scale length of 5 Mpc. }
\label{fig:skelcosmos}
\end{center}
\end{figure*}
%
%
\begin{figure*}
\begin{center}
{  \includegraphics[scale=0.65,clip]{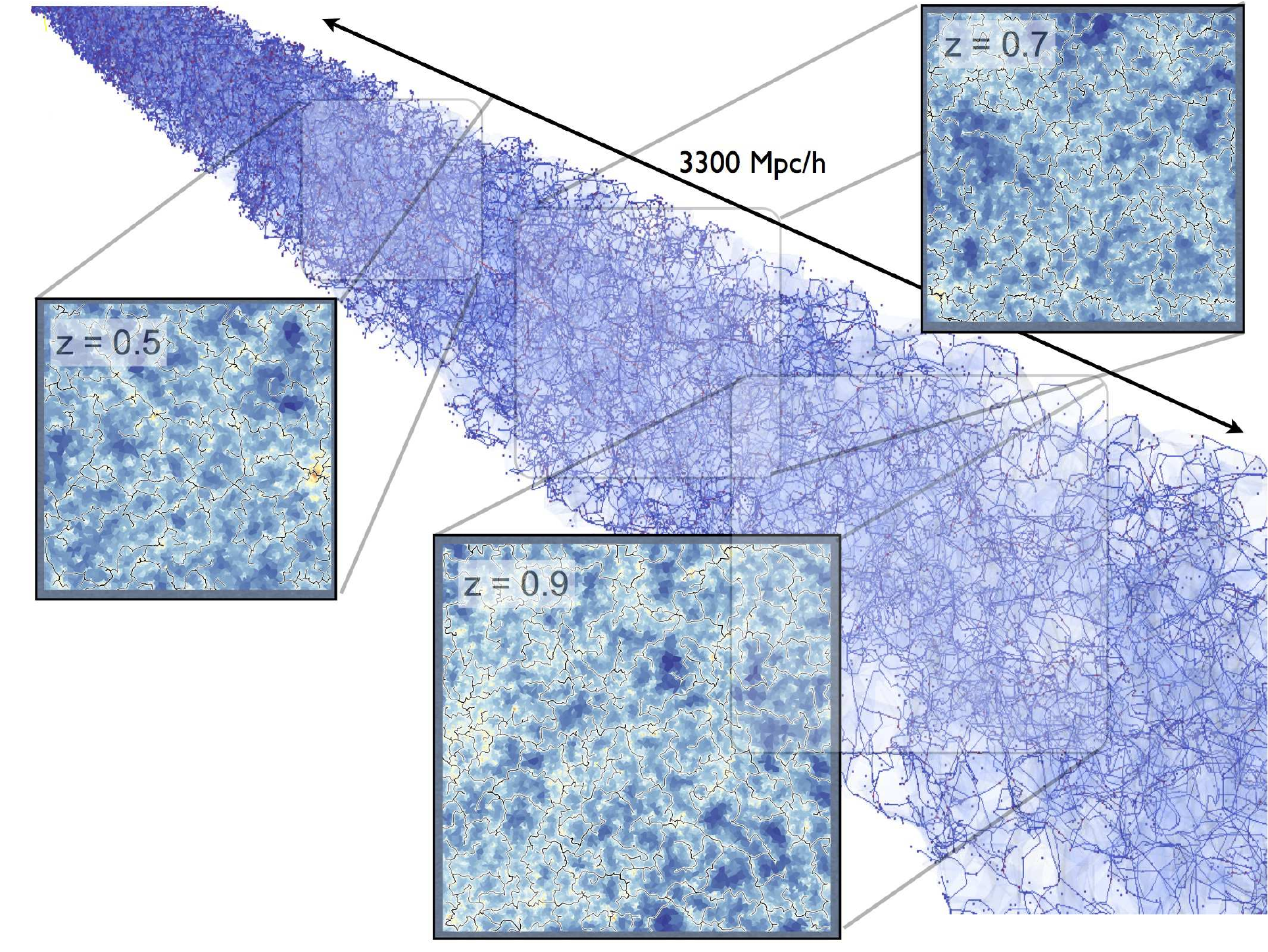} }
 \caption{ A three-dimensional view of the simulated {\sc Horizon-AGN} lightcone (\url{http://lightcone.horizon-simulation.org}). The whole lightcone shows the persistent filaments (blue solid line) and walls (light blue surfaces) extracted with {\sc DisPerSE}.  The three two-dimensional slices show the projection of the density field within 75 Mpc around $z\sim$0.5, $z\sim$0.7 and $z\sim$0.9 and are 2.25 degree a side. The skeleton extracted from the density field in two dimensions is overplotted in black.}
\label{fig:3d}
\end{center}
\end{figure*}
%
%
\subsection{Density and cosmic network identification}
\label{subsec:skeleton}
%
The filaments of the cosmic web are identified from the density field, which refers either to the number density of dark matter particles (in the case of the three-dimensional reference skeleton in the simulation) or to the number density of galaxies (for the observed and simulated two-dimensional skeletons). When computed from the galaxy distribution, the density is estimated from the Delaunay tessellation of the particles. Note that the density is not weighted by  galactic mass.
\\
To identify the cosmic network from the density, we use the  persistence based filament tracing algorithm   \citep[{\sc DisPerSE},][]{sousbie111,sousbie112}. This method identifies ridges  from the density field  as the special lines connecting topologically robust saddle points to peaks.
The identified filament network is by construction multi-scale,  while the extraction is robust to noise. 
 The set of all segments defining these ridges is called the skeleton  \citep{pogo09}. {\sc DisPerSE}  has been already successfully used in two dimensions to characterise stellar filaments in the Milky Way \citep[e.g.][]{arzoumanianetal2011,panopoulouetal2016}.
 Each filament is defined to be a set of connected small segments  linking extrema together. Persistence is defined as the  ratio of the density value at the two critical points in a topologically significant  pair of critical points: maximum-saddle, saddle-saddle (in three dimensions only) and saddle-minimum. This ratio quantifies the robustness  of the underlying topological feature characterised by this pair (the fact that the connected critical points are responsible for the appearance of a new topological feature such as connected component, tube, ball, in the corresponding excursion set). Expressed in terms of numbers of $\sigma$,  persistence quantifies the significance of the critical pairs  in the Delaunay tessellation of a random discrete Poisson distribution. In the following, a skeleton with a $2\sigma$ persistence level  is a skeleton for which all the persistence pairs which have their probability to be found in a random discrete Poisson distribution below $2\sigma$ of the mean have been removed. Removing low-persistence pairs is a  multi-scale non-local method to filter noise/low significance filaments.  This method is particularly well adapted to noisy datasets such as redshift catalogues \citep{sousbie111}.  
 A complementary notion is robustness \citep{weinkauf09b}, which is a local measure of how contrasted the critical points and skeleton segments are with respect to their local background.  
 \begin{table*}
\begin{center}
\def\arraystretch{1.5}
\begin{tabular}{|c|ccccccc|}
\hline
Name   &  catalogue &   Dim &   & $\sigma$ $^{*}$ & $M_{\rm lim}$ $^{*}$ & $D_{\Delta z}$ $^{*}$ & comments \\
  \hline
  {\sc skl}$_{\rm 2D}^{\rm obs}$   & {COSMOS2015} & 2D & galaxies with photo-$z$  & 2 & 10$^{10}$M$_{\odot}$ & 75 Mpc &   density weighted by the photo-$z$ errors\\
 {\sc skl}$_{\rm 2D}^{\rm phot}$ & {\sc Horizon-AGN} & 2D & galaxies with photo-$z$ & 2 & 
 10$^{10}$M$_{\odot}$ & 75 Mpc &   density weighted by the photo-$z$ errors \\
 {\sc skl}$_{\rm 2D}^{\rm spec}$ & {\sc Horizon-AGN} & 2D & galaxies with exact $z$ & 2 & 
 10$^{10}$M$_{\odot}$ & 75 Mpc &   - \\ 
  {\sc skl}$_{\rm 3D}^{\rm DM}$ & {\sc Horizon-AGN} & 3D & dark matter particles & - & 
$8\times10^{7}$M$_{\odot}$  & - &   density computed on a grid \\
{\sc skl}$_{\rm 3D}^{\rm gal}$ & {\sc Horizon-AGN} & 3D & galaxies & 5 & 
 10$^{10}$M$_{\odot}$ & - &   - \\
     \hline
\end{tabular}
\end{center}
\small
$^{*}$ unless specified otherwise.
  \caption{
The skeletons used in this work  to  assess the quality of the reconstruction or for  gradients measurements. $M_{\rm lim}$ refers to the mass limit of the galaxy sample used for the computation, $\sigma$ to the persistence threshold and $D_{\Delta z}$ is the thickness of the slice (for two-dimensional reconstructions). For {\sc skl}$_{\rm 3D}^{\rm DM}$ the persistence threshold is not expressed in terms of $\sigma$. The quoted values are those used in the main text, but  Appendices  investigate how the reconstruction changes when these values vary.
}
\label{Tab:skeleton}
\end{table*}
 Table~\ref{Tab:skeleton} summarises the extracted skeletons both on observations and simulations and their main features that we detail also below.
 \\
To identify filaments in the two-dimensional density field from the COSMOS2015 catalogue, we proceed as follows. We first divide the galaxy sample in slices of constant thickness, the choice of which is justified in Section~\ref{subsec:slice}. To estimate the density in each slice, we compute  the two-dimensional Delaunay tessellation from the galaxy distribution. To take into account the errors on  photometric redshifts in this calculation, we associate to each galaxy in each slice a weight $p_{{\rm gal},i}$ which is the probability for this galaxy to be in the slice $i$ given the probability function distribution of its redshift $P_{\rm gal}(z)$ (as computed by {\sc LePhare}): 
\begin{equation}
p_{{\rm gal},i}=\int_{z_{1,i}}^{z_{2,i}}{P_{\rm gal}(z)dz}\,/\int{P_{\rm gal}(z)dz}\,,
\end{equation}
where $z_{1,i}$ and $z_{2,i}$ are the lower and upper redshifts of the slice. This weighting allows to reduce the pollution effects of foreground and background galaxies for which  the probability to be in the slices is low. To correctly estimate the density and the topology close to boundaries, a surface of ``guard" particles is added outside the bounding box and new particles are added by interpolating the estimated density computed on the boundary of the distribution.
Note that the Delaunay tessellation is by construction a robust method to reconstruct reliably the density in a field which contains masks. 
The triangulation automatically connects galaxies from across the masked regions and thus interpolates the density \citep[e.g.][]{aragoncalvoetal2015}. 
Filaments are then identified with {\sc DisPerSE} from the two-dimensional Delaunay tessellation with a 2$\sigma$ persistence threshold. The properties of the noise are not necessarily the same in two dimensions as in three dimensions. This implies that although a persistence threshold above 3$\sigma$ would be more conservative in three dimensions (which guarantees that less than 0.3\% of the critical pairs are spurious), the two-dimensional skeleton extracted with a 3$\sigma$ persistence would miss a large number of robust three-dimensional filaments. To compensate this effect, choosing a two-dimensional persistence threshold of 2$\sigma$ enables to drastically decrease the number of unrecovered three-dimensional filaments (by a factor $\sim$1.7), without sensibly increasing the number of  spurious two-dimensional filaments \citep[see also][for a motivation behind this mapping; this choice is further justified in Appendix~\ref{Ap:persistence} using the simulation]{Malavasi2016b}. Figure~\ref{fig:skelcosmos} shows the density of the COSMOS field in four slices at redshifts $z\sim 0.52$, $z\sim 0.60$, $z\sim 0.67$ and $z\sim 0.73$ estimated from the Delaunay tessellation and their corresponding skeletons in red. The intensity of the red colormap reflects the robustness of the segments, the more robust segments being darker. 
\\
We extract the two-dimensional skeletons in the simulation with mock photometric redshifts following the exact method used for the COSMOS field. In addition, a two-dimensional skeleton is also extracted in each slice with true (``spectroscopic" in the following) redshifts (hence without including the weighting with the photometric uncertainties). Finally, a three-dimensional skeleton ({\sc skl}$_{\rm 3D}^{\rm DM}$) is extracted from the dark matter distribution. To compute this skeleton, in order to save computational time, dark matter particles are first projected on a three-dimensional grid with a resolution of 250 comoving kpc using a cloud-in-cell algorithm. We perform a first extraction with a very low persistence threshold. Then the adopted persistence threshold is estimated to be $\sim 5$ times the rms level of density fluctuations for the low density filaments in the previous extraction. Note that our results are robust against the exact value chosen for this persistence threshold. This skeleton becomes  the reference skeleton. It is indeed the most suitable, as the galaxy distribution itself is already a biased tracer of the underlying density field. In addition, dark matter particles are more numerous in the simulation, allowing us to extract  more accurate filament positions, which will be all the more important when computing  distances between two- and three-dimensional skeletons in Section~\ref{Sec:Qualification}. A three-dimensional view of the lightcone and the corresponding  skeleton is shown in Figure~\ref{fig:3d}.  
\begin{figure*}
\begin{center}
  \includegraphics[scale=0.5]{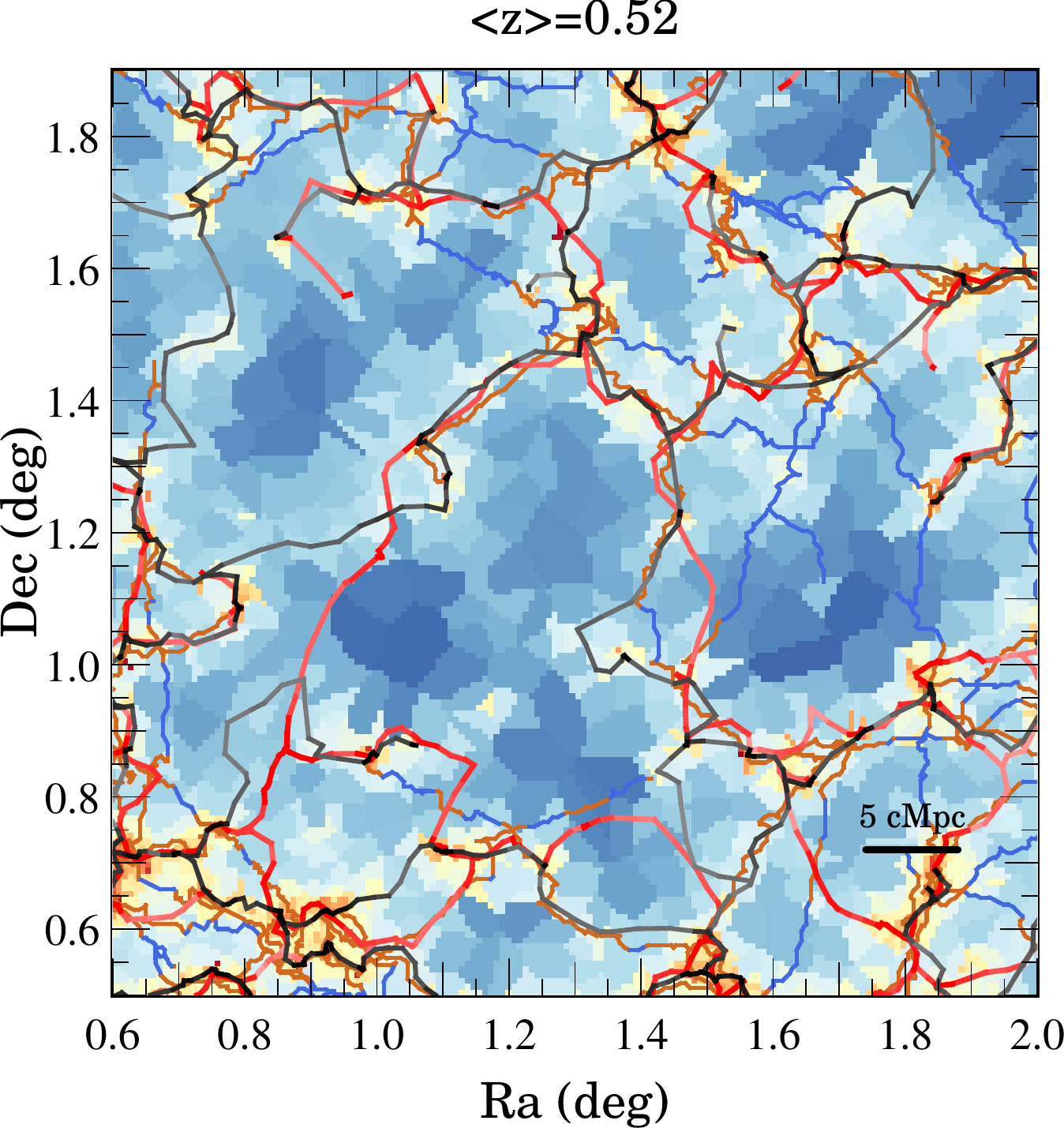}
   \hspace{0.2cm}
  \includegraphics[scale=0.5]{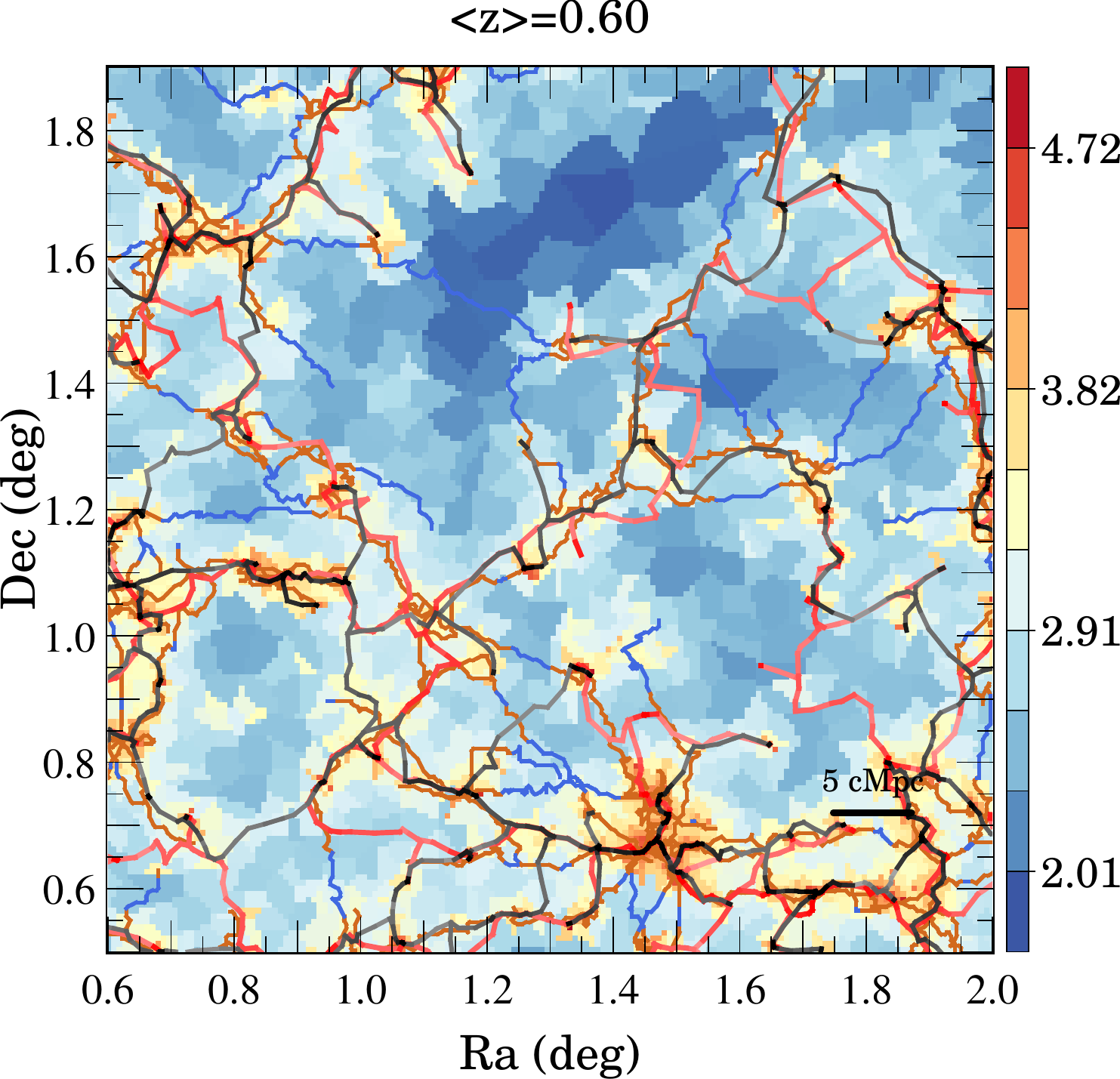}
    \par
   \vspace{0.4cm}
   \includegraphics[scale=0.5]{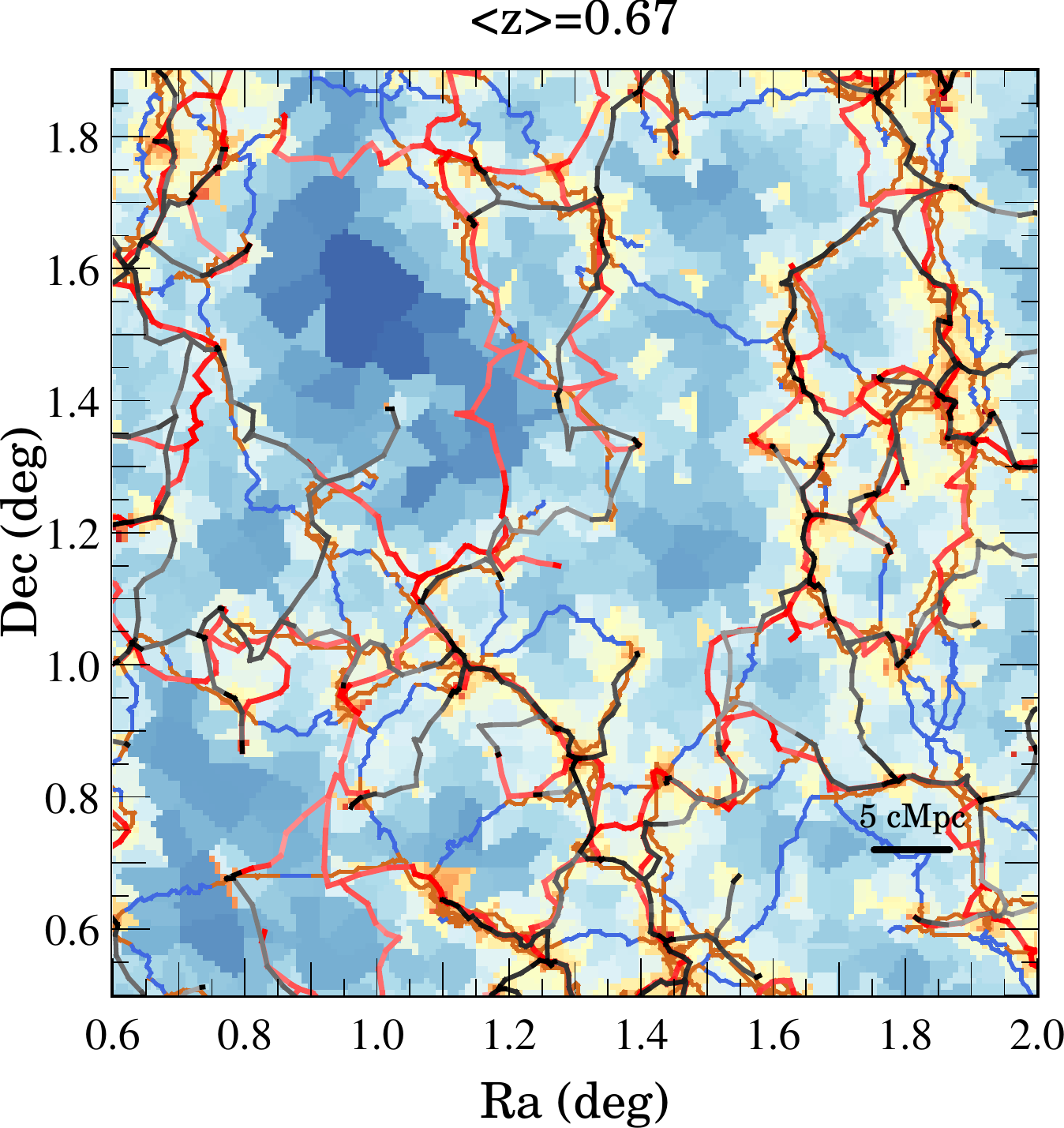}
    \hspace{0.2cm}
    \includegraphics[scale=0.5]{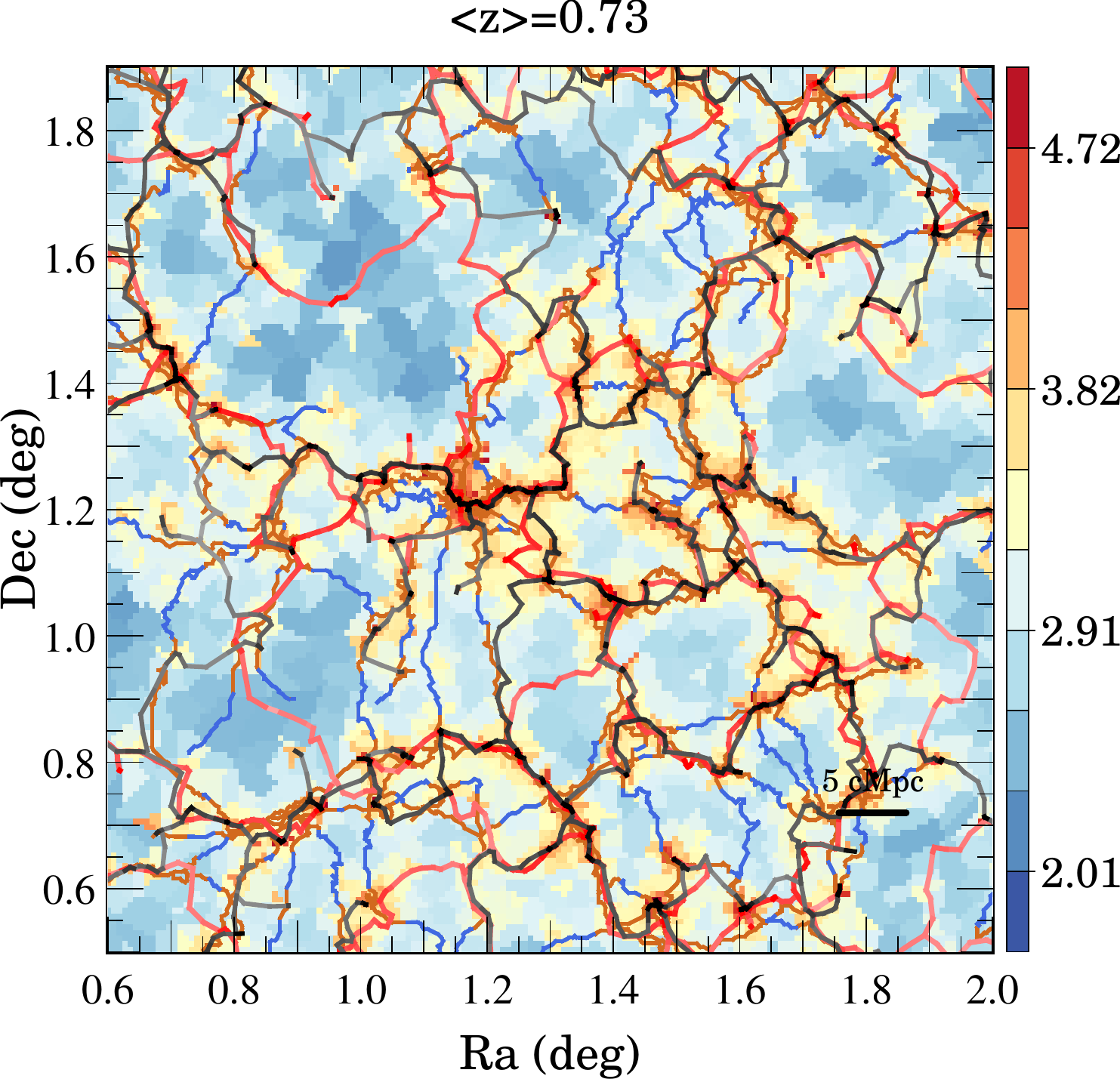}
 \caption{The density  field estimated from the Delaunay tesselation at different redshifts in slices of thickness 75 Mpc from the \hagn  lightcone. The orange/blue skeleton is {\sc skl}$_{\rm 3D}$ computed from the full dark matter particles distribution.  The segments are blue when they do not have a counterpart in {\sc sk}$_{\rm 2D}^{\rm spec}$ or {\sc skl}$_{\rm 2D}^{\rm phot}$ within 0.5$\times D_{\rm int}$, where $D_{\rm int}$ is the mean inter-particle distance. The black skeleton is {\sc skl$_{\rm 2D}^{\rm spec}$} and the red skeleton is  {\sc skl$_{\rm 2D}^{\rm phot}$} computed with a 2$\sigma$ persistence threshold from the galaxy distribution, the intensity of the colour reflecting the robustness of the segments. The transverse width of the slices (1.4~deg), is  comparable to the COSMOS field. The background colour codes the galaxy number density computed from the Delaunay tesselation with exact redshifts. Most of the filaments in {\sc skl}$_{\rm 2D}^{\rm phot}$ have a close counterpart in {\sc skl}$_{\rm 3D}$. We note that some three-dimensional filaments are recovered in {\sc skl}$_{\rm 2D}^{\rm phot}$ but not in {\sc skl}$_{\rm 2D}^{\rm spec}$ and conversely. The
horizontal black bar in the lower right indicates the comoving scale length of 5 Mpc.}
\label{fig:slice}
\end{center}
\end{figure*}
\begin{figure}
\begin{center}
 \includegraphics[scale=0.41]{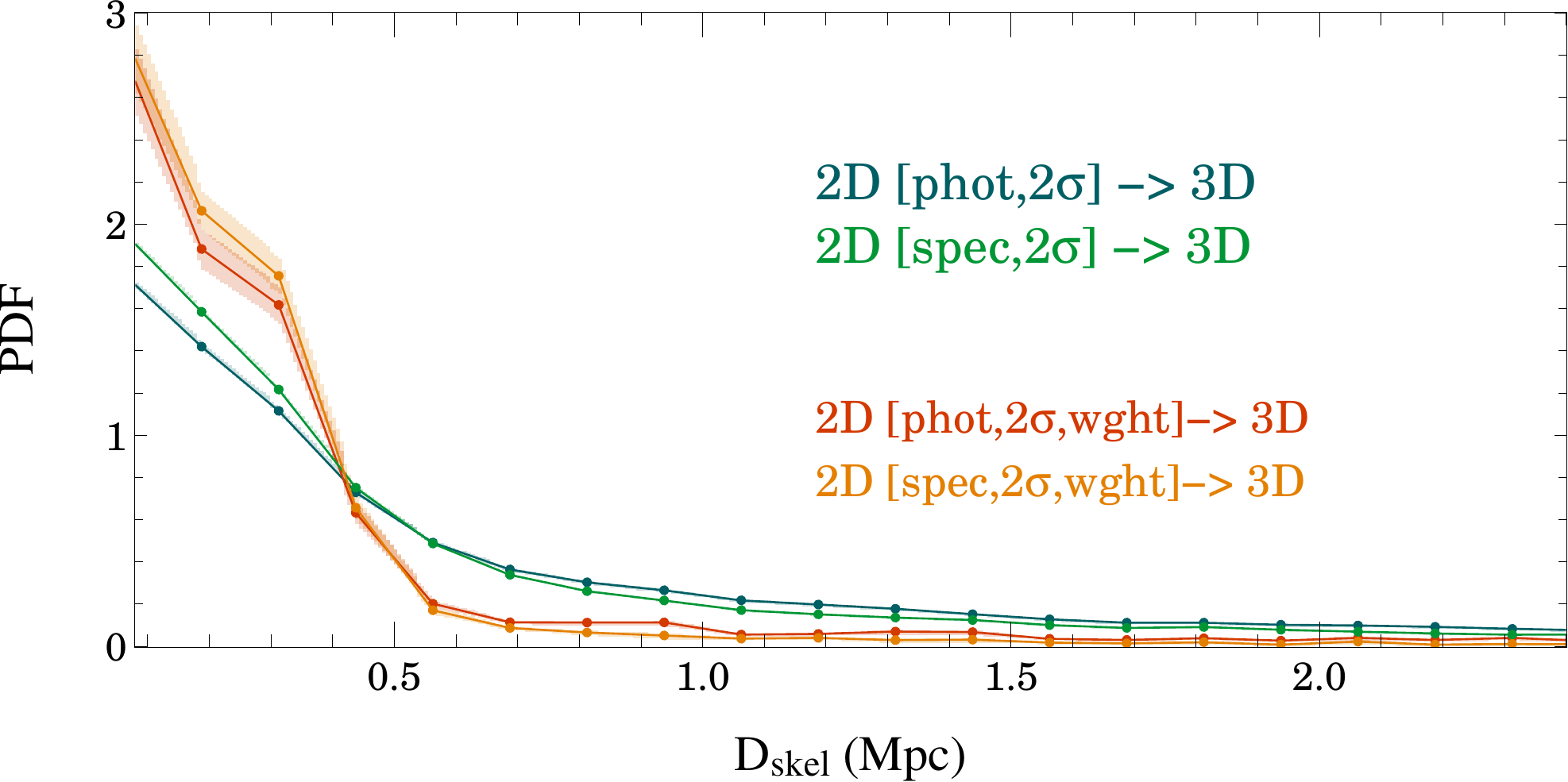} 
 \caption{Distribution of the distances between {\sc skl$_{\rm 2D}$} and {\sc skl$_{\rm 3D}^{\rm DM}$} in the \hagn lightcone.  For each segment  in {\sc skl$_{\rm 2D}$} we measure the distance to the closest segment in {\sc skl$_{\rm 3D}$} (blue line with photometric redshifts and green line with spectroscopic redshifts). The distances are measured in all slices between $z\sim 0.5$
 and $z \sim 0.9$. The red and orange lines correspond to the distributions weighted by  robustness.} 
\label{fig:dist2d3d}
\end{center}
\end{figure}
%
%
\subsection{Choice of  slices}
\label{subsec:slice}
%
The thickness of the slices on which the projected density is computed and the skeleton extracted will impact our analysis. 
A volume-limited sample with a constant stellar mass cut and a constant slice thickness is required, in order to avoid possible systematics in the measurements due to different quality of reconstruction (because of increasing slice thickness with increasing redshifts).
 In two dimensions, there can be some  confusion between nodes and projected three-dimensional filaments, or filaments and projected three-dimensional walls. Furthermore, projected filaments may  link some overdensities which are in fact not truly linked in three dimensions.  Hence if the slices are too thick, the skeleton derived from the projected density will have little correspondence  to  its three-dimensional counterpart, with a large number of fake two-dimensional filaments and non-recovered three-dimensional filaments.  However, if the chosen thickness of the slices is too small, the two-dimensional skeleton will trace only fragments of the three-dimensional filaments and will be in turn a poor representation of the fully connected three-dimensional skeleton. 
 Following Appendix~\ref{Ap:thickness}, we choose to work up to redshift 0.9  with slices of thickness 75 comoving Mpc, and by considering all galaxies more massive than $10^{10}$~M$_\odot$. This final choice is guided both by the required thickness for an optimal reconstruction when exact redshifts are available and by the current redshift accuracy of our sample. As it is shown in Table~\ref{Tab:thick}, with a thickness of 75 Mpc, all galaxies of the final sample have their redshift uncertainties smaller than the thickness of the slice, where uncertainties  are estimated from the $2\times1\sigma$ errors\footnote{The $1\sigma$ error  is defined as the value enclosing half of 68\% of the probability distribution function of the photometric redshift of the galaxy.} given by {\sc LePhare} for each galaxy.  
 \\
 Tests on the virtual photometric catalogue detailed below  confirm that this choice is good enough to  reconstruct the skeleton.
 We do not consider galaxies below redshift 0.5 as the comoving transverse width of the field starts to be too small to  extract reliably the large-scale structure. 
 The adjacent redshift slices are overlapping and spaced by half of the thickness of the slices. There are 30 slices on the redshift range $0.5<z<0.9$. 
%
%
\section{Extraction assessment in mocks}
\label{Sec:Qualification}
%
%
\subsection{Two-dimensional versus three-dimensional skeleton}
%
Let us now use the {\sc Horizon-AGN} simulated catalogue to estimate the quality of the skeleton reconstructed with photometric redshifts. For this assessment, we compare the two-dimensional photometric skeleton with the projected three-dimensional reference one. 
Figure~\ref{fig:slice} displays slices of thickness 75~Mpc around redshifts $z\sim 0.52$, $z\sim 0.60$, $z\sim 0.67$ and $z\sim 0.73$ in the {\sc Horizon-AGN} simulation. The orange skeleton corresponds to the projection of our reference skeleton {\sc skl$_{\rm 3D}^{\rm DM}$} computed from the dark matter distribution. The segments are blue when they do not have a two-dimensional counterpart within 0.5$\times D_{\rm int}$ where $D_{\rm int}$  is the mean inter-galaxy distance in the slice. The black skeleton corresponds to {\sc skl}$_{\rm 2D}^{\rm spec}$ while the red skeleton is {\sc skl}$_{\rm 2D}^{\rm phot}$. The two-dimensional skeletons are  visually in good agreement, hence confirming by eye that photometric redshifts are sufficient to reconstruct the two-dimension skeleton. 
In more details, we see that most filaments of {\sc skl}$_{\rm 2D}^{\rm phot}$ have a reliable counterpart in  {\sc skl$_{\rm 3D}^{\rm DM}$}. However some filaments of the former are not recovered in the latter  and some filaments of the latter have no counterpart in the former. Hence the choice of the persistence threshold is important to mitigate the number of filaments in {\sc skl}$_{\rm 2D}^{\rm phot}$ which have no counterpart in {\sc skl$_{\rm 3D}^{\rm DM}$} and which are thus potentially noisy detection. In addition, filaments detected in {\sc skl}$_{\rm 2D}^{\rm phot}$ but not in {\sc skl$_{\rm 3D}^{\rm DM}$} may potentially be walls seen in projection. We note that the two-dimensional segments which have generally no counterpart in {\sc skl$_{\rm 3D}^{\rm DM}$} have a low robustness.   
\\
Let us quantify this statistically by measuring the distribution of distances, $d_{\rm 2D\rightarrow3D}$, between the segments of {\sc skl}$_{\rm 2D}^{\rm phot}$ and {\sc skl}$_{\rm 2D}^{\rm spec}$ skeletons and their nearest segments in  the projected {\sc skl$_{\rm 3D}^{\rm DM}$}  skeleton \citep[following the method introduced by][]{sousbie111}.  These distributions give us an estimation of the reliability of the skeleton.
These distributions are shown on Figure~\ref{fig:dist2d3d}.  Green and blue lines  correspond to the distributions of the distances $d_{\rm 2D\rightarrow3D}$ using spectroscopic and photometric redshift respectively for the computation of {\sc skl}$_{\rm 2D}$. 
Although the photometric skeleton has a slightly larger tail, the curves computed with photometric and spectroscopic redshifts are quite similar, which suggests that the photometric redshift uncertainties have little impact on our global ability to recover the three-dimensional structures, at least under the chosen parameters. 
 However, the tail of the   distribution implies that a certain number of two-dimensional filaments have no three-dimensional counterpart at all, i.e. they correspond to spurious two-dimensional filaments connecting points in two dimensions that are spatially disconnected in three dimensions. We find that, on average, 21.3\%  and 15.2\% of segments have no counterpart closer than 1.5 projected Mpc in the photometric and spectroscopic skeleton respectively. If we keep only the 60\% more robust segments in the two-dimensional skeleton, these numbers fall { by a factor $\sim$2}. Weighting the distribution of $d_{\rm 2D\rightarrow3D}$ by the robustness allows to considerably reduce this tail, as shown by the red and yellow curves for the photometric and spectroscopic skeletons respectively, implying that the fake filament detections correspond generally to low-robustness filaments. 
 \\ 
Note that {\sc skl}$_{\rm 2D}^{\rm phot}$ is degraded or biased compared to the projected {\sc skl$_{\rm 3D}^{\rm DM}$} skeleton for three reasons. 
First, the photometric skeleton {\sc skl}$_{\rm 2D}^{\rm phot}$  will be noisier than the spectroscopic one {\sc skl}$_{\rm 2D}^{\rm spec}$ due to the galaxy redshift uncertainties. Secondly, even with exact redshifts, {\sc skl}$_{\rm 2D}^{\rm spec}$ is a biased reconstruction of the projected three-dimensional skeleton computed from galaxies due to the projection effect. Finally, it should be acknowledged that the three-dimensional galaxy skeleton {\sc skl}$_{\rm 3D}^{\rm gal}$ is biased compared to the three-dimensional dark matter skeleton {\sc skl}$_{\rm 3D}^{\rm DM}$. This bias, although relatively negligible, is dependent on the limiting mass of the galaxy catalogue and is investigated in Appendix~\ref{Ap:Dmgal}. 
%
%
\subsection{3D measurement from 2D  at  {COSMOS2015} accuracy}
\begin{figure}
\begin{center}
 \includegraphics[scale=0.5]{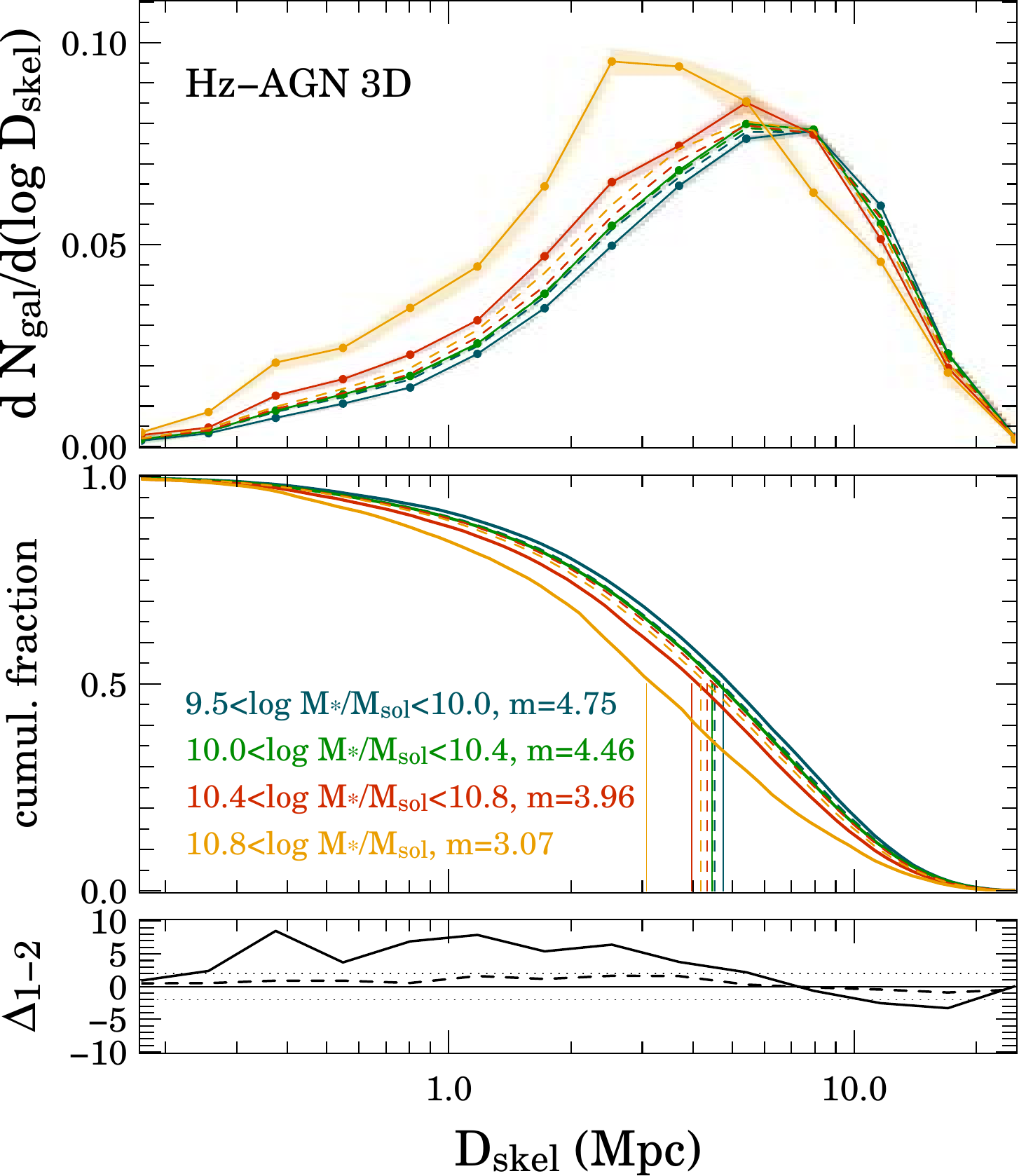} 
   \caption{Differential (\textit{top}) and cumulative (\textit{middle}) distributions of the distances to the {\it three}-dimensional (not projected) dark matter skeleton {\sc skl}$_{\rm 3D}^{\rm DM}$ for galaxies as a function of their masses in the {\sc Horizon-AGN} lightcone for $0.5<z<0.9$. The dashed distributions correspond to a random signal \textit{preserving the mass-density relation}: galaxy masses in the catalogue have been randomised with respect to their distances to the filament in given density bins, then the same mass cuts have been applied.   The $m$ values give the medians of each distribution, which are also indicated by vertical lines. The contribution of nodes to the gradient signal has been removed from the analysis  to highlight an effect specifically related to the filaments. The \textit{bottom} row shows the residuals between the two intermediate  mass bins ($10.4<\log M_*/{\rm M}_{\odot} <10.8$ and $10<\log M_*/{\rm M}_{\odot} <10.4$) for the original (solid line) and reshuffled (dashed line) samples expressed in number of $\sigma$: $\Delta_{1-2}=\Delta(D_1-D_2)/\sqrt(\sigma_1^2+\sigma_2^2)$
   }
\label{fig:distancesTest}
\end{center}
\end{figure}

%
Although the three-dimensional filaments can be reliably traced from two-dimensional slices at the {COSMOS2015} redshift accuracy, it is not obvious that a distance-to-filament trend will be statistically recovered in two dimensions. The projection will inevitably blur the measurement: galaxies far from the filament but on the same line-of-sight will be counted as very close  by  projection. We will therefore test in the {\sc Horizon-AGN} simulation if a signal existing in three dimensions may be recovered in two dimensions. Following the recent results from \cite{Malavasi2016b}, we expect to find mass gradients towards three-dimensional filaments. We first measure this trend in three dimensions. In a second step, we verify that the signal is also measurable in two dimensions. \\
Figure~\ref{fig:distancesTest} presents the distributions of the three-dimensional distances to {\sc skl}$_{\rm 3D}^{\rm DM}$  for simulated galaxies in four different bins of masses in the {\sc Horizon-AGN} lightcone. The uppest mass bin is chosen to contain galaxies above the knee of the observed galaxy mass function, which is found to be $\sim 10^{10.8}$M$_{\odot}$  in $0.5<z<0.8$ from the fit of the mass function with a double Schechter profile \citep{davidzonetal2017}. The most massive galaxies are found closer to the filament center than the low mass ones, in agreement with \cite{Malavasi2016b}.
 \\
  The same measurements are then carried in two dimensions with {\sc skl}$_{\rm 2D}^{\rm phot}$ on the \hagn lightcone, in all slices between $0.5\!<z<\!0.9$. The projected transverse distances to the filament, measured in degrees, are then converted in comoving Mpc. The cumulative distributions in all slices are stacked and the mean distribution is presented in the left panel of Figure~\ref{fig:massdistance2filament}. The galaxy mass segregation is also found in two dimensions with photometric redshifts demonstrating our ability to recover the signal in observed data.
 \\
 Let us now stress  that we are looking for environmental effects driven specifically by filaments. As galaxies are more massive and passive in clusters than in the field, we expect  mass gradients towards nodes  driven by the effect of cluster  alone. These gradients could contaminate the mass gradient-trend towards filaments. To minimise node contributions, we remove from the analysis galaxies closer to a node than 3.5 and 0.8 projected Mpc  (see Appendix~\ref{Ap:robdens}) in three and two dimensions respectively and we reshuffle galaxy masses with respect to their position along the filaments. We show in Appendix~\ref{Ap:robdens} that this allows to correctly remove their contribution to the gradients towards filaments. We still recover  mass gradients towards  filaments both in three and two dimensions: more massive galaxies are more confined to filaments than less massive ones. These gradients are driven by filaments alone. 
 \\
However, it is well known  that denser environments contain more massive, redder and much less star-forming galaxies \citep[e.g.][]{dressler80,baloghetal1997,blantonetal2003,kauffmannetal2004,davidzonetal2016,Cucciatietal2016}, even if we exclude cluster regions. Since the density is anisotropically distributed in filaments, such local mass-density relation could naturally lead to mass gradients towards filaments.
To disentangle the contribution of the anisotropic large-scale environment from the effect of the local mass-density relation, we proceed following \cite{Malavasi2016b}. The galaxy catalogue is split in local density bins (40 density bins logarithmically distributed), where the density is computed from the three-dimensional galaxy Delaunay tessellation smoothed with a Gaussian kernel of 3 Mpc and 1 \textit{projected} Mpc in three and two dimensions respectively  (we checked that varying this smoothing length does not significantly change the conclusions). Then stellar masses are reshuffled between the galaxies in each bin. The distributions after reshuffling are shown as dashed lines in Figures~\ref{fig:distancesTest} and~\ref{fig:massdistance2filament}. Because it preserves the mass-density relation, this reshuffling partially preserves the mass gradients towards filaments, however the reshuffled and original distributions are not identical. The median values of both original and reshuffled distributions are presented in  Table~$\ref{Tab:results}$. The gradient signal is much larger with the original distribution than the reshuffled one. In two dimensions, the median of the lowest mass bin distribution is higher than that  of the other bins, but there is otherwise no significant gradient. 
\\
Let us summarise these results. First, mass gradients are found towards filaments  in the {\sc Horizon-AGN} simulation.  These gradients are specifically induced by  filaments and  not driven by nodes. In addition, they cannot entirely  be explained by the mass-density relation alone. The particular geometry of the large-scale environment itself is what drives the gradient. Secondly, we demonstrate that  at the accuracy of the {COSMOS2015} catalogue, mass gradients found in  the three-dimensional mocks can also be identified in the two-dimensional ones. The method can therefore robustly be applied to the observed {COSMOS2015} catalogue.
%
%
\section{Galaxy properties in  filaments} 
\label{Sec:Results}
%
Having confirmed the feasibility to trace the filaments in two-dimensional slices with a thickness of $75$ Mpc and to recover the corresponding three-dimensional signal, let us now study in more details the distribution of galaxy masses and colour-types as a function of their distance to  filaments using both the observed {COSMOS2015} and simulated {\sc Horizon-AGN}  catalogues.
%
%
\subsection{Galaxy mass gradients towards filaments}
\begin{figure*}
 \includegraphics[scale=0.5]{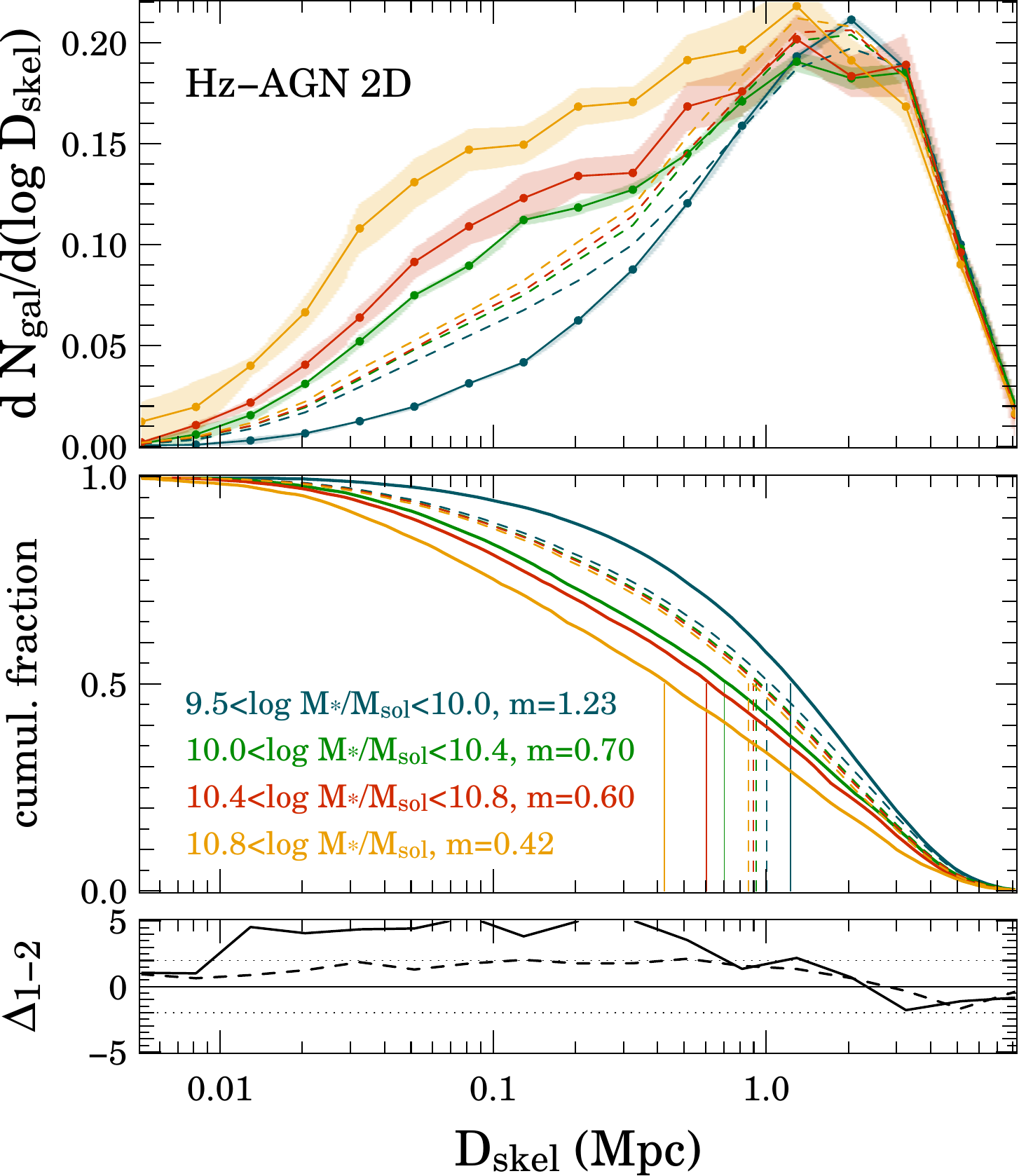} 
  \hspace{0.2cm}
   \includegraphics[scale=0.5]{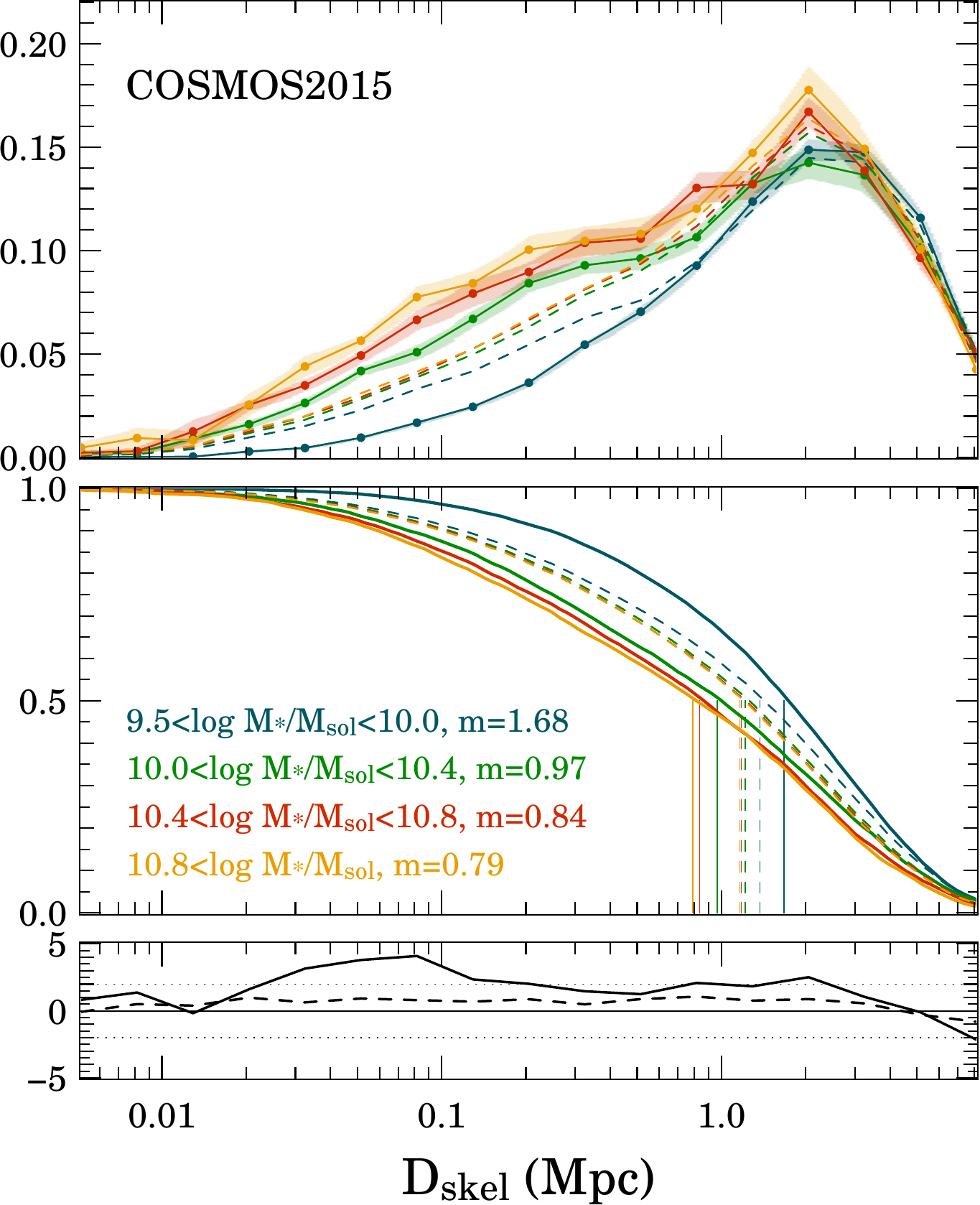} 
 \caption{Differential (\textit{top}) and cumulative (\textit{middle}) distributions of the distances to the two-dimensional photometric skeleton {\sc skl}$_{\rm 2D}^{\rm phot}$ for galaxies as a function of their masses in \hagn (\textit{left}) and {COSMOS2015} (\textit{right}). The dashed lines correspond to a random signal \textit{preserving the mass-density relation}: galaxy masses in the catalogue have been randomised with respect to their distances to the filament in given density bins, while the same mass cuts are applied. Note that the lowest mass bin $9.5<\log M_*/{\rm M}_{\odot} <10.0$ should be considered with caution because the thickness of the slices has been calibrated for galaxies with a mass $\log M_*/{\rm M}_{\odot} >10$. Here, the signal measured in all slices between $z\sim 0.5$ and $z \sim 0.9$ is stacked. The $m$ values give the medians of each distribution, which are also indicated by vertical lines. The contribution of nodes to the gradient signal has been removed from the analysis  to highlight an effect specifically related to the filaments. The \textit{bottom} row shows the residuals between the two intermediate ($10.4<\log M_*/{\rm M}_{\odot} <10.8$ and $10<\log M_*/{\rm M}_{\odot} <10.4$) mass bins for the original (solid line) and reshuffled (dashed line) samples.  
}
\label{fig:massdistance2filament}
\end{figure*}

%
The right panel of Figure~\ref{fig:massdistance2filament} presents the distributions of the distance to the filaments for the four bins of mass in the \hagn lightcone (after including photometric redshift errors) and in {COSMOS2015} respectively. The signal measured in all slices in the redshift range $0.5<z<0.9$ has been stacked. We remove the contribution of the nodes following the above-described prescription. Note that the lowest mass bin  $9.5<\log M_*/{\rm M}_{\odot} <10.0$ should be considered with caution. These galaxies are not used to compute the skeleton, and their redshift uncertainties (2$\times$ 1$\sigma$) are generally higher than the thickness of the slices. In COSMOS, we also find that that massive galaxies are closer  to the center of filaments compared to less massive ones, in agreement with the measurement in {\sc Horizon-AGN}. We note that the galaxy mass gradients signal are stronger in the simulation in two dimensions than  in {COSMOS2015}. A possible explanation is that uncertainties on galaxy masses are not  modelled in the simulation, and not taken into account in the measurement from {COSMOS2015}. These uncertainties are likely to dilute the gradients. As low mass galaxies are more numerous and with larger uncertainties than high mass galaxies, they are likely to pollute the higher mass bins. This bias should dilute the signal especially in the highest mass bins, as we indeed observe in Figure~\ref{fig:massdistance2filament}.
Node contribution is removed and we proceed with the above described reshuffling to disentangle the contribution of the anisotropic large-scale environment from the effect of the local mass-density relation. The distributions after reshuffling are shown as dashed lines in Figure~\ref{fig:massdistance2filament}. 
Consistently to the {\sc Horizon-AGN} result, we find that the reshuffling almost completely dilute the gradient signal: the values of the median of each bin are consistent within the errobars, except the one of the lowest mass bin which is slightly higher (see Table~$\ref{Tab:results}$). Varying the number of density bins used for the reshuffling does not change the trend. We conclude that, after removing the significant contribution of the nodes, most of the gradient signal towards filament is driven by the geometry of the large-scale environment itself.
%
%
\subsection{Colour-type segregation towards filaments}
\begin{figure*}
\begin{center}
  \includegraphics[scale=0.5]{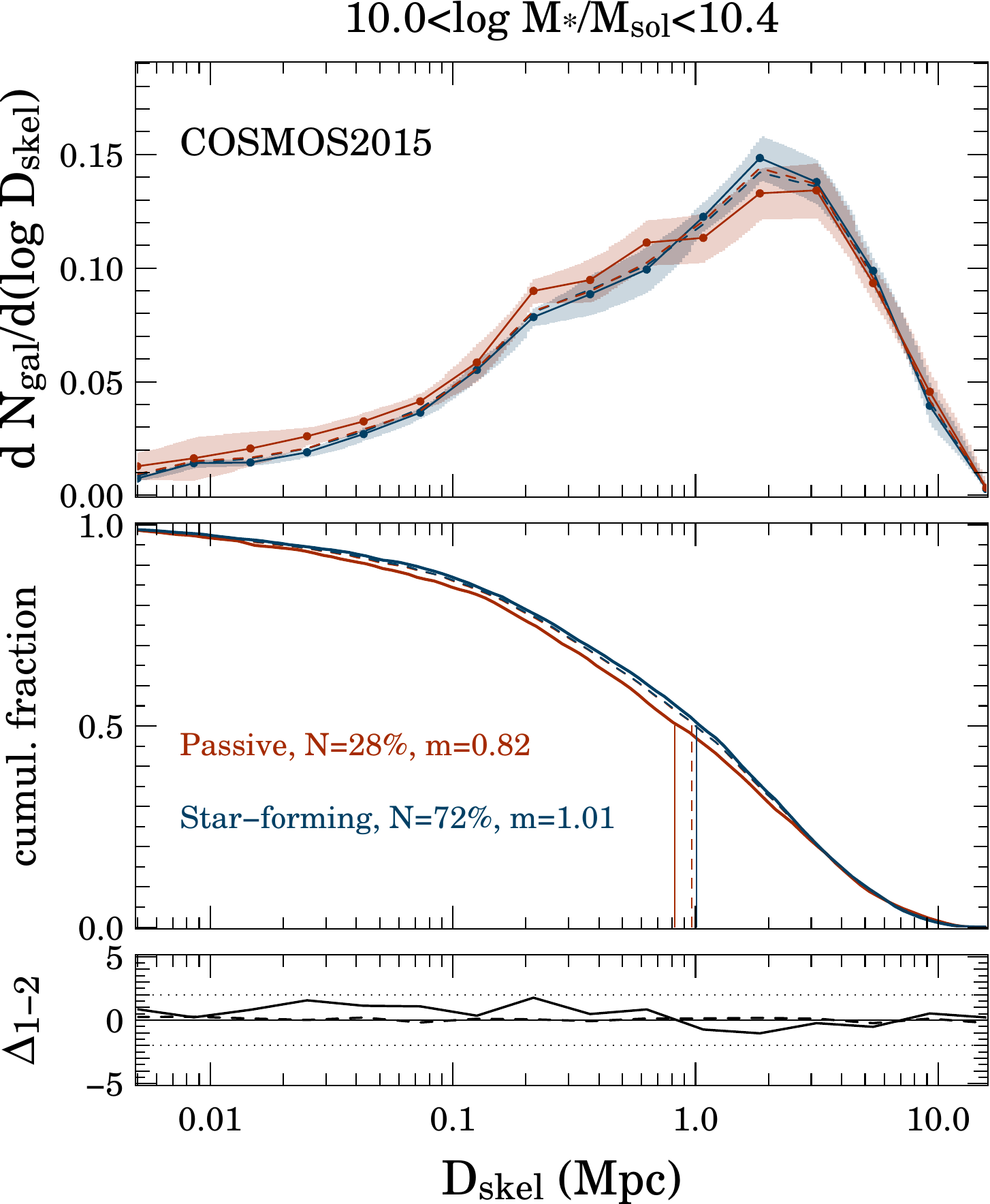} 
     \hspace{0.3cm}
  \includegraphics[scale=0.5]{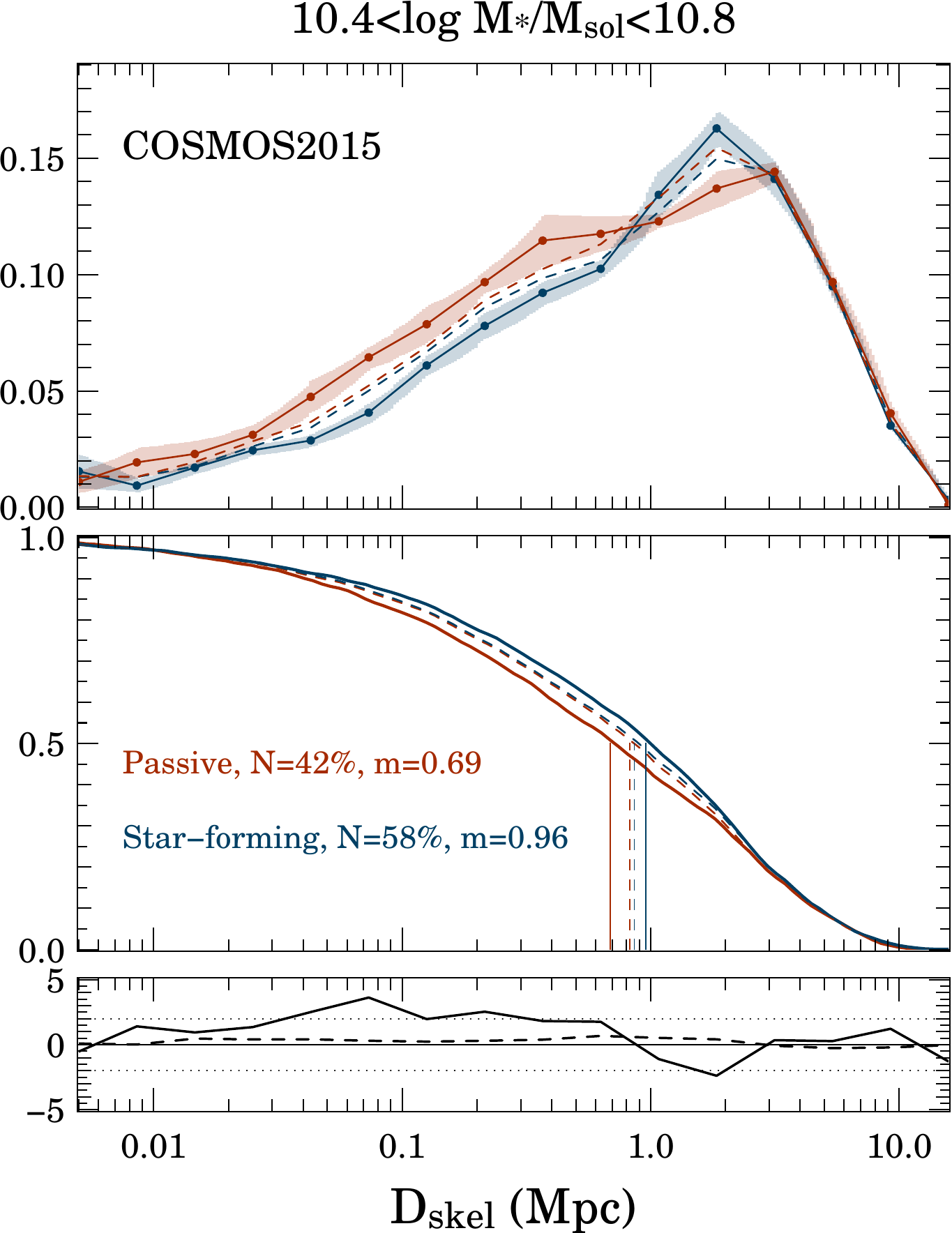}
  \par
   \vspace{0.4cm}
  \includegraphics[scale=0.5]{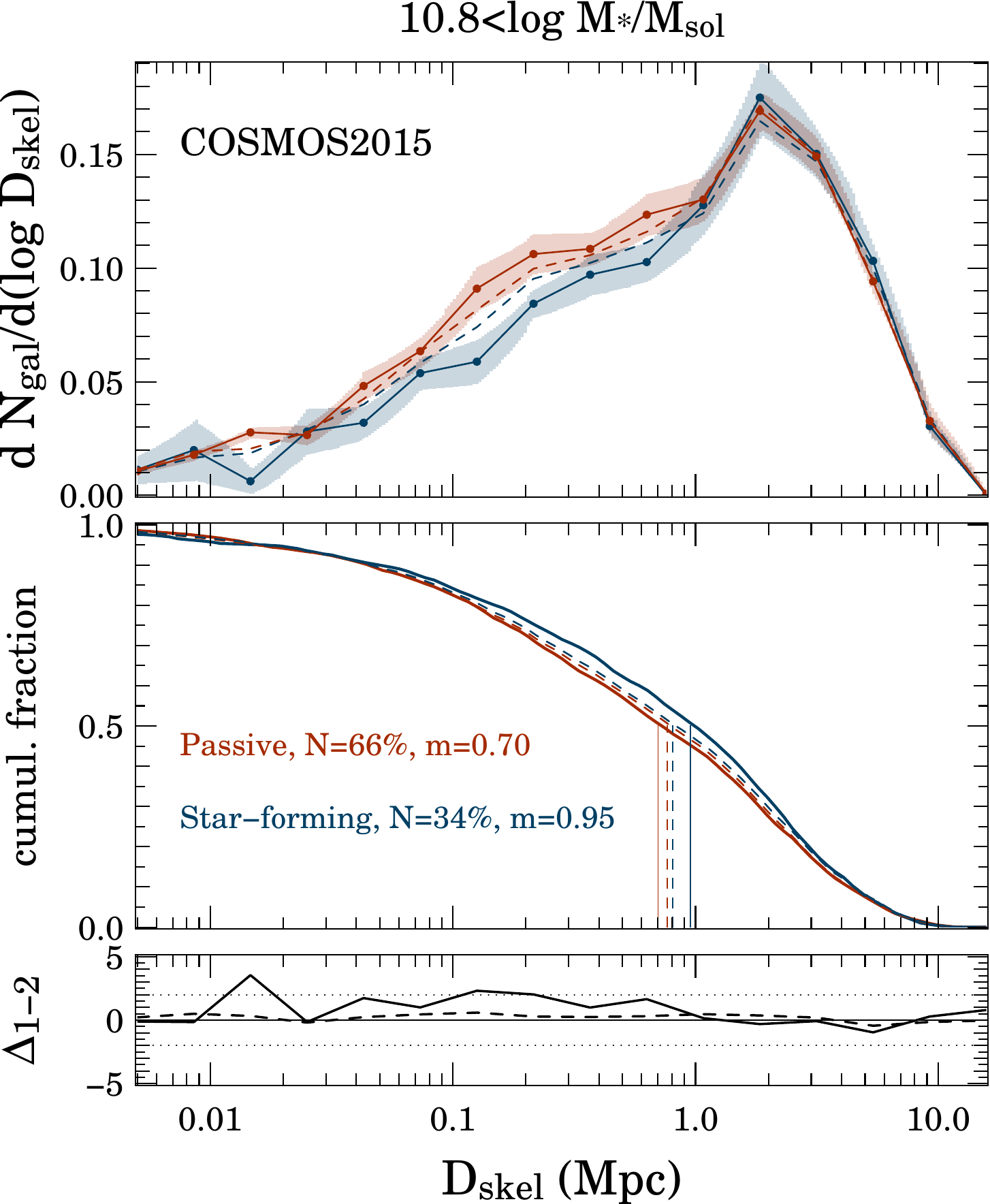}
   \hspace{0.3cm}
    \includegraphics[scale=0.5]{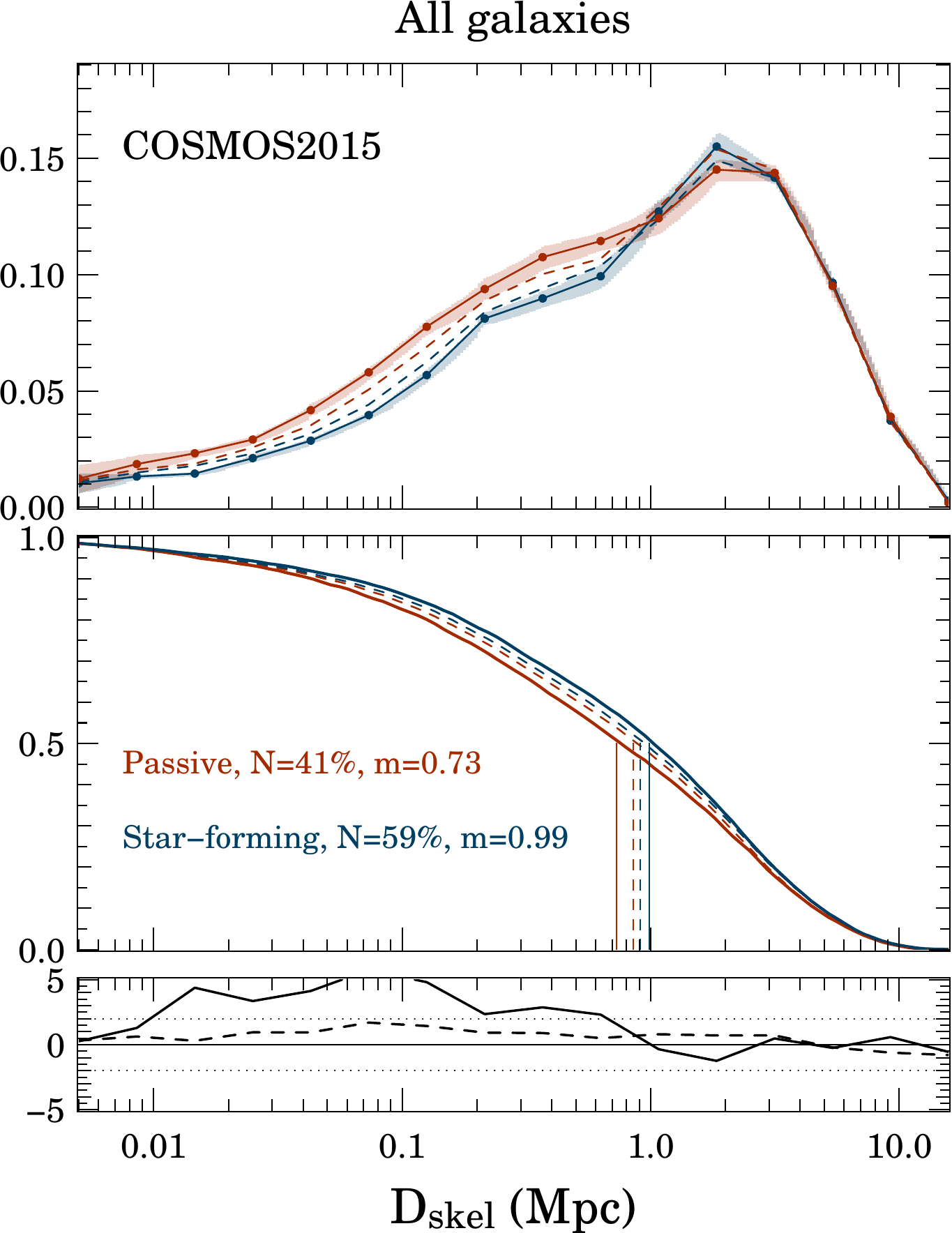}  
   \caption{
   Differential (\textit{top}) and cumulative (\textit{middle}) distributions of the distances to the two-dimensional photometric skeleton {\sc skl}$_{\rm 2D}^{\rm phot}$ for  passive (red) and star-forming (blue) galaxies in {COSMOS2015} for $10.0 <\log M_*/{\rm M}_{\odot} <10.4$ (\textit{top left}), $10.4 <\log M_*/{\rm M}_{\odot} <10.8$ (\textit{top right}), $10.8 <\log M_*/{\rm M}_{\odot} $ (\textit{bottom left}) and all galaxies such that $10. <\log M_*/{\rm M}_{\odot}$  (\textit{bottom right}).
   The value $m$ is the median of each distribution, which is also indicated by  vertical lines. $N$ gives the percentage of galaxies of each type in the considered mass bin.
   The dashed lines correspond to a random signal \textit{preserving the galaxy type-density relation} in each bin of mass: galaxy types in the catalogue have been randomised with respect to their distances to the filament in given density and mass bins.  Here, the signal measured in all slices between $z\sim 0.5$ and $z \sim 0.9$ is stacked. The contribution of nodes to the gradient signal has been removed from the analysis. The \textit{bottom} rows show the residuals between the passive and star-forming distributions for the original (solid line) and reshuffled (dashed line) samples expressed in   number of $\sigma$: $\Delta_{1-2}=\Delta(D_1-D_2)/\sqrt(\sigma_1^2+\sigma_2^2)$ where the subscripts 1 and  2 refer to the passive and star-forming distributions respectively.  }
\label{fig:passdistance2filament}
\end{center}
\end{figure*}

\begin{figure}
\begin{center}
\includegraphics[scale=0.5]{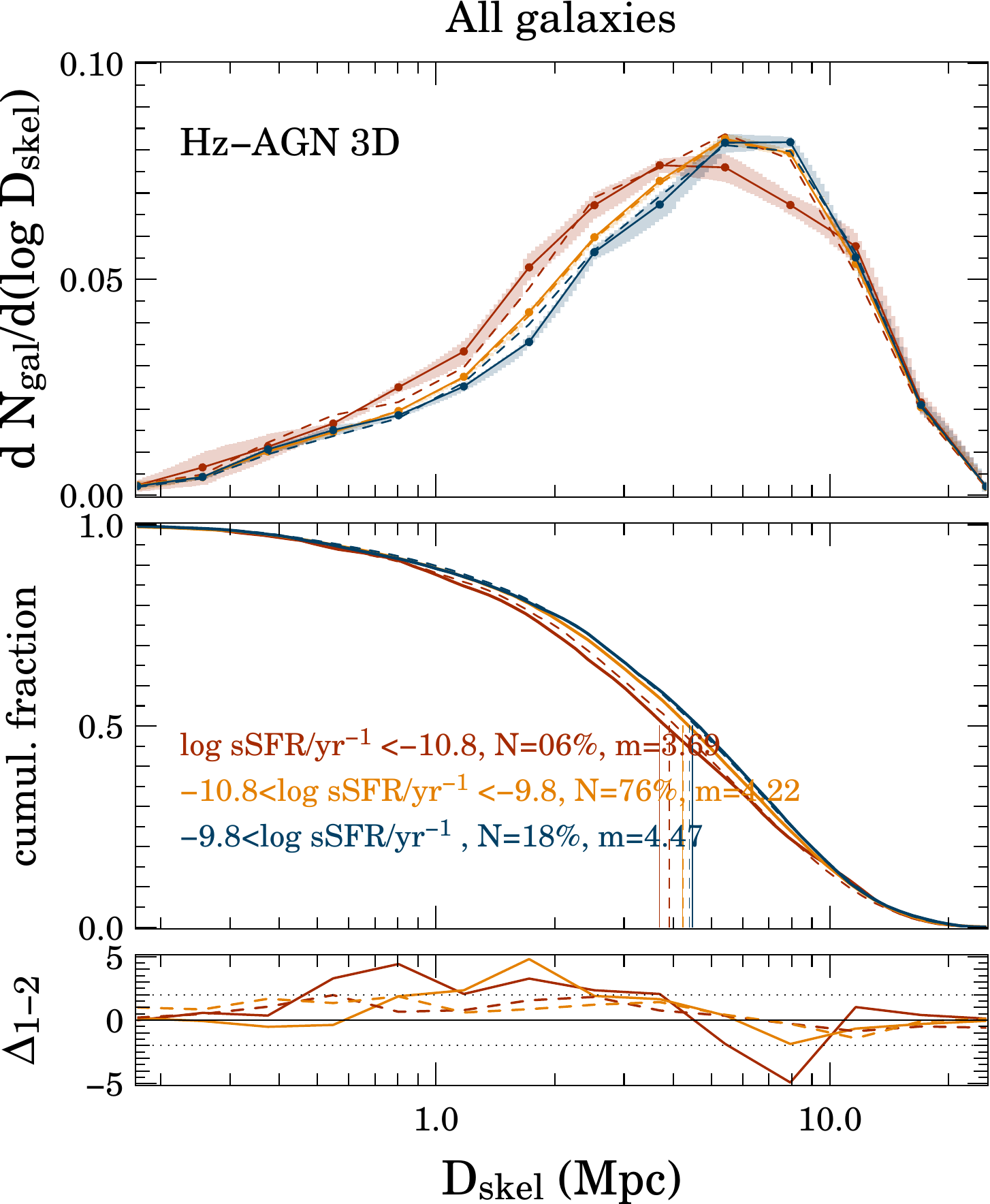}
   \caption{
   Differential (\textit{top}) and cumulative (\textit{middle}) distributions of the distances to the three-dimensional  skeleton {\sc skl}$_{\rm 3D}^{\rm DM}$ for  galaxies in {\sc Horizon-AGN} as a function of sSFR  using all galaxies such that $10. <\log M_*/{\rm M}_{\odot}$ and $0.5<z<0.9$.
   The value $m$ is the median of each distribution, which is also indicated by  vertical lines. $N$ gives the percentage of galaxies of each type in the considered mass bin.
   The dashed lines correspond to a random signal \textit{preserving the galaxy sSFR-density relation} in each bin of mass.    
   The \textit{bottom} row shows the residuals between the lowest  and intermediate sSFR bins (red line), and  between the intermediate and highest sSFR bins (orange line) for the original (solid line) and reshuffled (dashed line) samples expressed in terms of number of $\sigma$.  }
\label{fig:sSFR3D}
\end{center}
\end{figure}

%
Let us now investigate the effect of the anisotropic environment on galaxy star formation. Our COSMOS2015 galaxy sample has been divided in passive and star-forming populations based on the colour diagram, as explained in Section~\ref{Sec:COSMOS}.
Figure~\ref{fig:passdistance2filament} presents the distributions of the distances to the filaments for star-forming and passive galaxies for different mass bins in {COSMOS2015}. All galaxies with $M_*>10^{10}~{\rm M}_{\odot}$ are considered. Once again, we remove the contribution of the nodes as explained previously. At  fixed mass in all mass bins  we find that passive galaxies are statistically closer to the filament center than star-forming ones. As we minimised the contribution of galaxies in nodes, this effect is specifically related to filaments. 
However, in accordance  with the above mentioned  mass-density relation, it has long been established that at low redshift galaxies  are found to be  much less star-forming in high-density regions relative to low-density regions. As the density is anisotropically distributed in filaments, the density itself could explain these colour-driven gradients towards filaments. To disentangle the effect of the geometry of the large-scale environment  from the galaxy type-density relation, we proceed as previously. In each slice, the galaxy catalogue is split in local density bins (15 density bins logarithmically spaced) and stellar mass bins (4 stellar mass bins logarithmically spaced). Then galaxy types are randomised within these bins preserving the fraction of passive and star-forming galaxies in each bin. After reshuffling, no significant gradients are found whatever the considered mass bins. To better quantify the significance of the result, we look at the residual between the passive and the star-forming distributions ($D_1$ and $D_2$), expressed in number of $\sigma$: $\Delta_{1-2}=\Delta(D_1-D_2)/\sqrt(\sigma_1^2+\sigma_2^2)$, where the subscripts 1 and  2 refer to the passive and star-forming distributions respectively. We find  that, for the original sample, the deviation generally exceeds $1 \sigma$, and $2 \sigma$ for the intermediate mass bin ($10^{10.4}\!<\!{M_*/\rm M}_{\odot}\!<\!10^{10.8}$) and the whole galaxy population. However the reshuffling generally completely cancels the signal.  We conclude that the colour segregation within the filaments is not driven by the local density but by the geometry of the environment itself. Finally, we want to assess that these gradients are not dominated by the different mass distributions of the star-forming and passive populations even within a given mass bin. Therefore we reproduce the same measurements by selecting one galaxy sub-sample in each population with the same mass distribution. The results remain unchanged for the original sample while the residual gradients for the reshuffled one disappear.
\\
 The same measurement cannot be reproduced exactly in {\sc Horizon-AGN}, since the  separation between star-forming and passive galaxies is less straightforward in the simulation than in data. Nevertheless, we can compute the distribution of the distance to the closest filaments for galaxies in different sSFR bins.  The result for all galaxies more massive than $10^{10}$ M$_{\odot}$ are presented in Figure~\ref{fig:sSFR3D}. The measurement is computed in three dimensions, and the skeleton used is the reference dark matter skeleton {\sc skl}$_{\rm 3D}^{\rm DM}$. Three sSFR bins are chosen, such as to isolate an undoubtedly passive fraction (log sSFR/yr$^{-1}<-$10.8) and a very star-forming ones ($-9.8<{\rm log \,sSFR/yr}^{-1}$). The intermediate bin corresponds to the remaining galaxies ($-10.8<{\rm log\, sSFR/yr}^{-1}<-$9.8).
When considering together all the galaxies more massive than $10^{10}$ M$_{\odot}$, we  recover in the simulation the same trend as in COSMOS2015:  passive galaxies are closer to the filament center than star-forming ones. The gradient signal is not completely cancelled by the reshuffling, but it is stronger for the original sample than for the reshuffled one. The results for individual bins of mass are presented in Table~\ref{Tab:3dsfr}. Interestingly, depending on the mass bin, gradients are not only found between passive and star-forming populations, but also within the star-forming population. Thus there are significant gradients between the intermediate and star-forming populations for  $10.0 <\log M_*/{\rm M}_{\odot} <10.4$, while the local density alone does not drive such gradients. We also note that we do not find significant gradients for $10.4 <\log M_*/{\rm M}_{\odot} <10.8$, although that mass bin displays the strongest signal in COSMOS2015. Finally, for the highest mass bin $10.8 <\log M_*/{\rm M}_{\odot} $ a strong signal is also observed in the original sample, but the reshuffled one is consistent with no gradient. These measures are comforting the results from COSMOS2015 and highlight once again the role played by the geometry of the anisotropic environment in driving galaxy properties. Although not shown here, we find that, in the simulation, the strength of the signal is  dependent on the choice of the sSFR bins as well as the robustness of the filaments included in the analysis (which indicates that the signal is probably multi-scale). A more detailed analysis of the galaxy property gradients in the simulation taking into account these variations will be the topic of future work. 
 \begin{table*}
\begin{center}
\def\arraystretch{1.35}
\begin{tabular}{c|c|c|c}
 bin   &data &  original  &  reshuffled  \\
  \hline
   & {\sc Hz-AGN} 3D & 4.75$\pm$ 0.01 [54]  & 4.53$\pm$ 0.02    \\
  9.5$<$log$M_*/{\rm M}_{\odot}<$10 & {\sc Hz-AGN} 2D & 1.23$\pm$ 0.01 [54] & 1.01$\pm$ 0.01   \\
   & COSMOS2015 & 1.68$\pm$0.02 [47] &  1.37$\pm$ 0.02       \\
     \hline     
     
 & {\sc Hz-AGN} 3D & 4.46$\pm$ 0.01 [27] &     4.48$\pm$ 0.03   \\
     10$<$log$M_*/{\rm M}_{\odot} <$10.4& {\sc Hz-AGN} 2D & 0.70$\pm$ 0.01 [27] &  0.92$\pm$ 0.01    \\
   & COSMOS2015 & 0.97$\pm$ 0.02 [23] &   1.21$\pm$ 0.02   \\

     \hline  
      & {\sc Hz-AGN} 3D & 3.96$\pm$ 0.03 [14]  & 4.34$\pm$ 0.03   \\
       10.4$<$log$M_*/{\rm M}_{\odot}<$10.8 & {\sc Hz-AGN} 2D & 0.60$\pm$ 0.02 [14] & 0.89$\pm$ 0.02  \\
   & COSMOS2015 & 0.84$\pm$ 0.02 [18] & 1.18$\pm$ 0.03 \\
   
     \hline  
  & {\sc Hz-AGN} 3D & 3.07$\pm$ 0.08 [05]  &   4.18$\pm$ 0.08    \\
       10.8$<$log$M_*/{\rm M}_{\odot}$  & {\sc Hz-AGN} 2D & 0.42$\pm$ 0.03 [05] & 0.86$\pm$ 0.03    \\
   & COSMOS2015 & 0.79$\pm$ 0.05 [12] &  1.16$\pm$ 0.04     \\
     \hline    
\end{tabular}
  \caption{
Summary of the results from the mass gradients towards filaments. The given values are the medians of the distributions and the standard errors from bootstrap resampling with 100 realisations expressed in comoving Mpc. Note that the median values in three- and in two-dimensions are not directly comparable because of projection effects. The bracketed values indicate the percentage of galaxies in each mass bin. The percentages are the same for the original and reshuffled samples. The reshuffled sample corresponds to a random signal preserving the galaxy mass-density relation. 
}
\label{Tab:results}
\end{center}
\end{table*}
%
 \begin{table*}
\begin{center}
\def\arraystretch{1.35}
\begin{tabular}{c|c|c|c|c|c}
        &  \multicolumn{3}{c|}{ original sample}  & \multicolumn{2}{c}{ reshuffled samples } \\
       \hline
 bin  & All &  SF   & passive  &  SF   & passive \\
        \hline
     10$<$log$M_*/{\rm M}_{\odot} <$10.4  & 0.97$\pm$ 0.02  &  1.01$\pm$ 0.04 [72] & 0.82$\pm$ 0.05 [28]  &  0.96$\pm$ 0.03 & 0.96$\pm$ 0.05   \\
     \hline 
       10.4$<$log$M_*/{\rm M}_{\odot}<$10.8   & 0.84$\pm$ 0.02  &  0.96$\pm$ 0.04 [58] & 0.69$\pm$ 0.03 [42]  &  0.86$\pm$ 0.04 & 0.82$\pm$ 0.04   \\
     \hline  
  10.8$<$log$M_*/{\rm M}_{\odot}$     & 0.79$\pm$ 0.05  &  0.95$\pm$ 0.07 [34] & 0.70$\pm$ 0.04 [66]  &  0.81$\pm$ 0.08 & 0.76$\pm$ 0.05   \\
      \hline  
       10$<$log$M_*/{\rm M}_{\odot}$  [All]    & -- &  0.99$\pm$ 0.03 [59] & 0.73$\pm$ 0.03 [41] &   0.91$\pm$ 0.02 & 0.86$\pm$ 0.03   \\
     \hline    
\end{tabular}
  \caption{
Summary of the results from the colour-type gradient signal towards filaments in COSMOS2015. The given values are the medians of the distributions and the standard errors from bootstrap resampling with 100 realisations expressed in projected comoving Mpc.   The bracketed values indicate the percentage of galaxies  within a mass bin in each population. The percentages are the same for the original and reshuffled samples. The reshuffled sample corresponds  to a random signal preserving the galaxy type-density relation. 
}
\label{Tab:results2}
\end{center}
\end{table*}
%
 \begin{table*}
\begin{center}
\def\arraystretch{1.35}
\begin{tabular}{c|c|c|c|c|c|c|c}
        &  \multicolumn{4}{c|}{ original sample}  & \multicolumn{3}{c}{ reshuffled samples } \\
       \hline
 bin (log sSFR/yr$^{-1}$) & All & $<-10.9$ &  $\left[-10.9,-9.8\right]$   & $-9.8<$  & $<-10.9$& $\left[-10.9,-9.8\right]$  & $-9.8<$ \\
        \hline
     10$<$log$M_*/{\rm M}_{\odot} <$10.4  & 4.46$\pm$ 0.01 & 4.38$\pm$ 0.12 [03] &  4.45$\pm$ 0.03 [75] & 4.70$\pm$ 0.05 [22]  &  4.54$\pm$ 0.14 & 4.50$\pm$ 0.03 & 4.52$\pm$ 0.05   \\
     \hline 
       10.4$<$log$M_*/{\rm M}_{\odot}<$10.8 & 3.96$\pm$ 0.03 & 3.87$\pm$ 0.26  [07] &  4.06$\pm$ 0.04 [79] & 3.94$\pm$ 0.18 [14]  & 3.99$\pm$ 0.11 & 4.02$\pm$ 0.04 & 4.02 $\pm$0.16   \\
     \hline  
  10.8$<$log$M_*/{\rm M}_{\odot}$   & 3.07$\pm$ 0.08 & 2.89$\pm$ 0.10 [21]  & 3.23$\pm$ 0.13 [74] & 3.24$\pm$ 0.13 [06]  &  3.09$\pm$ 0.10 & 3.16$\pm$ 0.07 & 3.10 $\pm$ 0.22   \\
      \hline  
       10$<$log$M_*/{\rm M}_{\odot}$ [All]    & -- & 3.69 $\pm$ 0.05 [06] & 4.22$\pm$ 0.03 [76] & 4.47$\pm$ 0.05 [18] &   3.91$\pm$ 0.01 & 4.23$\pm$ 0.02 & 4.39 $\pm$ 0.07   \\
     \hline    
\end{tabular}
  \caption{
Summary of the results from the sSFR gradient signal towards filaments in {\sc Horizon-AGN} using {\sc skl}$_{\rm 3D}^{\rm DM}$. The given values are the medians of the distributions and the standard errors from bootstrap resampling with 100 realisations expressed in comoving Mpc. Note that the median values found in three dimensions are not directly comparable to the two-dimensional ones because of the projection. The bracketed values indicate the percentage of galaxies  within a mass bin in each population. The percentages are the same for the original and reshuffled samples. The reshuffled sample corresponds  to a random signal preserving the galaxy sSFR-density relation in each mass bin. Galaxies with log sSFR/yr$^{-1}$  $<-10.9$ are considered as passive, while those with log sSFR/yr$^{-1}$  $>-9.8$ are star-forming.
}
\label{Tab:3dsfr}
\end{center}
\end{table*}
%
\subsection{Interpretation and discussion}
\label{Sec:Discussion}
%
Our results are quantitatively summarised in Tables~\ref{Tab:results},~\ref{Tab:results2} and~\ref{Tab:3dsfr} which report the median values of each distribution. The normalisation of these values are quite sensitive to cosmic variance and to the chosen thickness of the slices for the two-dimensional reconstruction. Hence they have to be taken cautiously  when comparing to other datasets. However the significant differences between the medians of the different populations, either divided in mass bins, sSFR bins or in terms of colour-types, is a robust result which allows us to report   mass gradients towards filaments in COSMOS2015 and {\sc Horizon-AGN}, and colour-type gradients towards filaments in COSMOS2015. In {\sc Horizon-AGN}, we also find sSFR gradients in agreement with COSMOS2015.  They corroborate results found in other fields by independent methods of analysis \citep{alpaslanetal2016,Malavasi2016b,poudeletal2016,chenetal2017}. 
\\
 Theoretical considerations, together with  measurements both in observations and simulations give us clues to   how galaxy properties depend on the geometry of their environment. 
    Galaxy harassment \citep{mooreetal1996}, ram-pressure stripping of the gas \citep{Gunn1972,quilisetal2000}, strangulation of low-mass satellite galaxies \citep{larsonetal1980,baloghetal2000} or cumulative galaxy-galaxy hydrodynamic/gravitational interactions \citep{ParketHwang2009} may explain the depletion or heating of the cold gas in  quenched galaxies in group or cluster environment, i.e. mainly at the node of the cosmic web.  At intermediate density ranges, galaxy mergers, while temporarily modestly increasing star formation \citep[e.g.][]{jogeeetal2009,Kaviraj15}, may also halt star-formation and turn  star-forming galaxies into passive ones~\citep[e.g.][]{duboisetal16}. But none of these processes can be  responsible for  galaxy segregation  depending only on  distance to  filaments  taking place in regions of low to intermediate densities, where galaxy interactions are limited. 
    
    Since  the  signal found in this work  is sensitive to the large-scale geometry of the matter distribution itself, it must be driven by the large scale (traceless part of the) tidal tensor, which  quantifies this anisotropy.
  The physical properties of  dark matter halos \citep{codisetal2012} and their host galaxies~\citep{duboisetal14,welkeretal14} indeed correlate with the {\sl dynamics} of their embedding  anisotropic large scale structure. This has  identified kinematic signatures: angular momentum advection  \citep{codisetal2015}
 and internal kinetic anisotropy \citep{Faltenbacher:2009fl} are also set by the larger scale tides  of the environment.  
 The induced tides  not only impact the merger tree and accretion history of the host, but also the filamentary flow of cold gas connecting to the host,  hence its coherent gas supply.   
 The past and present efficiency of star formation, as traced by the mass, colours and specific star formation rate, depends critically on the infalling rate and impact parameter of cold gas. It 
does not only depend on the timing of infall (in connection with the accretion history of the host), but also on its geometry and physical content, in connection with the dynamics of the matter flows
(e.g. in plane gas-rich and co-rotating, or dry mergers with random orbital parameters). 
 As discussed in the introduction, the vorticity-rich large-scale filaments  are indeed the locus where low-mass galaxies steadily grow in mass via quasi-polar cold gas accretion \citep{welkeretal2015b}, with their angular momentum  aligned with the host filament.  
Galaxies are expected to accrete more efficiently cold gas when their angular momentum is aligned with the preferential direction of the gas infall, \textit{i.e.} aligned with the filament   \citep{pichonetal11}. 
   Hence one  expects star formation efficiency to be strongest wherever the alignment is tightest.   Maximal alignment occurs in the highest vorticity regions \citep{laigle2015}, i.e at the edge of  filaments  (whereas haloes terminate their mass assembly in the filament core while drifting towards nodes  following the large-scale flow). 
The locus of this induced excess of star formation  should therefore have measurable signatures in observations when quantified in the metric of the filament, as discussed e.g. in \cite{codisetal2015}.  This is fully consistent with 
what is observed in Figure~\ref{fig:passdistance2filament} and~\ref{fig:sSFR3D}.

In the present study we have attempted to disentangle the specific signature {of} this anisotropy while relying on the physical 
properties of galaxies at fixed mass.
The various tests presented in this paper  confirm that the anisotropy of the environment  traced by transverse gradients 
with respect to  filaments  is detected in COSMOS2015 catalogue, as anticipated and quantified from mocks derived from the light-cone of {\sc Horizon-AGN}. 
We provide observational  evidence that what has been a  pillar of galaxy formation, i.e. the assumption that the average effects of the environment  can be approximated  by the effect of the  (in particular azimuthally)  average environment is only valid at first order.
This assumption  is obviously strictly true for linear dynamic and linear observables, which does not capture the full complexity of what builds up the properties of galaxies. 
This  past simplification was partially driven by the community's strong focus on two-point functions, which cannot capture the anisotropy of the cosmic web.
Yet it had long been known that for instance angular momentum -- which is  undoubtedly built from anisotropic tides -- is  a key underlying  property   driving morphology, which correlates strongly with colour and star formation efficiency. 
Hence one could attempt to extend, e.g. assembly bias theory \citep{shethettormen2004} to try and explain the observed diversity at a given  mass, so long as it is understood that mass assembly is in fact also driven by anisotropic large scale tides, which will  impact gas inflow towards galaxies, hence their properties
\citep{Yan2013,Tramonte2017}. 

More elaborate -- but still restricted in scope --  bias theories such as excursion set peak theories involving a moving barrier account for the other invariants of the  tidal tensor \cite[see][and reference therein]{Desjacques2016}, but fail to incorporate the effect of the orientation of its eigen-vectors. Semi-analytical models \citep[e.g.][]{LaceySilk1991} do account for the 
timings encoded in merging trees  measured in simulations-- which reflect in part  the cosmic web -- but  are restricted to following dark halo only and typically leave out the configuration of their mergers. 
\\
An improved model for galaxy properties should therefore explicitly integrate the diversity of the  topology of the large environment on multiple scales \citep[following, e.g. ][]{Hanami2001}
and quantify the impact of its anisotropy on galactic mass assembly history, and more generally on the kinematic history of galaxies \citep[e.g. extending][]{codisetal2015}.
What also remains to be understood is  what process dominates at what redshift:   does the  offset of merger and accretion rate     imposed by the large scale turbulent flow  explain most of the environment dependence in observed physical properties \citep{CWD2016}, or is   the geometry of  gas inflow within filaments  prevalent in 
feeding galactic discs coherently  \citep{pichonetal11,stewart11}?  The kinematic of the large scale flow is neither strictly  coherent  nor fully turbulent. Both processes will in turn modify the efficiency of star formation and feedback, which is  critically non-linear, currently poorly understood and fine tuned. In order to disentangle these effects, the  next step  will involve studying the connection between galaxy properties, angular momentum and distance to filaments in the {\sc Horizon-AGN} simulation and in connection with  the COSMOS2015 catalogue.

%

\section{Conclusion}
\label{Sec:Conclusion}

Using the photometric catalogue {COSMOS2015}, we  investigated how galaxy properties depend on their anisotropic  environment. A realistic photometric catalogue extracted from the {\sc Horizon-AGN} hydrodynamical simulation was used to quantify our ability to recover the filamentary structures from two-dimensional slices at the same photometric redshift precision, and to predict  the expected gradients. The main findings of this work are the following:
\begin{itemize}
\item Photometric surveys with a photometric precision similar to {COSMOS2015} offer the prospect to reliably study the 3D properties of cosmic web from projected two-dimensional slices;
\item We observe both in {\sc Horizon-AGN} and in {COSMOS2015} a significant tendency for massive galaxies to be closer to the filament center than for less massive galaxies; 
\item We also observe segregation of passive and star-forming galaxies near the filaments both in {COSMOS2015} and {\sc Horizon-AGN}. At fixed mass, passive galaxies are more confined in the core of the filament than star-forming ones. These two signals persist when minimising the contribution of nodes and cannot be  explained by the local mass-density relation alone.
\end{itemize}
These findings are qualitatively consistent between simulation and data, and in agreement with   studies  relying on spectroscopic redshifts  \citep[e.g.][]{Malavasi2016b}. They underline  the specific  role played by the anisotropy of the large scale  cosmic web  in shaping galaxies properties. The geometry of the large scale environment  drives tides, which  impacts the  merger and accretion history of galaxies, their spin and in turn their observed properties. 
Mass and density alone are not sufficient to characterise the observed properties of galaxies aways from the cosmic web.

More observations are needed to disentangle the effects of all these competing processes,  while relying on a wider fields and  (photometric) redshift baseline, towards the peak of the cosmic star formation,  where cold streams are expected to be more efficient. 
This will be  challenging as one will need to carefully disentangle the evolutionary effects and selection biases.   An alternative observational strategy  is to rely on  ongoing \cite[SAMI and MANGA][]{sami2012,Manga2015} or upcoming \cite[Hector,][]{Hector2015} integral field spectroscopy surveys and explore spin properties of galaxies \citep[e.g.][]{harrisonetal2017} such as their physical and kinematic axis orientation as a function of the large scale environment. Simulated end-to-end catalogues  such as those produced here will remain  key to qualify these investigations.

\vskip 0.1cm
\section*{Acknowledgments}
{\sl 
CL was supported by the ILP LABEX (under reference ANR-10-LABX-63 and ANR-11-IDEX-0004-02) during the first part of this work and is now  supported by a Beecroft Fellowship. 
JD and AS acknowledges funding support from Adrian Beecroft, the Oxford Martin School and the STFC. OI acknowledges the funding of the French Agence Nationale de la Recherche for the project
``SAGACE''.  This work  relied on the HPC resources of CINES (Jade) under the allocation 2013047012 and c2014047012 made by GENCI
and on the Horizon Cluster hosted by Institut d'Astrophysique de Paris. We  warmly thank S.~Rouberol for running  the  cluster on which the simulation was  post-processed. 
This research is part of  Spin(e) (ANR-13-BS05-0005, \url{http://cosmicorigin.org}) and {\tt horizon-UK}.   
  This research  is also partly supported by the Centre National d'Etudes
Spatiales (CNES). This work is based on data products from
observations made with ESO Telescopes at the La Silla Paranal
Observatory under ESO programme ID 179.A-2005 and on data products
produced by TERAPIX and the Cambridge Astronomy Survey Unit on behalf
of the UltraVISTA consortium.
We thank D.~Munro for freely distributing his {\sc \small  Yorick} programming language and opengl interface (available at \url{http://yorick.sourceforge.net/}).
}
\vspace{-0.5cm}

\bibliography{author}

\appendix
\section{Choice of  persistence in 2D}
\label{Ap:persistence}
\begin{figure}
 \includegraphics[scale=0.41]{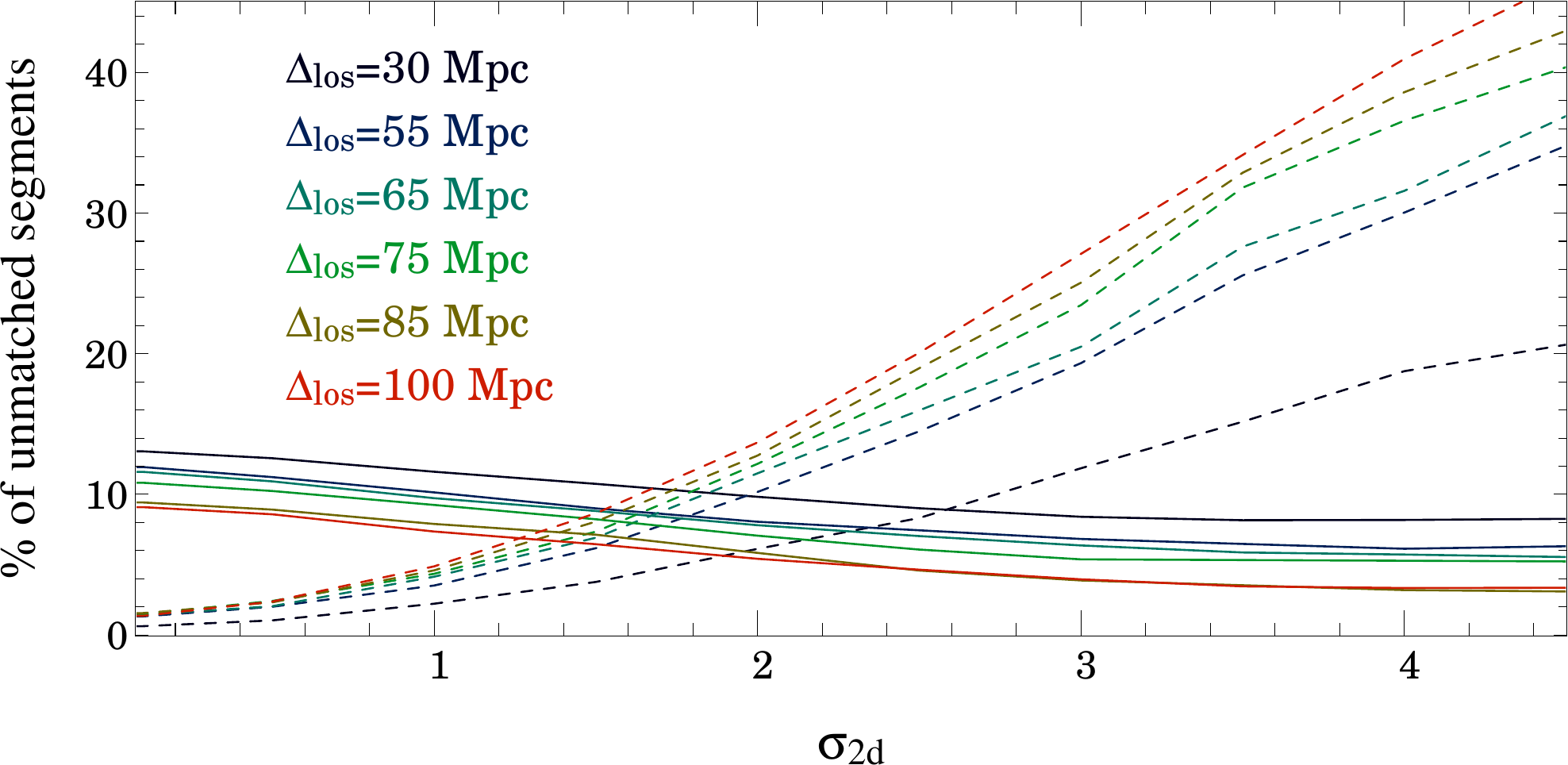} 
 \caption{Percentage of unmatched segments between {\sc skl}$_{\rm 3D}^{\rm DM}$ and {\sc skl}$_{\rm 2D}^{\rm spec}$ for different slice thicknesses as a function of the persistence threshold used for the computation of {\sc skl}$_{\rm 2D}^{\rm spec}$. The solid lines refer to the percentage of {\sc skl}$_{\rm 2D}^{\rm spec}$ segments which do not have a counterpart in {\sc skl}$_{\rm 3D}^{\rm DM}$, and the dashed line refer to the percentage of {\sc skl}$_{\rm 3D}^{\rm DM}$ segments which do not have a counterpart in {\sc skl}$_{\rm 2D}^{\rm spec}$.
 The mass limit used for the computation of {\sc skl}$_{\rm 2D}^{\rm spec}$ is 10$^{10}{\rm M}_{\odot}$, but the trend remains in overall the same when this limit varies.}
\label{fig:unmatchedseg}
\end{figure}

As stated in Section~\ref{subsec:skeleton},  the extraction of  skeletons with the same persistence thresholds $\sigma$ in two and three dimensions will typically not lead to the detection of the same filaments. 
Indeed, one must   significantly decrease $\sigma$ with respect to its three-dimensional value to recover in two dimensions most of the three-dimensional filaments. Let us estimate  the optimal two-dimensional $\sigma$ which  recovers most of the filaments without introducing  spurious detections. This optimal $\sigma$ is estimated from the simulated catalogue of {\sc Horizon-AGN}, by comparing the number of unmatched segments between {\sc skl}$_{\rm 2D}^{\rm spec}$ and  {\sc skl}$_{\rm 3D}^{\rm DM}$. We compute first the distances between the two skeletons. In order   to  compare equitably the percentage of unmatched segments for different slice thicknesses, the distances between the two skeletons are divided by the mean two-dimensional inter-galaxy distance $D_{\rm int}$ (which increases with the slice thickness), taken as the square root of the area divided by the number of galaxies in the slice. Then we define the percentage of unmatched segments as the fraction of segments such as their closest counterpart is further than $ D_{\rm int}$ (changing this threshold would move Figure~\ref{fig:unmatchedseg} vertically). The percentage of unmatched segments is shown in Figure~\ref{fig:unmatchedseg}. In order  to reduce as much as possible the number of {\sc skl}$_{\rm 3D}^{\rm DM}$ segments which do not have a counterpart in {\sc skl}$_{\rm 2D}^{\rm spec}$, without sensitively  increasing spurious filament extraction in {\sc skl}$_{\rm 2D}^{\rm spec}$, we find that a 2$\sigma$ persistent threshold for {\sc skl}$_{\rm 2D}^{\rm spec}$ offers a reasonable balance between the percentage of unmatched segments in  {\sc skl}$_{\rm 2D}^{\rm spec}$ and  {\sc skl}$_{\rm 3D}^{\rm DM}$, whatever the slice thickness. Although {\sc skl}$_{\rm 2D}^{\rm spec}$ is computed   for Figure~\ref{fig:unmatchedseg} with a mass limit of $10^{10} {\rm M}_{\odot}$, this result does not significantly change with a different mass limit.
\section{Galaxy versus DM skeletons}
\label{Ap:Dmgal}
\begin{figure}
 \includegraphics[scale=0.41]{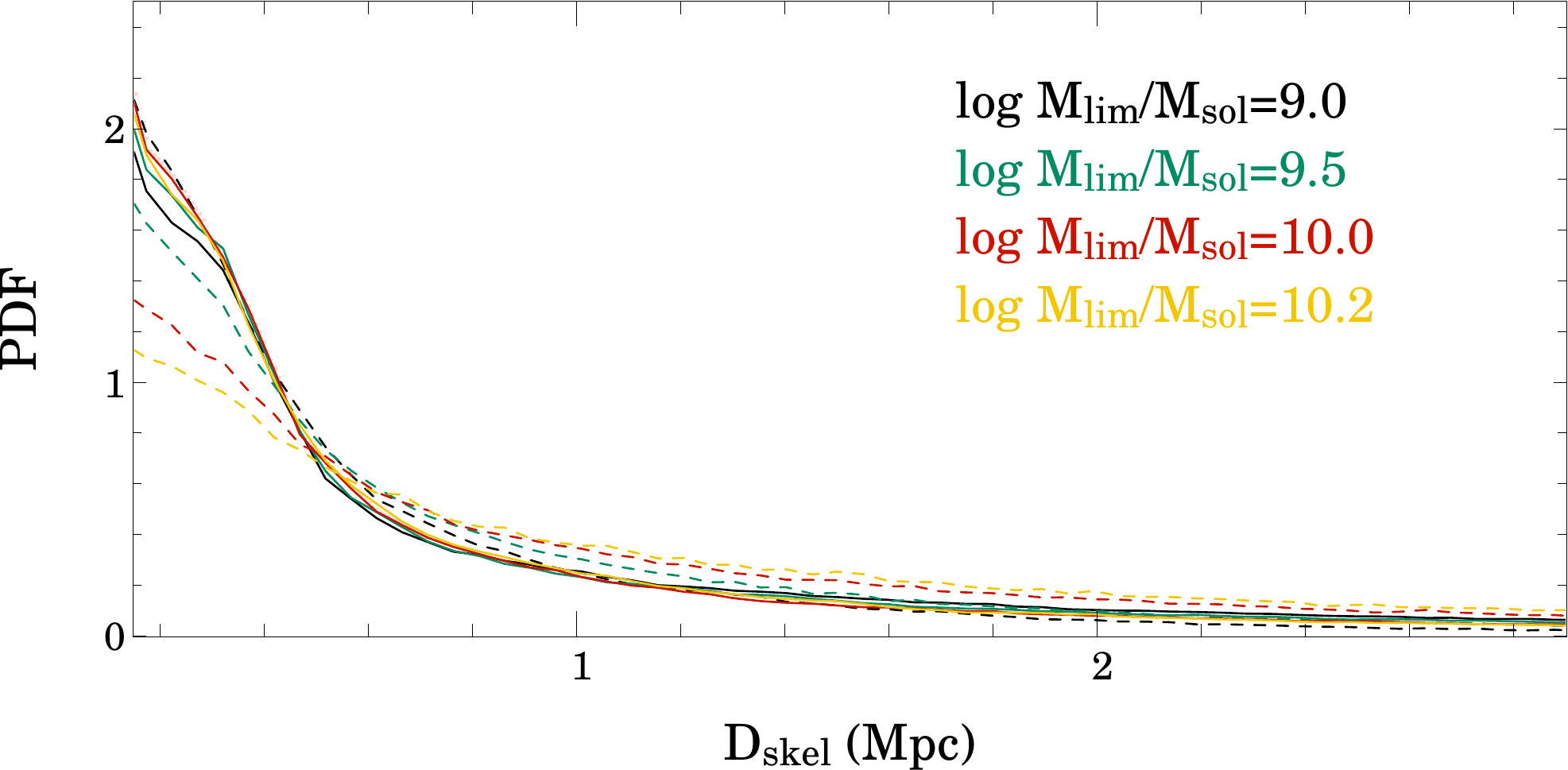} 
 \caption{Distribution of the distance from {\sc skl}$_{\rm 3D}^{\rm gal}$ towards {\sc skl}$_{\rm 3D}^{\rm DM}$ (solid lines) and conversely (dashed lines) calculated with different mass limits ranging from 10$^{9}$M$_{\odot}$ to 10$^{10.2}$M$_{\odot}$.}
\label{fig:distanceDMGal}
\end{figure}

The skeleton computed from dark matter particles is the most suitable to stand as the reference skeleton, since the galaxy distribution is a biased tracer of the underlying density field. In this Appendix, we compare the three-dimensional galaxy and dark matter skeletons and we highlight how their relative difference   depends on the mass limit of the galaxy sample. To get values directly comparable to Figure~\ref{fig:dist2d3d}, we measure the distances between the galaxy and dark matter skeletons in projection, although both skeletons are computed in three dimensions.\\
 The distribution of the distance between the dark matter and galaxy skeletons {\sc skl}$_{\rm 3D}^{\rm gal}$ and {\sc skl}$_{\rm 3D}^{\rm DM}$ are plotted in Figure~\ref{fig:distanceDMGal} for different mass limits. The solid lines refer to the distance from the galaxy skeleton towards the dark matter one, while the dashed lines refer to converse. All  solid lines are almost overlapping. These distribution are  peaked at very close separation, but still display a long tail. This suggests that whatever the chosen mass limits for the galaxy skeleton reconstruction, the galaxy filaments have generally a very close counterpart in the dark matter skeleton, although a small fraction has no counterpart at all.  We note a  slight improvement in  recovering  the DM skeleton when working with increasing masses. However, when decreasing the mass limit of the galaxy sample, the number of unmatched filaments in the dark matter skeleton decreases strongly (the tail of the dashed distribution decreases).  To summarise, working  only with massive galaxies allows us to recover the main dark matter filaments quite accurately, but at the price of missing a non negligible fraction of dark matter filaments. A reasonable balance involves choosing a mass limit between 10$^{9.5}$ and 10$^{10} {\rm M}_{\odot}$. 
Finally we note that these variations are relatively small; overall the width of these distributions are  smaller than those displayed in Figure~\ref{fig:dist2d3d}.

\balance

\section{Choice of  slice thickness}
\label{Ap:thickness}
We aim now to determine if there is an optimal slice thickness to extract the cosmic web  and  estimate transverse gradients. We proceed in two steps. First, given an ideal survey (without photometric redshift uncertainties), we  estimate how the accuracy of the reconstruction varies with the thickness of the slices. Secondly, we estimate how  the quality of the reconstruction varies as a function of  mass limits given a photometric redshift sample. Given the redshift accuracy of the {COSMOS2015} catalogue,we choose then the optimal slice thickness and mass limit. 
\subsection{Optimisation of  slice thickness with exact redshifts}
The optimal slice thickness should lead to the best  match between the two-dimensional skeleton and its projected three-dimensional counterpart. The agreement is quantified from the distribution of the distances between the projected {\sc skl}$_{\rm 3D}^{\rm DM}$ skeleton and  {\sc skl}$_{\rm 2D}$.  From the distribution, we measure the median and the width which encompasses 68\% of the distribution around the median. Good agreement between the skeletons requires both the median and the width to be  small. Let us first test the single effect of the projection for different thicknesses, independently of the redshift accuracy.
\begin{figure}
 \includegraphics[scale=0.41]{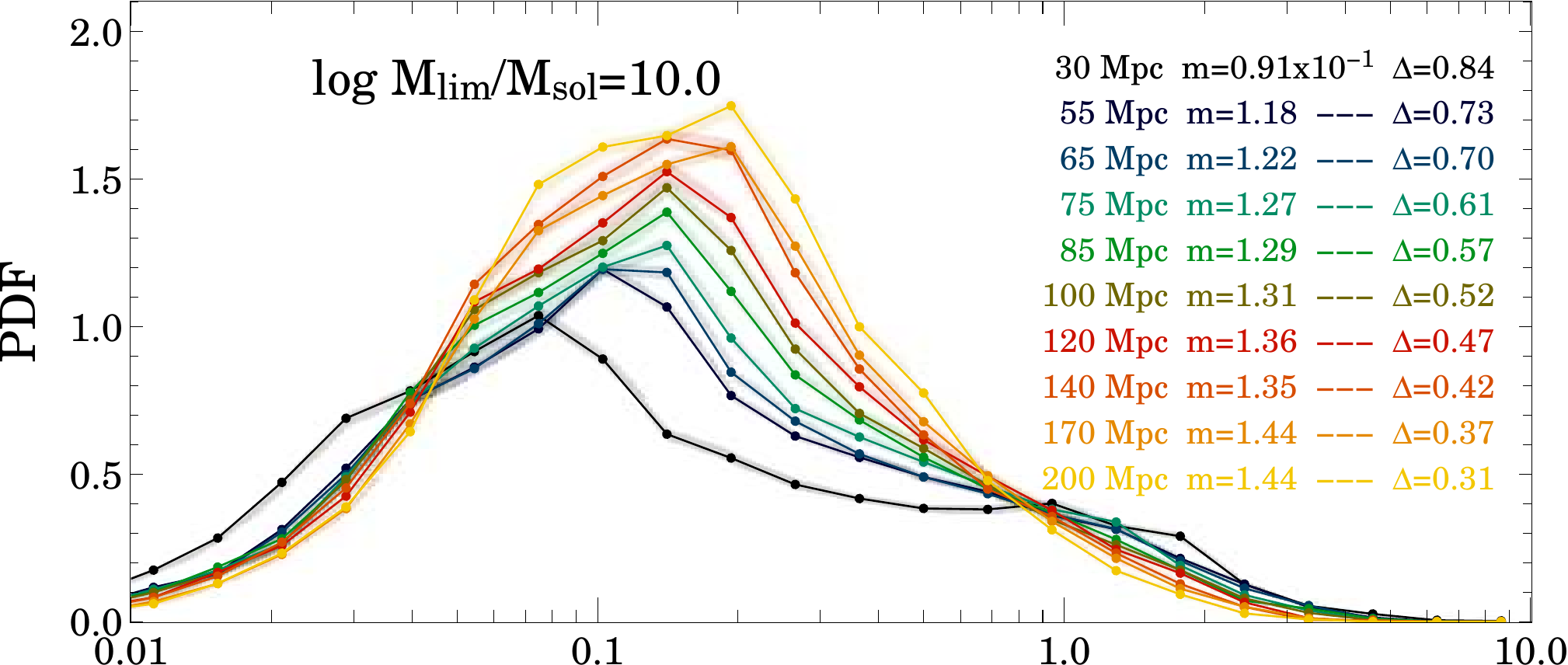}  
  \caption{Distribution of the distance between {\sc skl}$_{\rm 2D}^{\rm spec}$ and {\sc skl}$_{\rm 3D}^{\rm DM}$, for slices of different thicknesses ranging from 30 Mpc to 200 Mpc, with a limiting mass of 10$^{10}$M$_{\odot}$. Redshift uncertainties are not introduced here. The values $m$ and $\Delta$ indicate the median and the width (68\% around the median) of the distributions. The distances are measured in all slices between $z\sim 0.5$
 and $z \sim 0.9$. }
\label{fig:distanceslicea}
\end{figure}
The distances are normalised by the mean inter-particle distance $D_{\rm int}$.  Figure~\ref{fig:vis3thick}  shows three slices of {\sc skl}$_{\rm 2D}^{\rm spec}$ and the projected {\sc skl}$_{\rm 3D}^{\rm DM}$, corresponding to three slice thickness: 30 Mpc, 75 Mpc and 170 Mpc around redshift $z\sim0.59$. In overall the more robust two-dimensional filaments trace existing three-dimensional filaments. For the lowest thickness slice, most of the three-dimensional filaments appear to have a counterpart in the two-dimensional skeleton, but there are a lot of fake filaments, generally with a low robustness  because the skeleton traces only fragments of filaments. For the highest thickness slice, this fraction of fake two-dimensional filaments decreases strongly, but much more three-dimensional filaments have no counterpart. 
We statistically quantified it in Figure~\ref{fig:distanceslicea}, which shows the distribution of the distances between {\sc skl}$_{\rm 2D}^{\rm spec}$ (M$_{\rm lim}=$10$^{10}$M$_{\odot}$) and {\sc skl}$_{\rm 3D}^{\rm DM}$, for slices of different thicknesses ranging from 30 Mpc to 200 Mpc. 
It therefore appears that for exact redshifts, there is no optimal slice thickness within the considered range. Working with thin slices ensures accurate  filaments identification, but a non negligible fraction of two dimensional filaments  have no counterpart at all in the three dimensional skeleton. This fraction diminish significantly when increasing the slice thickness, but at the price of less accurate filament identifications. In addition, as displayed by the black line in Figure~\ref{fig:distanceslice}, the fraction of  unmatched three-dimensional segments increases significantly with the thickness. Hence a reasonable choice could lie between 60 and 100 Mpc.  But our final choice of the slice thickness will be in any case limited by the redshift accuracy of the sample.
Finally, Figure~\ref{fig:vis3thick} indicates that filtering low robustness segments would be a way to remove the filaments which have probably no counterpart in three dimensions. This could be   of interest when working with  thicker slices. However, as the robustness is an estimation of the density of the filament with respect to the local background, keeping only the more robust filament would bias the measure towards the denser filaments. In addition, keeping only segments above a certain robustness threshold would destroy the connectivity of the cosmic web by removing entire sections of filaments. It would bias  statistical measurements on the cosmic web, hence  prevent accurate  cosmological measurements based on such statistics. 
Note that although only the results for a mass limit of 10$^{10}$M$_{\odot}$ are shown in this Section, the  trend is  preserved  when the mass limit of the sample varies.   
\begin{figure*}
 \includegraphics[scale=0.41]{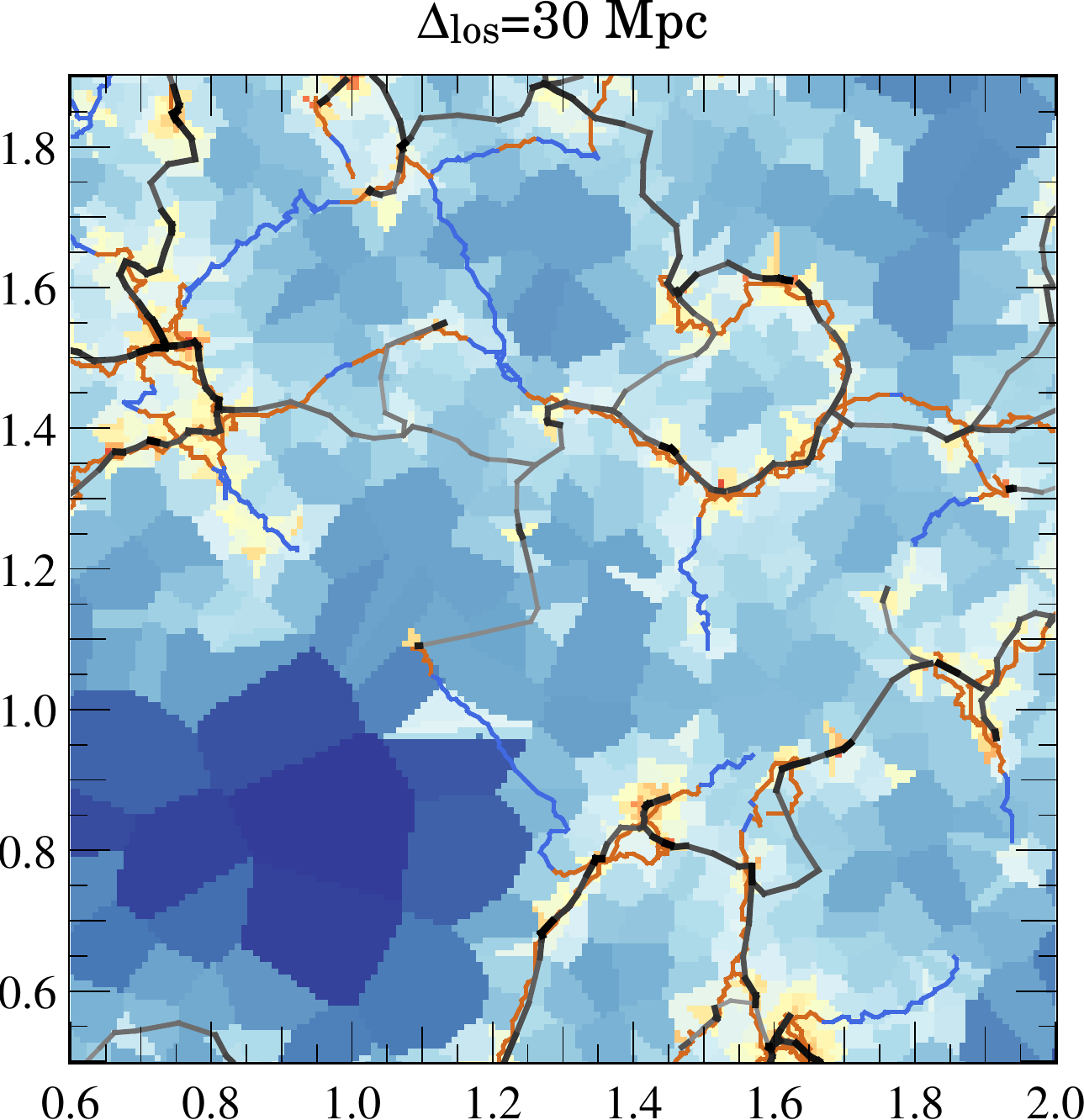} 
  \includegraphics[scale=0.41]{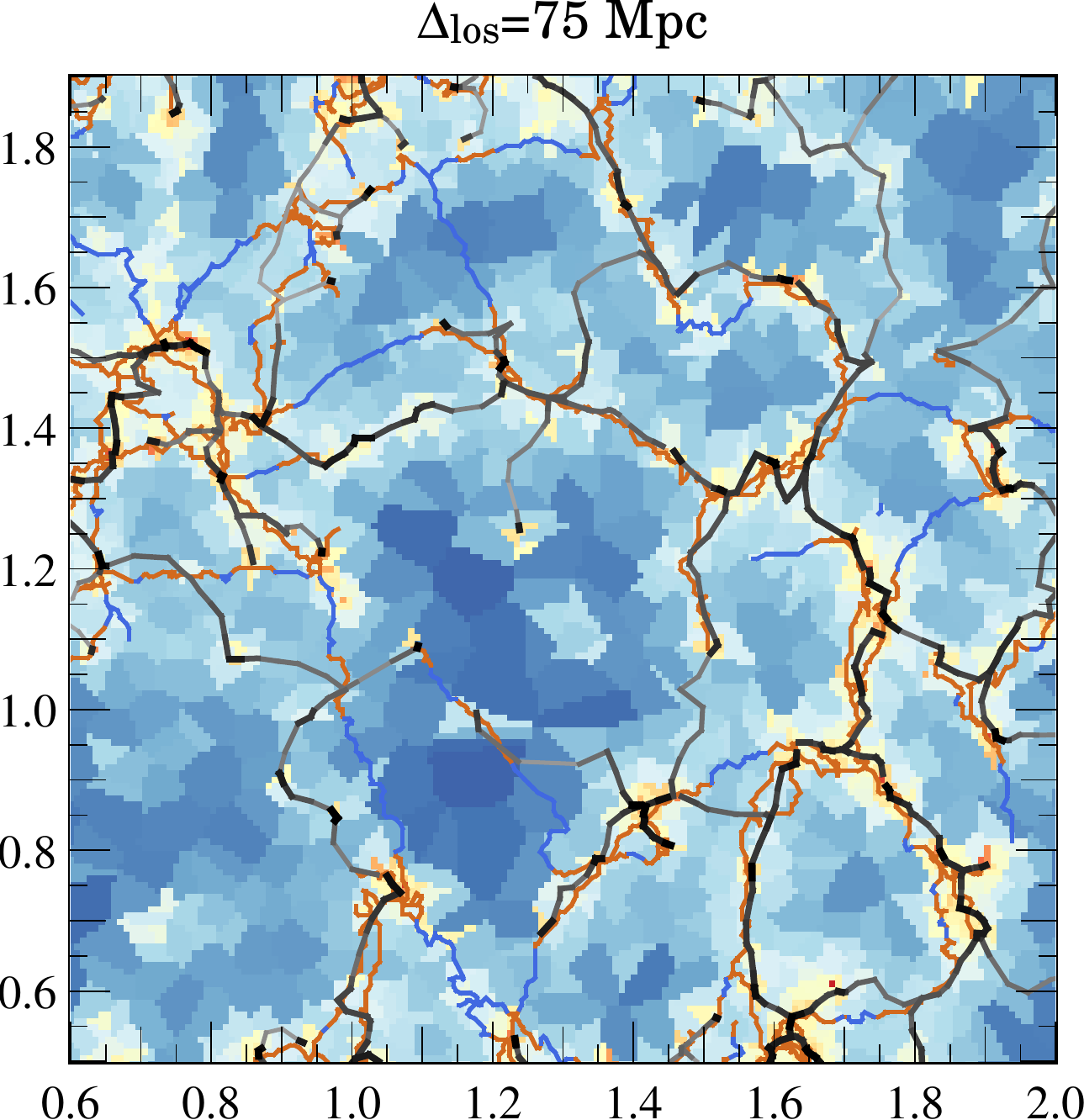} 
   \includegraphics[scale=0.41]{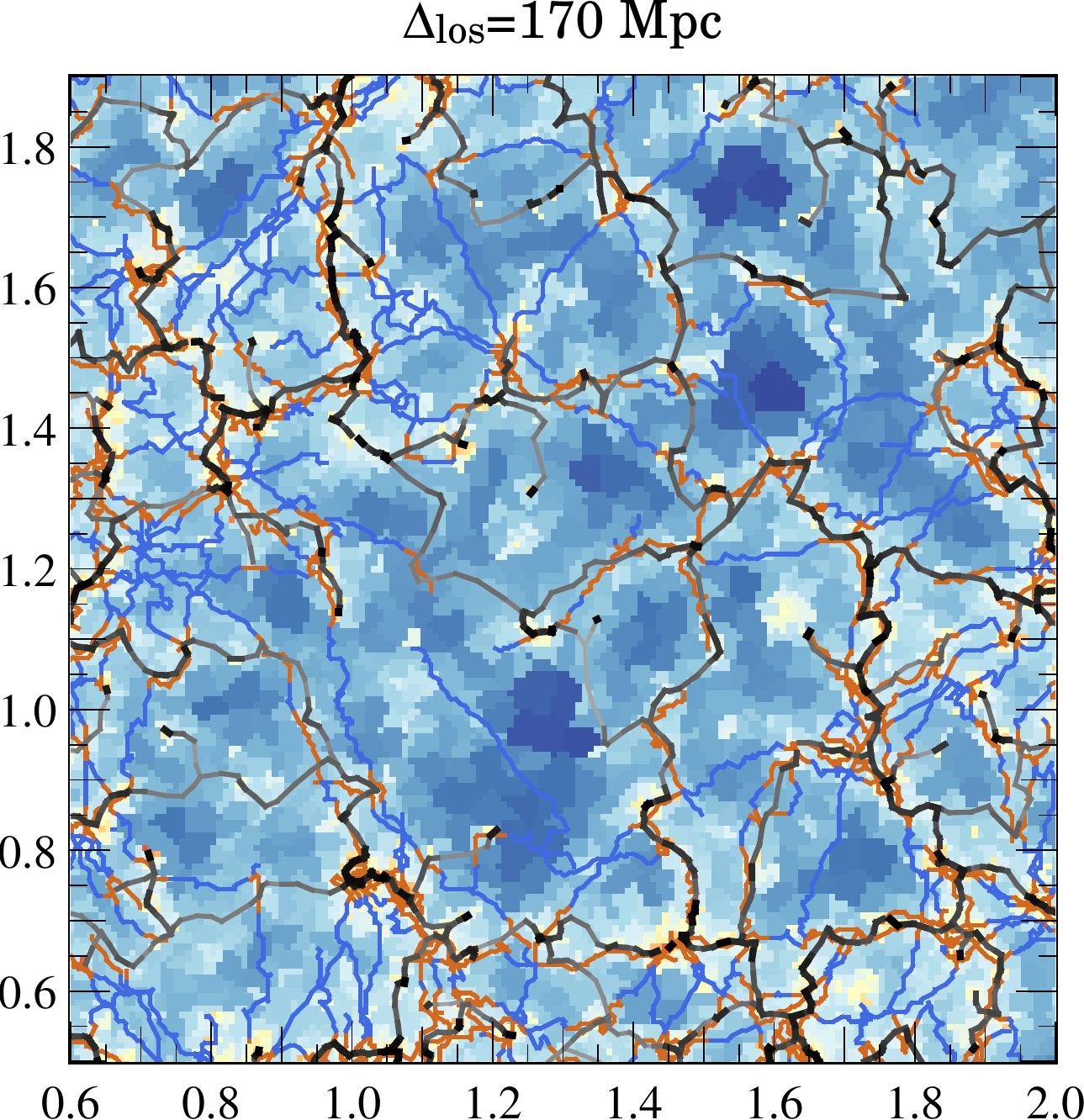} 
 \caption{Skeleton reconstruction for different thicknesses ($\Delta_{\rm los}$) in the {\sc Horizon-AGN} simulation around $z\sim0.59$. The background density is computed from the Delaunay tessellation in two dimensions. The orange/blue skeleton is {\sc skl}$_{\rm 3D}^{\rm DM}$, while the grey/black skeleton is {\sc skl}$_{\rm 2D}^{\rm spec}$. The blue segments in {\sc skl}$_{\rm 3D}^{\rm DM}$ are those which do not have a counterpart in {\sc skl}$_{\rm 2D}^{\rm spec}$ within 0.5$\times D_{\rm int}$, where $D_{\rm int}$ is the mean inter-particle distance.  The grey scale for {\sc skl}$_{\rm 2D}^{\rm spec}$ encodes the robustness of the segments. More robust filaments are darker. In overall, the more robust segments in {\sc skl}$_{\rm 2D}^{\rm spec}$ have a counterpart within 0.5$\times D_{\rm int}$. Even with a large thickness, the robust filaments in {\sc skl}$_{\rm 2D}^{\rm spec}$ represent existing three-dimensional filaments.}
\label{fig:vis3thick}
\end{figure*}
\subsection{Robustness  w.r.t.  photometric redshift accuracy}
\begin{figure}
    \includegraphics[scale=0.41]{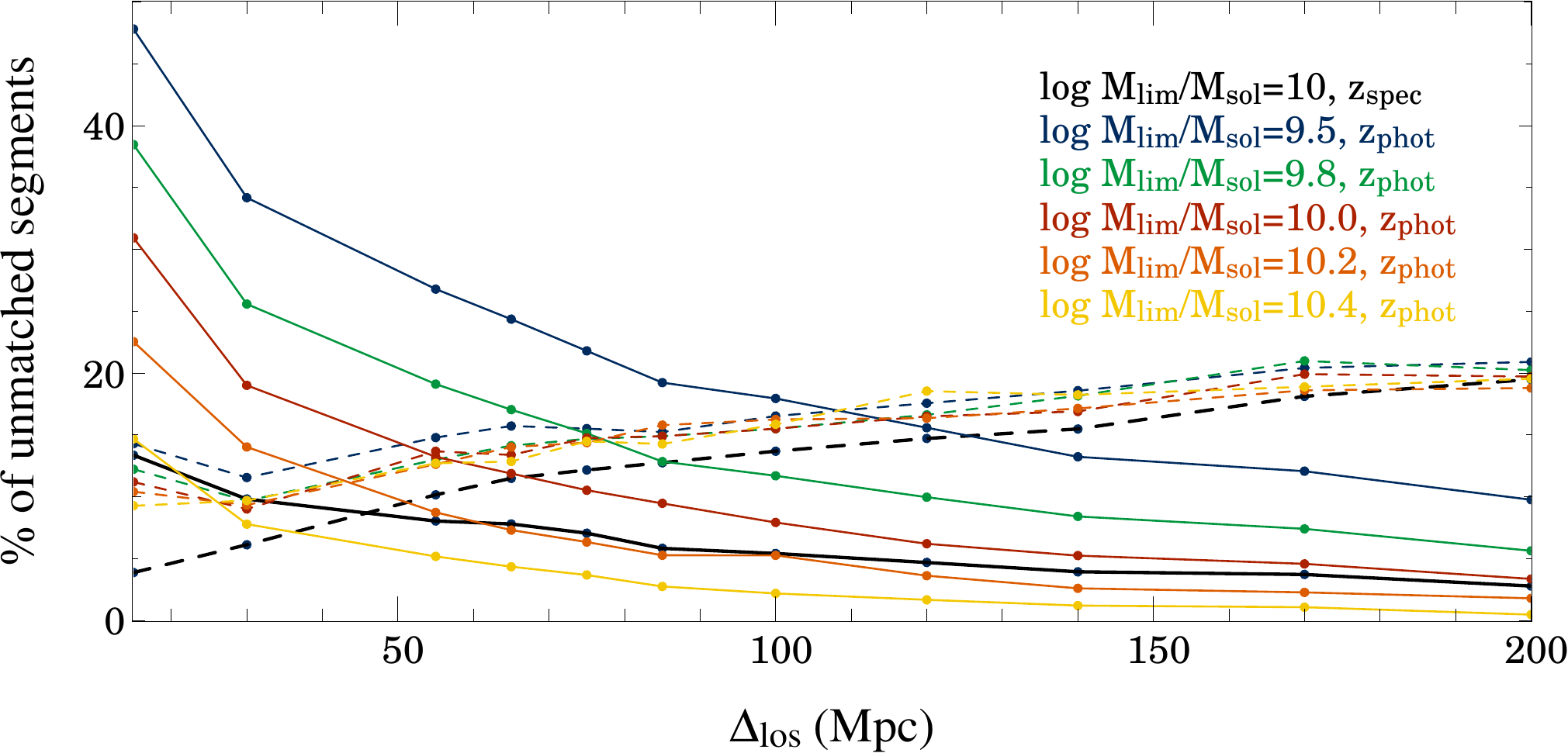} 
 \caption{Percentage of unmatched segments within $D_{\rm int}$ between {\sc skl}$_{\rm 3D}^{\rm DM}$ and {\sc skl}$_{\rm 2D}^{\rm phot}$ as a function of the slice thicknesses used for the computation of {\sc skl}$_{\rm 2D}^{\rm phot}$, for different mass limits (galaxies with lower mass have higher redshift uncertainties) for $0.5<z<0.9$. The solid lines refer to the percentage of {\sc skl}$_{\rm 2D}^{\rm phot}$ segments which do not have a counterpart in {\sc skl}$_{\rm 3D}^{\rm DM}$, and the dashed lines refer to the opposite.
 For comparison, the black line displays the distribution for {\sc skl}$_{\rm 2D}^{\rm spec}$ computed with a mass limit of 10$^{10}{\rm M}_{\odot}$.}
\label{fig:distanceslice}
\end{figure}

We found (Figure~\ref{fig:distanceDMGal}) that decreasing the mass limit of the sample is desirable to bring a better agreement between the galaxy and dark matter skeletons. At fixed mass limits, the optimal choice of the slice thickness broadly stands between 60 and 100 Mpc (Figure~\ref{fig:distanceslice}), although increasing the slice thickness is possible at price of missing a larger number of three-dimensional filaments. Let us now consider how the accuracy of the reconstruction is affected by the redshift uncertainties of the sample. This will drive our final choice both for the mass limit and the slice thickness, knowing that lower mass galaxies have higher redshift uncertainties. Figure~\ref{fig:distanceslice} shows the percentage of unmatched segments within $D_{\rm int}$ between {\sc skl}$_{\rm 3D}^{\rm DM}$ and {\sc skl}$_{\rm 2D}^{\rm phot}$ as a function of the slice thicknesses for different mass limits. For comparison, the black line displays the distribution for {\sc skl}$_{\rm 2D}^{\rm spec}$ computed with a mass limit of 10$^{10}{\rm M}_{\odot}$. When the mass limit decreases, more galaxies with high redshift uncertainties are included in the computation. 
It appears therefore that it is not reasonable to use a mass limit of 10$^{9.5}$M$_{\odot}$, as the fraction of unmatched two-dimensional filaments is very consequent (always greater than 20\% for thickness lower than 100 Mpc). Increasing the mass limit to 10$^{10}$M$_{\odot}$ brings this fraction  below 10\%, while the fraction of unmatched three-dimensional segments does not increase significantly. Given the redshift uncertainties of COSMOS2015, we finally choose to work with galaxies more massive than 10$^{10}$M$_{\odot}$, which guarantees both that the galaxy skeleton is not strongly biased compared to the dark matter one, and that the fraction of unmatched two-dimensional segments is not too high. Within the allowing thickness range, we pick the value of 75 Mpc: it ensures that all the galaxies in the sample have their redshift uncertainties (2$\times$1$\sigma$) lower than the thickness of the slice. 
\section{Filaments or nodes gradients?}
\label{Ap:robdens}
Let us finally  assess the robustness of the gradients towards  filaments after removing  the nodes. 
Nodes do indeed also drive mass gradients. These gradients could be either isotropic around the overdensity, or anisotropic along the filament, or  a combination of both. In the first case, gradients towards nodes could mimic gradients towards filaments, even if they play no role in driving them. 
\begin{figure*}
   \includegraphics[scale=0.38]{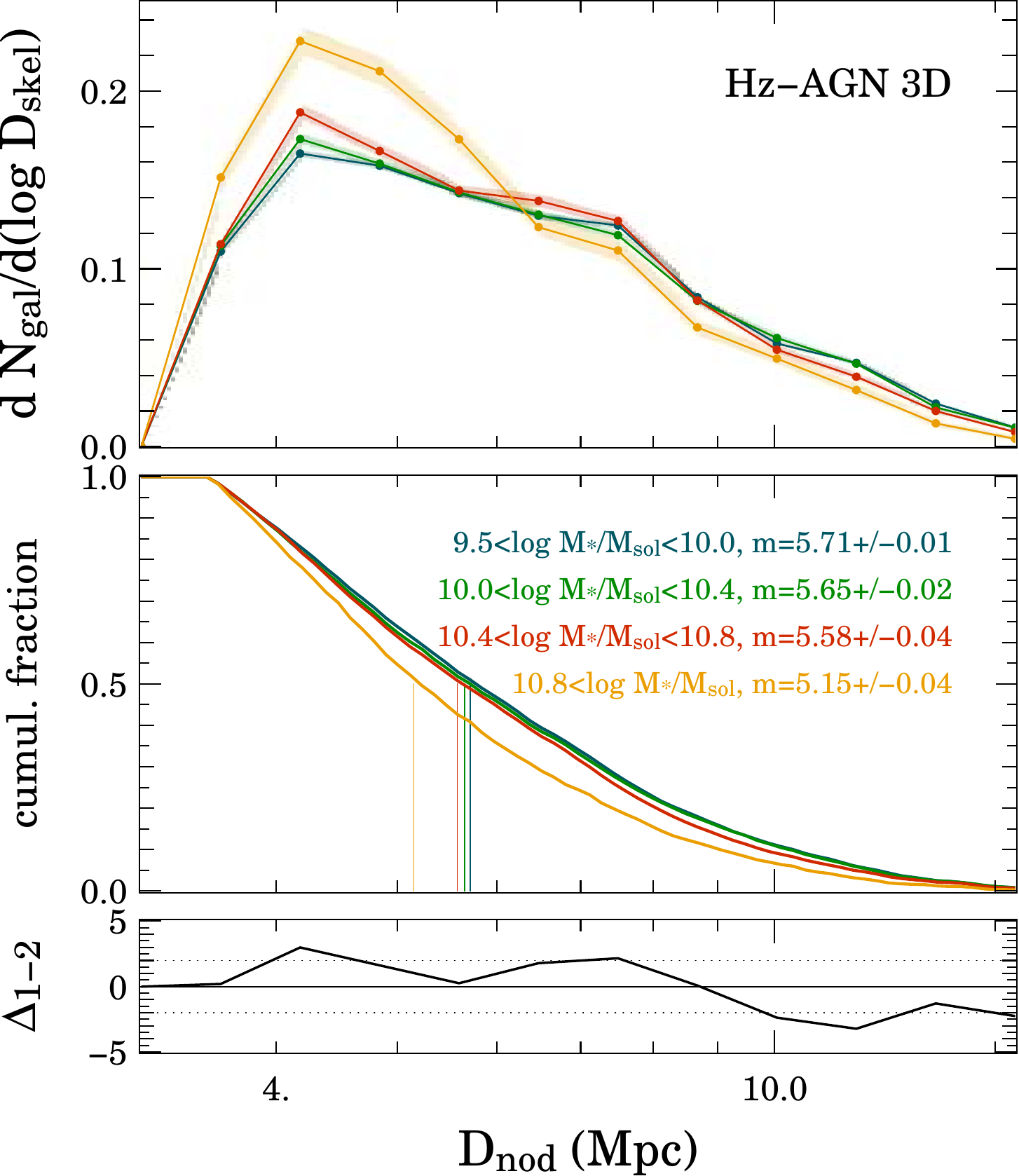} 
  \hspace{0.1cm}
     \includegraphics[scale=0.38]{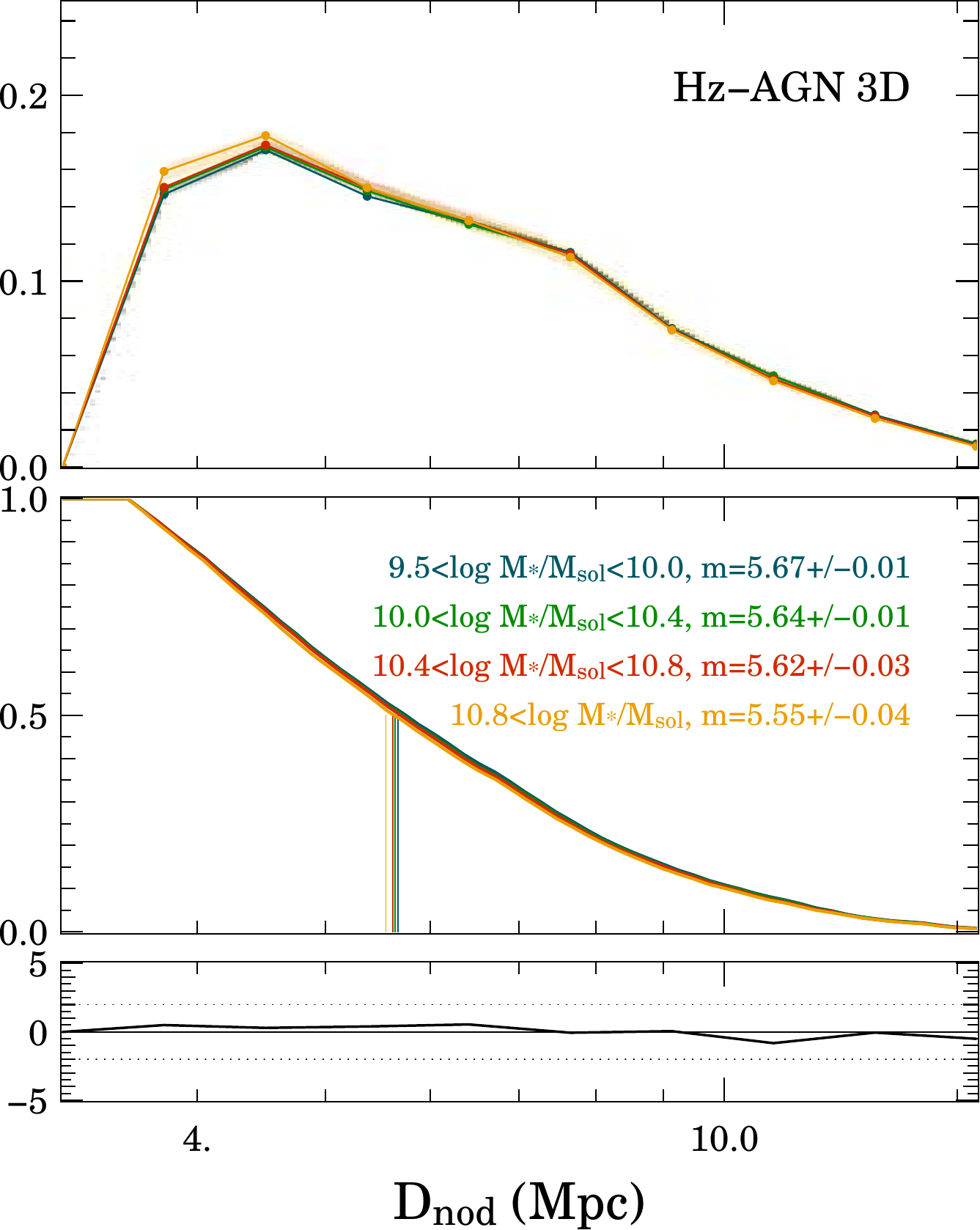} 
       \hspace{0.1cm}
     \includegraphics[scale=0.38]{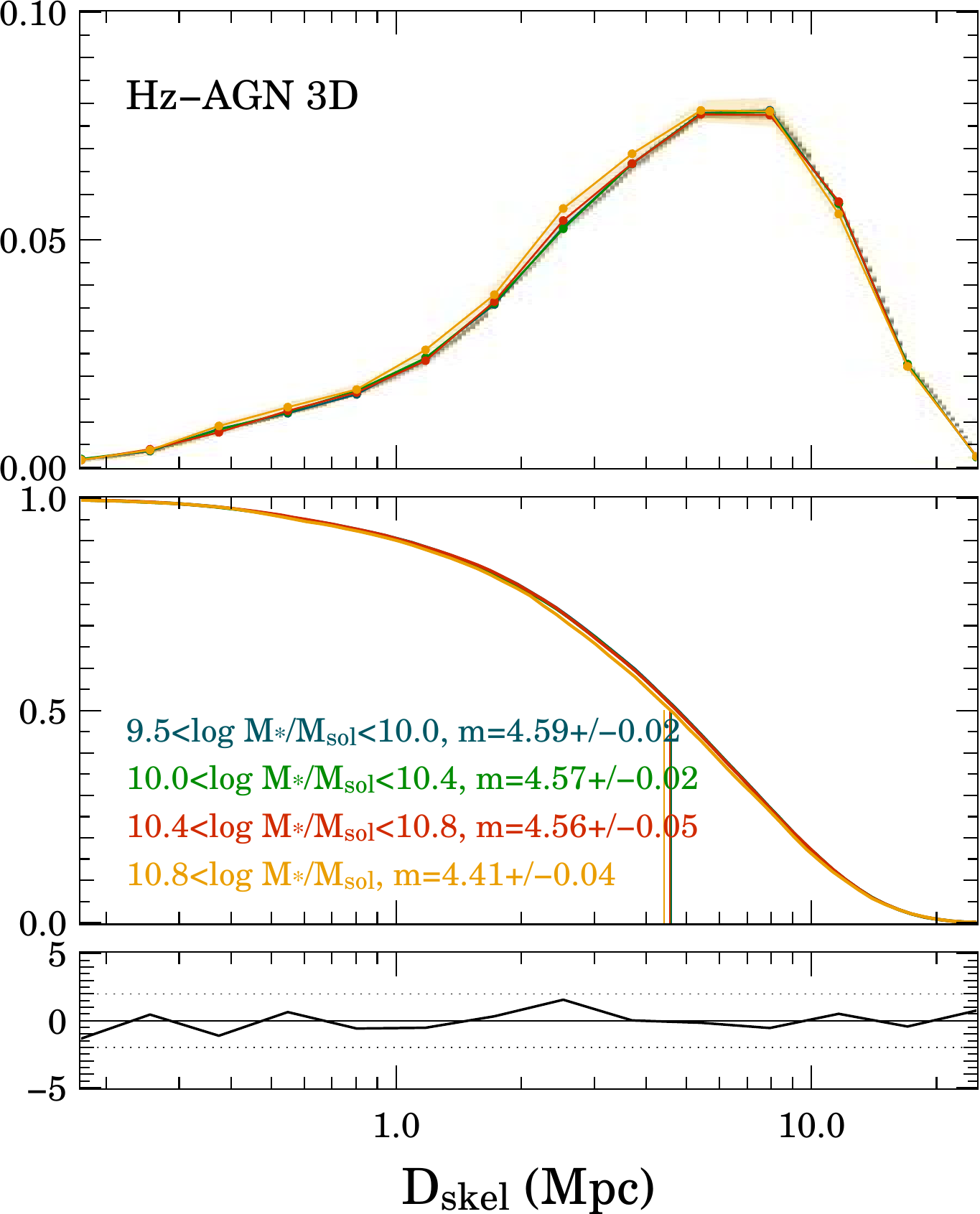}
 \caption{Differential and cumulative distributions displaying mass gradients towards nodes  without (\textit{left}) and with reshuffling (\textit{middle}) galaxy masses with respect to their distance to the nodes but preserving their distances to the filament. The \textit{right} panel displays mass gradients towards filaments with reshuffling galaxy masses with respect to their distance to filaments but preserving their distances to the node.  The signal is computed for $0.5<z<0.9$ in the {\sc Horizon-AGN} simulated ligthcone in three dimensions using {\sc skl}$_{\rm 3D}^{\rm DM}$. Vertical lines indicate the median of each distribution. The bottom row shows the residual between the two intermediate mass bins ($10.4<{\rm log} M_{*}/M_{\odot}<10.8$ and $10<{\rm log} M_{*}/M_{\odot}<10.4$).
 }
\label{fig:weight3d}
\end{figure*}
To probe independently  gradients towards nodes and gradients towards filaments,  we will show in the following that we are able to cancel gradients towards nodes without canceling gradients towards filaments, and vice-versa. 
For each galaxy, the distance to the closest node is defined along the filament (i.e. the distance to the closest node is taken as that of its closest segment in the skeleton\footnote{The distance to the node is chosen to be that along the filament instead of the (euclidian) shortest distance to the node, because pure gradients towards filaments could otherwise also mimic gradients towards nodes: galaxies closer to  filament would be also on average geometrically closer to the nodes. This cannot be the case if the distance to the node  is taken as the one computed along the filament.}). \\
As stated in the main text, in order to minimise the effect of the nodes in driving  gradients, we first exclude   from the sample  galaxies which are closer to the nodes than a given distance $d_{\rm min}$. This distance has to be  large enough to effectively mimimise the effect of the nodes, but small enough in order to keep a significant number of galaxies in the sample.  We find that $d_{\rm min}=3.5$ Mpc in three dimensions and $d_{\rm min}=0.8$ (projected) Mpc in two dimensions offer a good compromise.
The left panel of Figure~\ref{fig:weight3d} shows  the mass gradients towards nodes  in three dimensions after excluding the regions around nodes. Even after excluding a large region around nodes, significant gradients are found in the direction of the nodes. \\
We first try to cancel these gradients by preserving the gradients towards filaments. We therefore reshuffle galaxy masses with respect to their position along the filament (in given bins of  distance to the filament). Hence the gradients towards filaments are preserved by construction (Figure~\ref{fig:distancesTest}). However, we find that, after reshuffling, no significant gradients towards nodes are measured (middle panel of Figure~\ref{fig:weight3d}): we are able to cancel gradient towards nodes without cancelling gradients towards filaments. In a second step we try to cancel filament  gradients by preserving the gradients towards nodes. We therefore reshuffle galaxy masses with respect to their distance to the filaments (in given bins of  distance to the nodes). Hence the gradients towards nodes are preserved by construction. However, we find that, after reshuffling, no significant gradients towards filaments are measured (right panel of Figure~\ref{fig:weight3d}): we are able to cancel gradient towards filaments without cancelling gradients towards nodes. This would not have been possible without excluding galaxies within $d_{\rm min}$ around the nodes. From these tests we conclude that gradients towards nodes and gradients towards filaments outside these exclusion regions are independent. \\
In two dimensions, it is not straightforward \textit{a priori} that the nodes extracted from the two-dimensional skeleton are reliable tracers of their three dimensional counterparts. In the observed dataset we have only access to the two-dimensional nodes, but in the simulation, we  also have access to the projected three-dimensional nodes. We use first the mock catalogue from the simulation to exclude regions around two-dimensional nodes and to reshuffle the position of the galaxies along the filaments as it would be possible to do with the observed catalogue. We compute then the distribution  of galaxy distances by bin of mass around the projected three-dimensional nodes, in order to check that this method effectively cancel the signal.   Figure~\ref{fig:weight2d} shows  the mass gradients towards projected three-dimensional nodes, after excluding the regions around two-dimensional nodes, before and after the reshuffling along the two-dimensional filaments. After reshuffling, no significant gradients towards three-dimensional nodes are found, while the gradients towards filaments  are preserved by construction (Figure~\ref{fig:massdistance2filament}). Similarly to the three-dimensional case, we also check that we can suppress gradient towards filaments but preserving those towards nodes. We therefore reshuffle galaxy masses with respect to their distance to the filaments in given bins of distance to the nodes. This operation partially preserves gradients towards three-dimensional nodes (as two-dimensional and projected three-dimensional nodes are not completely identical). However, it completely cancels gradients towards filaments. We conclude that gradients towards projected three-dimensional nodes and gradients towards two-dimensional filaments are independent.
\begin{figure*}
   \includegraphics[scale=0.412]{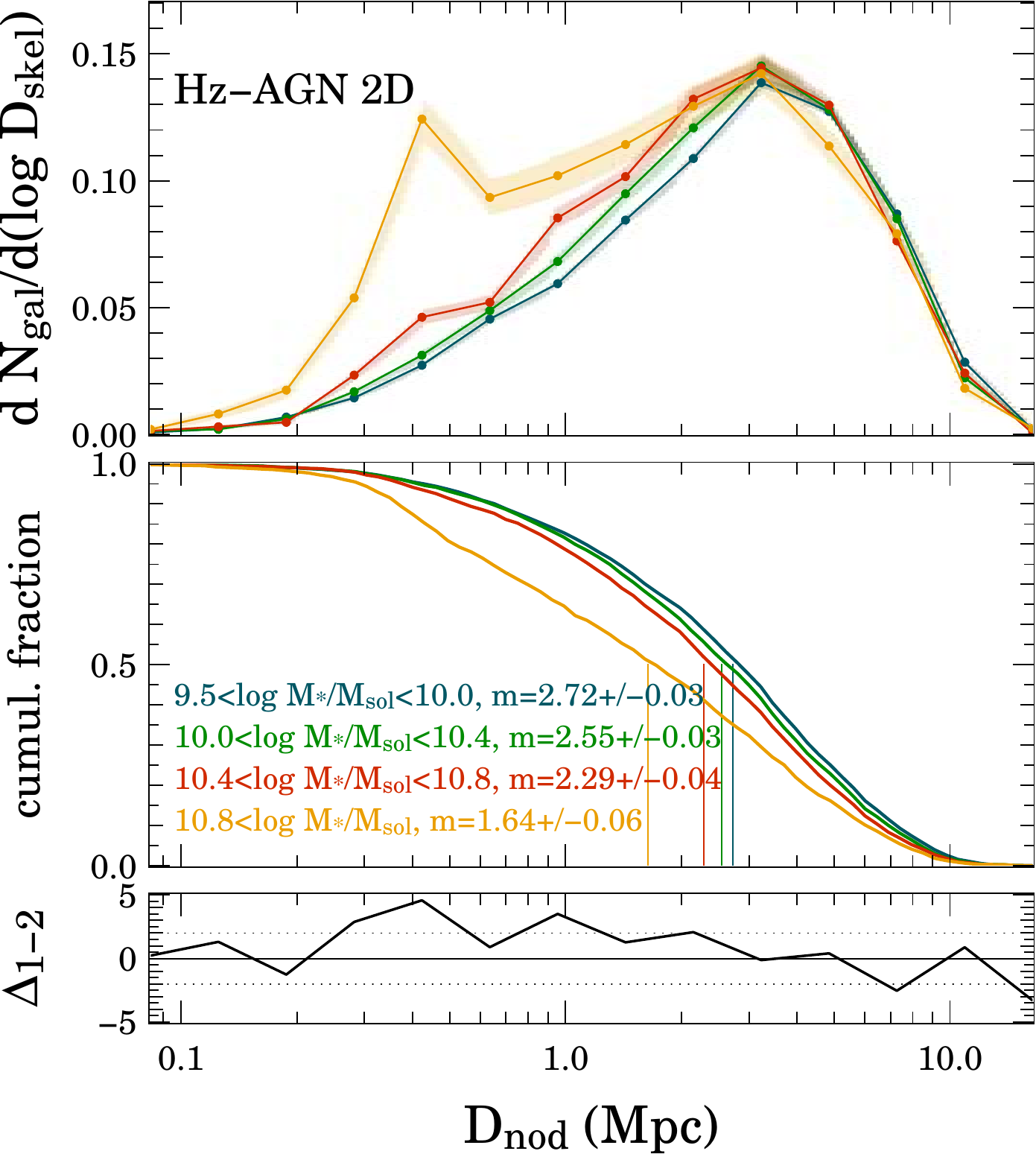} 
  \hspace{0.1cm}
     \includegraphics[scale=0.38]{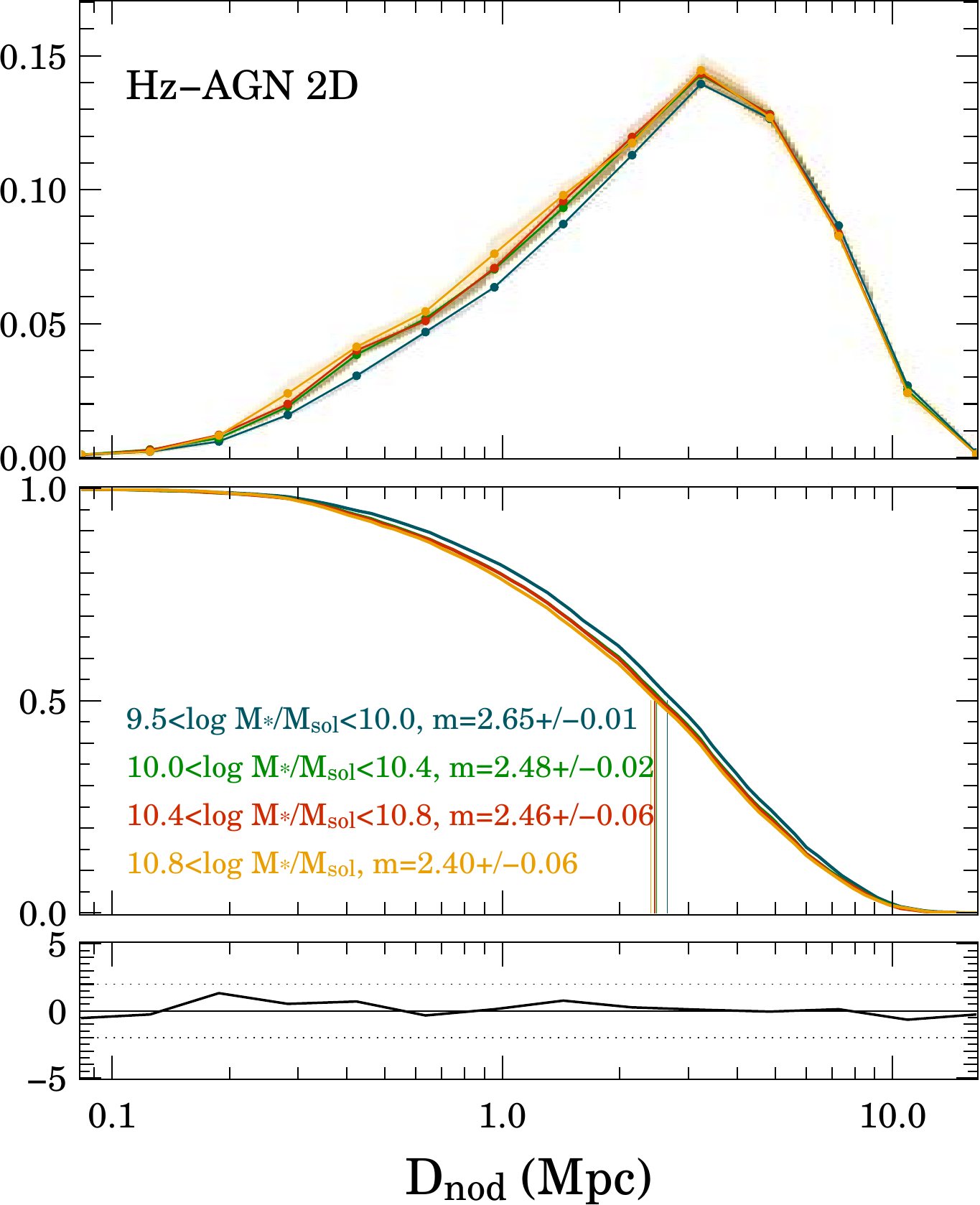} 
       \hspace{0.1cm}
     \includegraphics[scale=0.38]{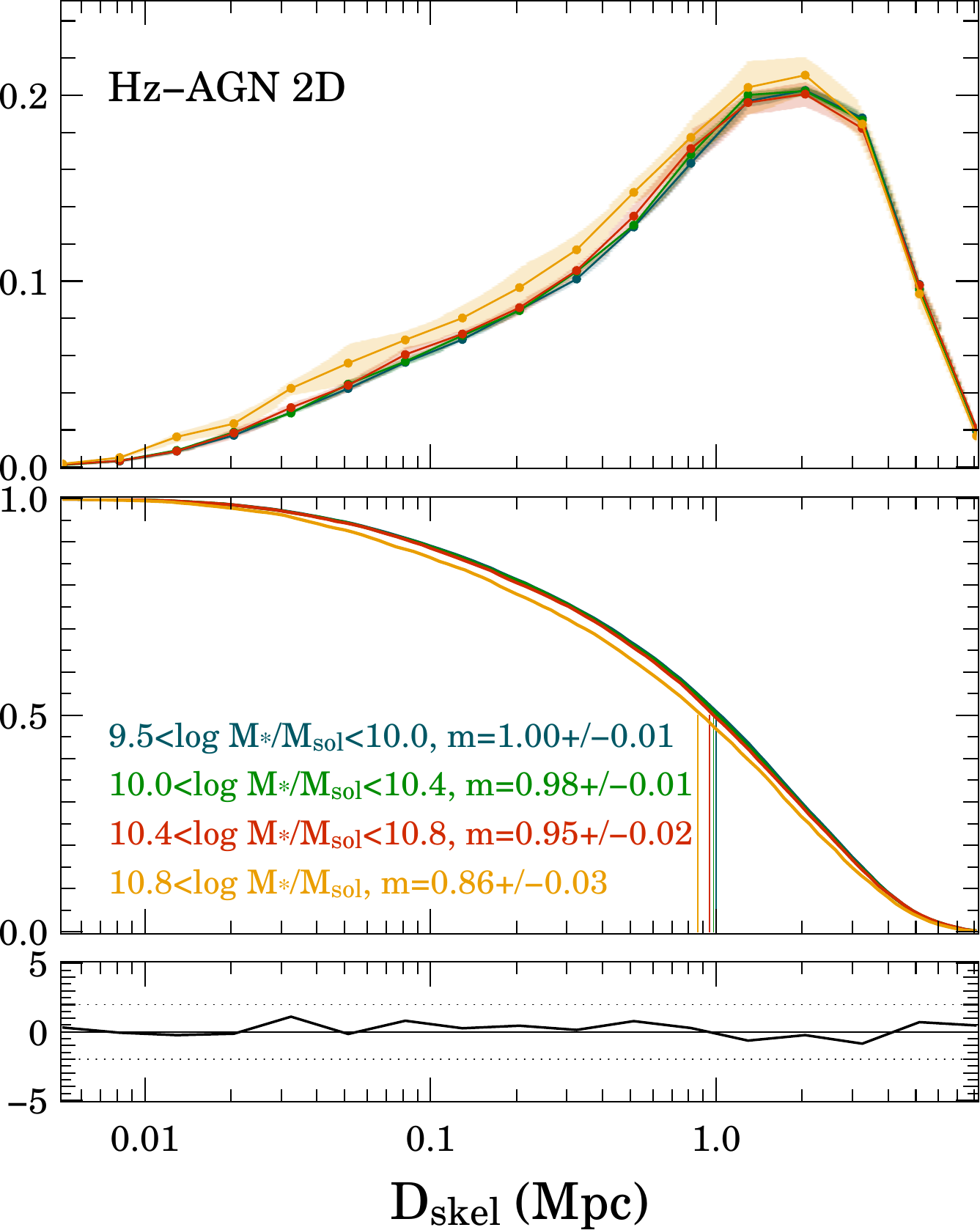}
      \caption{      Differential and cumulative distributions displaying two-dimensional mass gradients towards projected three-dimensional nodes  without (\textit{left}) and with reshuffling (\textit{middle}) galaxy masses with respect to their distances to the nodes but preserving their distances to the filament. The \textit{right} panel displays mass gradients towards filaments with reshuffling galaxy masses with respect to their distances to filaments but preserving their distances to the nodes.
  The signal is computed in the {\sc Horizon-AGN} simulated ligthcone in two dimensions, but nodes are identified in {\sc skl}$_{\rm 3D}^{\rm DM}$ for $0.5<z<0.9$. Vertical lines indicate the median of each distribution. The bottom row shows the residuals between the two intermediate mass bins ($10.4<{\rm log} M_{*}/M_{\odot}<10.8$ and $10<{\rm log} M_{*}/M_{\odot}<10.4$).}
\label{fig:weight2d}
\end{figure*}

\end{document}